\documentclass[mnsc,nonblindrev]{informs3-YCC} %

\OneAndAHalfSpacedXI %

\usepackage{longtable}
\usepackage{booktabs}
\usepackage{mathrsfs}
\usepackage{placeins}
\usepackage[]{multirow}
\usepackage{amsmath}
\usepackage{breqn}
\usepackage{bbm}
\usepackage{natbib}
\usepackage{physics}
\usepackage{pdfpages}
\usepackage{endnotes}
\usepackage{bm}
\usepackage{mathrsfs}
\usepackage{comment}
\usepackage{algorithm}
\usepackage{algorithmic}
\usepackage{hyperref}
\hypersetup{hidelinks}
\usepackage{longtable}
\usepackage{dcolumn}
\usepackage{ctable}
\usepackage{multirow}
\usepackage{subcaption}
\usepackage{xr-hyper} 
\usepackage{hyperref} 

\usepackage{multibib}
\usepackage{makecell}

\theoremstyle{definition}

\def\Ab{\mathbf{A}}

\def\oneb{\mathbf{1}}
\def\Bb{\mathbf{B}}
\def\Eb{\mathbf{E}}

\def\ub{\mathbf{u}}

\def\vb{\mathbf{v}}

\def\bb{\mathbf{b}}
\def\xb{\mathbf{x}}
\def\zb{\mathbf{z}}

\def\zerob{\mathbf{0}}

\def\tb{\mathbf{t}}
\def\qb{\mathbf{q}}
\def\hb{\mathbf{h}}

\def\Pbb{\mathbb{P}}

\def\Rbb{\mathbb{R}}

\def\Bcal{\mathcal{B}}

\def\Lcal{\mathcal{L}}
\def\Acal{\mathcal{A}}
\def\Ccal{\mathcal{C}}

\def\Scal{\mathcal{S}}

\def\Rcal{\mathcal{R}}
\def\Ucal{\mathcal{U}}
\def\Tcal{\mathcal{T}}
\def\Lcal{\mathcal{L}}

\def\Mcal{\mathcal{M}}

\def\RBM{\text{RBM}}

\def\CSM{\text{CSM}}

\def\Rev{\text{Rev}}

\def\alphab{\boldsymbol{\alpha}}

\def\chib{\boldsymbol{\chi}}
\def\pib{\boldsymbol{\pi}}
\def\lambdab{\boldsymbol{\lambda}}

\def\alphab{\boldsymbol{\alpha}}

\def\mub{\boldsymbol{\mu}}

\newcommand{\YC}[1]{{\color{black} #1}}

\newcommand{\dmr}[1]{{\color{black} #1}}

\newcommand{\YCR}[1]{{\color{blue} #1}}

\newcommand{\YCRminor}[1]{{\color{black} #1}}

\usepackage{natbib}
 \bibpunct[, ]{(}{)}{,}{a}{}{,}%
 \def\bibfont{\small}%
 \def\bibsep{\smallskipamount}%

\TheoremsNumberedThrough     %
\ECRepeatTheorems

\EquationsNumberedThrough    %

\begin{document}

\RUNAUTHOR{Akchen and Mitrofanov}

\RUNTITLE{Consider or Choose?}

\TITLE{Consider or Choose? The Role and Power of Consideration Sets}

\ARTICLEAUTHORS{%
\AUTHOR{Yi-Chun Akchen}
\AFF{School of Management, University College London, London E14 5AB, United Kingdom, \EMAIL{\tt yi-chun.akchen@ucl.ac.uk}} %
\AUTHOR{Dmitry Mitrofanov}
\AFF{Carroll School of Management, Boston College,  \EMAIL{\tt dmitry.mitrofanov@bc.edu}}
} %

\ABSTRACT{Consideration sets play a crucial role in discrete choice modeling, where customers often form consideration sets in the first stage and then use a second-stage choice mechanism to select the product with the highest utility. While many recent studies aim to improve choice models by incorporating more sophisticated second-stage choice mechanisms, this paper takes a step back and goes into the opposite extreme. We simplify the second-stage choice mechanism to its most basic form and instead focus on modeling customer choice by emphasizing the \emph{role} and \emph{power} of the first-stage consideration set formation. To this end, we study a model that is parameterized solely by a distribution over consideration sets with a bounded rationality interpretation. Intriguingly, we show that this model is characterized by the axiom of symmetric demand cannibalization, enabling complete statistical identification. The latter finding highlights the \emph{critical role} of consideration sets in the identifiability of two-stage choice models. We also examine the model’s implications for assortment planning, proving that the optimal assortment is revenue-ordered within each partition block created by consideration sets. Despite this compelling structure, we establish that the assortment problem under this model is NP-hard even to approximate, highlighting how consideration sets contribute to nontractability, even under the simplest uniform second-stage choice mechanism. Finally, using real-world data, we show that the model achieves prediction performance comparable to other advanced choice models. Given the simplicity of the model's second-stage phase, this result showcases the enormous \emph{power} of first-stage consideration set formation in capturing customers' decision-making processes.
}%

 \HISTORY{First version: February 8, 2023; Second version: June 12, 2024; This version: February 19, 2025}

\maketitle

\section{Introduction}
\label{sec:intro}

The rise of the digital economy and technological advancements has fundamentally changed how we shop and make purchasing decisions. With an overwhelming variety of products and services available online, consumers, when making a choice, must navigate an enormous amount of information, including detailed product descriptions, customer reviews, and ratings. In an ideal scenario, fully rational individuals would make their purchase decisions by thoroughly assessing the features of each alternative, calculating its utility, and selecting the option with the highest utility, provided they have unlimited time and resources to perform such an evaluation. In reality, individuals have physical and cognitive limitations and thus only consider a subset of the available alternatives. Economists and psychologists term such smaller sets of alternatives as ``consideration sets'' \citep{wright1977phased} or ``evoked sets'' \citep{howard1969theory,brisoux1981evoked}. The notion of consideration sets has been well-documented in the marketing literature when studying consumer behavior \citep{hauser2014consideration} and various heuristics have been proposed to model the formation of the consideration sets, such as screening rules based on product prices or other features \citep{pras1975comparison,gilbride2004choice,jedidi2005probabilistic}.
The concept of consideration sets is further supported by the psychology literature, which questions consumers' ability to consistently evaluate every product within the offer set \citep{miller1956magic,hauser1990evaluation,iyengar2000choice}.

\begin{figure}
	\centering
	\includegraphics[scale=0.4]{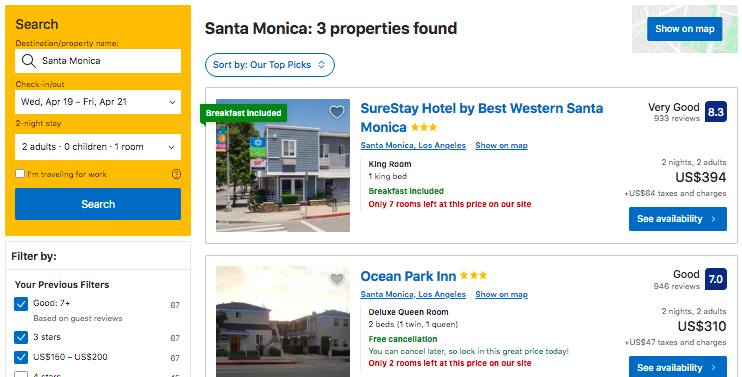}
	\caption{An example of a two-stage decision-making process, where a customer uses screening rules to form a consideration set.}
	\label{fig:booking_hotel}
\end{figure}

In fact, the concept of consideration set formation goes beyond merely interpreting consumer behavior; it also has substantial practical implications, particularly in developing more comprehensive and accurate choice models.
To this end, the seminal work by \cite{hauser1978testing} uses a goodness-of-fit statistic to demonstrate that consideration sets can explain nearly three-quarters of the variation in choice data, whereas a logit model, which is based solely on consumer preferences, explains only a quarter. This insight has contributed to the widespread adoption of \emph{two-stage} choice models that integrate consideration sets into consumers' decision-making processes and many papers provide evidence of their superior prediction performance \citep{silk1978pre,hauser1984application,gensch1987two}. Within this two-stage framework, consumers initially form consideration sets, often using screening rules and simple heuristics \citep{hutchinson2005simple}. In the second stage, they apply a \emph{choice mechanism} to select and purchase the product that maximizes their utility from the options in the consideration set \citep{ben1995discrete,shocker1991consideration}. Specifically, we can illustrate two-stage choice models with an example of online hotel booking, where customers face an abundance of alternatives and information. On a travel website, they encounter details such as the hotel's location, services, amenities, quality, view, ratings, photos, and reviews, among other characteristics. Analyzing all this information for every hotel within a limited time is impractical. Instead, customers might use simple heuristics to narrow their options. Figure~\ref{fig:booking_hotel} shows a customer applying screening criteria -- rating (7+), price (\$150-\$200 per night), and three-star quality -- to reduce five hundred options to a consideration set of three alternatives. A customer then is expected to use a choice mechanism to evaluate these three considered alternatives and select the one offering the highest utility.

Given the evidence showing the importance of capturing customers' consideration set formation, it is unsurprising that two-stage choice models incorporating this process have gained significant popularity over the past several decades. Many of these models leverage advanced choice mechanisms to further enhance their predictive accuracy and data-fitting capabilities \citep{aouad2021assortment, jagabathula2022demand}. A common feature of these models is their foundation in nonparametric choice modeling principles, drawing inspiration from machine learning models and algorithms. Their primary characteristic is an advanced and refined second-stage choice mechanism which can be as sophisticated as the choice mechanisms proposed by \cite{farias2013nonparametric} and \cite{chen2022decision}. In this paper, we adopt an opposing approach: instead of exploring more complex {second-stage choice mechanisms}, we take a step back to examine the fundamental properties of decision-making processes driven solely by consideration sets, where the second-stage choice mechanism is kept as simple as possible. More specifically, we examine a class of nonparametric models defined solely by a distribution over consideration sets, which we refer to as the \emph{consideration set model (CSM)}. Through the lens of this generic consideration-based model, we examine the role and power of consideration sets in choice model representation, demand cannibalization, choice prediction, and assortment optimization. We also discuss the implications for other choice models that incorporate consideration set structures. Specifically, we make the following contributions:

\begin{itemize}
\item \emph{Considering rather than choosing.} We study a nonparametric choice model which is fully characterized by a distribution over the consideration sets. This distribution can be viewed as the probability mass of various customer types in the market, or it can indicate the stochasticity in how a customer forms a consideration set. After forming a consideration set, we assume that the customer does not rely on a particular choice mechanism to select the ``best'' product among those considered, but instead chooses any of them in a uniformly random way. Overall, this {consideration set model} represents a \emph{simplified} version of the general class of two-stage choice models, as it does not incorporate a specific second-stage choice mechanism.

\item \emph{The role of consideration sets in choice model identifiability.} Surprisingly, we show that the consideration set model can be fully identified from the collection of choice probabilities. We further provide a closed-form expression on how one can compute the model parameters by means of the choice probabilities. In other words, the consideration set model can be uniquely reconstructed from its observed choice probabilities on assortments, making it one of the identifiable choice models with the highest degree of freedom in the existing literature.
This finding on identifiability conveys an important insight: if a two-stage choice model cannot be uniquely identified -- having several solutions that fit the data equally well even with unlimited choice data -- it is not solely because of the first-stage consideration process. In other words, consideration sets themselves do not play a \emph{critical role} in choice model nonidentifiability. %

\item \emph{The role of the consideration sets in modeling demand cannibalization.} 
We also characterize the necessary and sufficient
conditions for the choice probabilities to be consistent with our model. To this end, we demonstrate that with a mild condition imposed on the no-purchase probabilities, a choice model is a consideration set model if, and only if, the introduction of a product symmetrically reduces the market share of another product in each assortment. We refer to this condition as the \emph{symmetric demand cannibalization axiom}. More specifically, this axiom is based on the assumption that for any two products $i$ and $j$, the decrease in demand for product $i$ when product $j$ is introduced is the same as the decrease in demand for product $j$ when product $i$ is introduced. This axiomatic characterization highlights a fundamental limitation of any choice model based solely on consideration set formation, as consideration sets alone are only capable of capturing symmetric demand cannibalization. It also demonstrates that the choice mechanism, rather than the consideration set formation in the first stage, plays a significant role in representing demand cannibalization comprehensively.

\item \emph{The critical role of consideration sets in driving the intractability of the assortment problem.} We also explore the impact of consideration sets on downstream applications. In particular, we focus on assortment optimization problems.
We first provide a precise characterization of the optimal assortment, showing that it is revenue-ordered \YCRminor{within each block belonging to the partition of the product universe induced by the consideration sets}. This structural property leads to a polynomial-time optimal algorithm when the number of customer types is fixed. However, even with this compelling optimal assortment structure in place, the problem of assortment optimization under the general consideration set model remains NP-hard to approximate within a factor of $O(n^{1/2 - \epsilon})$ for any $\epsilon>0$.
This result implies that accounting for consideration sets in choice models inherently complicates downstream applications, even if the second-stage choice mechanism is significantly simplified.
	
\item \emph{Empirical study: the prediction power of consideration sets.} Finally, we conduct numerical experiments based on a real-world dataset \citep{bronnenberg2008database} to show the prediction power of consideration sets when modeling consumer choice. We propose an estimation methodology to calibrate the consideration set model and empirically evaluate its predictive performance. Despite being a special case of the ranking-based and the mixed multinomial logit (mixed MNL) models, the consideration set model provides a predictive performance that is highly competitive with these more general models. This emphasizes that consideration sets \emph{alone} can be responsible for the strong predictive performance of the two-stage consider-then-choose models in real-world prediction tasks.

\end{itemize}

\subsection{Related Literature on Consideration Sets}
\label{subsec:consideration_set_literature}

The concept of consideration sets originated in the fields of marketing and psychology and has been studied extensively for several decades, with numerous studies exploring how consumers form and use consideration sets in their purchasing decisions. For a more detailed overview, we refer readers to the survey papers by \cite{roberts1997consideration} and \cite{hauser2014consideration}. Overall, it is widely accepted in the literature that consumers generally make decisions through a two-stage process \citep{swait1987incorporating}. Specifically, consumers first consider a subset of products (i.e., form a consideration set) and then select one item from that set to purchase. In fact, abundant empirical evidence supports the concept of consideration sets. For example, \cite{hauser1978testing} adapts an information-theoretic viewpoint and shows that a model which incorporates the consideration set concept can explain up to 78\% of the variation in the customer choice. Then, \cite{hauser1990evaluation} empirically investigate the consideration set size for various product categories and report that the mean consideration set size, which is the number of brands that a customer considers before making a choice, is relatively small: 3.9 for deodorant, 4.0 for coffee, and 3.3 for frozen dinner. 
To this end, our numerical analysis in Section~\ref{subsec:IRI_performance}, examining the size of consideration sets across various product categories using real-world grocery sales data, is consistent with the finding that consideration set sizes tend to be relatively small.

Over the past few decades, numerous theories and models have been developed to explain how consumers form their consideration sets. From a decision theory perspective, consumers are expected to continue searching for new products as long as the utility (or satisfaction) gained from finding a better option exceeds the cost of the search \citep{ratchford1982cost,roberts1991development}. Additionally, cognitive limitations may lead consumers to rely on simple heuristic rules to construct their consideration sets. These rules include elimination by aspects \citep{tversky1972elimination}, conjunctive and disjunctive rules \citep{pras1975comparison,brisoux1981evoked,laroche2003decision,gilbride2004choice,jedidi2005probabilistic}, and compensatory rules \citep{keeney1993decisions,hogarth2005simple}.
In this paper, we take a more general and flexible perspective on the formation of consideration sets. Rather than focusing on specific rules or heuristics that consumers use to construct their consideration sets, we model the possibility that a consideration set sampled by a customer could be any subset of products. To this end, our paper is related to the work of \cite{jagabathula2022demand}, where the authors model customer choice using the consider-then-choose (CTC) model, which is parameterized by the joint distribution over consideration sets and rankings. Different from the CTC model, which uses a mixture of rankings as its second-stage choice mechanism, our model employs the simplest possible choice mechanism. Consequently, it can be viewed as a special case of the model proposed by \cite{jagabathula2022demand}. Furthermore, our model is also a special case of the ranking-based model \citep{block1959random,farias2013nonparametric} and the mixed MNL model \citep{train2009discrete} as discussed in Section~\ref{subsec:connection_to_other_models}.

More recently, the concept of a consideration set has been gaining a lot of attention in the field of operations management. \cite{feldman2019assortment} propose an algorithm for determining the optimal assortment, incorporating the unique features of the ranking-based model and the assumption of small consideration sets. Similarly, \cite{aouad2021assortment} examine assortment optimization under the CTC choice model, where customers select the product with the highest rank from the intersection of their consideration set and the offered assortment. In \cite{wang2018impact}, the authors present a mathematical model that represents the trade-off between a product's expected utility and the search cost associated with it. \cite{mitrofanov2024choice} investigate the assortment optimization problem under a two-stage choice model, where customers initially use non-parametric dominance relationships to narrow down their options and then make a selection from the shortlisted products using the multinomial logit (MNL) model. 
\cite{chitla2023customers} use the consideration set concept in order to build the structural model and study the multihoming behavior of users in the ride-hailing industry. 
Interestingly, it was shown in the paper by \cite{jagabathula2022demand} that the consideration set approach is particularly advantageous in the online platform setting or in the grocery retail setting where we might expect the noise in the sales transaction data because of the stockouts. To this end, there is a lot of evidence in the literature that online platforms might not have real-time information on the product availability in the grocery retail stores which could be the major cause for the stockouts \citep{knight2022disclosing} despite the AI-enabled technology to alleviate this problem \citep{knight2023impact,kim2024ai}.

\section{Model Description}
\label{sec:model}
In this section, we begin with an overview of choice modeling before introducing the \emph{consideration set model}. Then, we connect this model to other choice models in the economics, marketing, and operations literature.

\subsection{Preview of Choice Modeling and Motivation}
\label{subsec:choice_modeling_preview}
We begin this subsection by introducing several notations. We consider a universe $N \equiv \{ 1,2,\ldots,n \}$ of $n$ products. Each assortment or offer set $S$ is a subset of $N$, i.e., $S \subseteq N$. When a set of products $S$ is offered, a customer or an agent chooses either a product in~$S$ or the ``default'' option~$0$. Depending on the context, the default option can be the no-purchase option or any product outside the universe $N$. Following the standard assumption in the literature, the default option is always assumed to be available to customers. Then, to simplify notation, we denote $N^+ = N \cup \{ 0 \}$ and $S^+ = S \cup \{ 0 \}$ for any $S \subseteq N$.
Throughout the paper, $\mathbb{I} \left[ A \right]$ or $\mathbb{I}_{A}$ denotes the indicator function \YCRminor{that} is equal to 1 if condition $A$ is satisfied and 0, otherwise.

A choice model can be described by a mapping or function $\mathbb{P}$ that takes as input an assortment $S$ and outputs a probability distribution over the elements of $S^+$. This probability distribution represents the likelihood of each product in the assortment being chosen by a consumer. 
Specifically, $\mathbb{P}_j(S)$ represents the probability that a customer selects product $j$ from the offer set $S$, while $\mathbb{P}_0(S)$ represents the probability of choosing the default option. Note that when the offer set is empty, the default option is always chosen, i.e., $\mathbb{P}_0(\emptyset)=1$. More formally, a choice model specifies the choice probabilities $\{\mathbb{P}_j(S) \colon j \in S^+, S \subseteq N\}$ that satisfy the standard probability laws: $\mathbb{P}_j(S) \geq 0$ for all $j \in S^+$ and $\sum_{j \in S^+} \mathbb{P}_j(S) = 1$, for all $S \subseteq N$. Note that discrete choice models are widely used to predict consumers' purchase decisions in operations, marketing, and economics research. For further details, we refer readers to \cite{ben1985discrete} and \cite{train2009discrete}. While early research on choice modeling primarily focused on parametric models such as the multinomial logit (MNL) model, the mixed MNL model, and the nested logit model \citep{train2009discrete}, recent years have seen a rapid surge in consumer choice data, driving the development of nonparametric choice models. These models aim to enhance the accuracy of consumer decision predictions and provide greater flexibility in fitting the data \citep{block1959random,farias2013nonparametric,aouad2021assortment,jagabathula2022demand, chen2022decision}. 

More specifically, recent research in nonparametric choice modeling has primarily focused on developing more advanced and sophisticated second-stage choice mechanisms, with or without incorporating a consideration set formation stage, to better fit consumer choice data. In contrast, our paper takes the opposite approach by employing the simplest possible second-stage choice mechanism. This approach enables us to explore the role and power of first-stage consideration set formation and examine how incorporating consideration sets influences the characteristics and applicability of choice models. To this end, we introduce and analyze a model, referred to as the \emph{consideration set model}, which is described in detail below.

\subsection{Consideration Set Model (CSM)}
\label{subsec:model_and_example}
In what follows, we introduce the consideration set model. As elaborated later in Section~\ref{subsec:connection_to_other_models}, this choice model is \emph{not} based on entirely new principles but is instead nested within several well-established choice models. This structure enables us to extend our findings to other choice models that implicitly or explicitly incorporate consideration set structures. In essence, the consideration set model represents a probability distribution over sets of products, with each set $C \subseteq N$ corresponding to a consideration set. Specifically, let $\Ccal$ be a collection of subsets of $N$ and $\lambdab$ is a probability distribution over $\Ccal$ such that $\lambda_C \geq 0$ for all $C \in \Ccal$ and $\sum_{C \in \Ccal} \lambda_C = 1$. Each $C \in \Ccal$ denotes a \emph{consideration set}, which is a subset of $N$. Furthermore, we assume that each consideration set $C$ specifies a set of preference relations between elements in $N^+$ as follows. 
\begin{definition}
    \label{def:preselected_set_preferences}
	Each consideration set $C \in \Ccal$ of the consideration set model $(\Ccal,\lambdab)$ represents a set of preference relations between items in $N^+$ as follows: (a) $i \sim j$, for all $i,j \in C$; (b) $i \succ 0$, for all $i \in C$; and (c) $0 \succ j$, for all $j \neq C$.
\end{definition}

In Definition~\ref{def:preselected_set_preferences}, following the convention, we use $i \succ j$ to denote ``$i$ is preferred to $j$'' and use $i \sim j$ to denote that $i$ and $j$ are equally preferred. Note that we do not need to specify preference relations between the items outside of the consideration set $C$, since those items are dominated by the default (i.e., no-purchase) option $0$ which is always available. However, without loss of generality, one can still assume that $i \sim j$ for all $i, j \notin C$. %

Following the standard interpretation in choice modeling, the consideration set model $(\Ccal,\lambdab)$ can be viewed either as an individual customer's stochastic decision rule or as a representation of customer segments in the market. In the former case, we assume that with probability $\lambda_C$, a customer samples a consideration set $C$ according to distribution $\lambdab$ before making a final choice. In the latter case, each consideration set $C \in \Ccal$ is associated with a customer type, and $\lambda_C$ represents the proportion of customers of type $C$ in the market. Before we show how to compute the choice probabilities under the consideration set model, we further provide an alternative parametrization of the model, which specifies a conditional distribution $\lambdab_{\cdot | S}$ given an offered assortment $S$.

\begin{definition}
	\label{def:random_set_distribution}
	A \emph{conditional set distribution} $\lambdab_{\cdot | S}$ is defined with respect to a distribution $\lambdab$ over subsets in $N$, for every $S \subseteq N$, as follows:
	\begin{align}
		\label{eq:conditional_set_distribution}
		\lambda_{C' | S} = \begin{cases}
			\sum_{C \subseteq \Ccal} \lambda_C \cdot \mathbb{I} \left[ C \cap S = C' \right], & \text{if $C' \subseteq S$ and $C' \neq \emptyset$},\\
			0, & \text{if $C' \not \subseteq  S$ and $C' \neq \emptyset$},\\
		\end{cases}
	\end{align}
	along with $\lambda_{\emptyset \mid S} = 1 - \sum_{C' \subseteq S: C' \neq \emptyset} \lambda_{C' \mid S}$, where $\mathbb{I} \left[ A \right]$ is an indicator function.
\end{definition}

In fact, Equation~\eqref{eq:conditional_set_distribution} specifies the distribution over the sets within the offered assortment $S$. For example, assume that $N = \{ 1,2,3,4,5 \}$ and an assortment $S = \{ 1,2,3 \}$ is offered. Then the consideration set $C = \{ 2,3,4 \}$ is equivalent to the consideration set $C' = C \cap S = \{ 2,3 \}$. Put differently, if a customer who is considering buying products $C = \{ 2,3,4\}$ enters a store that only offers products $S = \{1,2,3\}$, then the customer would only consider buying product $C \cap S = \{ 2,3 \}$, as product $4$, despite being considered, is not offered. Intuitively, Definition~\ref{def:random_set_distribution} specifies that the probability that the customer samples a conditional set $C'$ from the offer set $S$ is the sum of the probabilities of sampling consideration sets $C$ such that $C \cap S = C'$. With the conditional set distribution defined in Definition~\ref{def:random_set_distribution}, there are two ways to describe a consumer's decision-making process when selecting from an assortment $S$. In the first approach, the customer is assumed to sample a consideration set $C \subseteq N$ according to the distribution $\lambdab$ and then make a final choice from the intersection $C \cap S$. Alternatively, the customer can be assumed to sample a conditional set $C' \subseteq S$ directly, based on the conditional set distribution $\lambdab_{\cdot | S}$, and then make a final choice from $C'$. Both approaches lead to identical choice probabilities, making them equivalent.

It is straightforward to verify that the conditional set distribution $\lambdab_{\cdot | S}$ satisfies the standard conditions such as $\lambda_{C'|S} \geq 0$ for all $C' \subseteq N$ and $\sum_{C' \subseteq N} \lambda_{C'|S}=1$. In addition, $\lambdab_{\cdot|S}$ is consistent with the ``monotonicity property'' where for all $S_2 \subseteq S_1 \subseteq N$, we have that $\lambda_{C'|S_2} \ge \lambda_{C'|S_1}$ for all $C' \subseteq S_2$. More specifically, for any $C \subseteq N$, such that $C \cap S_1 = C'$ is satisfied for $C' \subseteq S_2$, we have $C \cap S_2 = C'$. Next, we formally define the probability of choosing an item $j$ from the assortment $S \subseteq N$ under the consideration set model $(\Ccal,\lambdab)$ by means of the conditional set distribution $\lambdab_{\cdot | S}$.

\begin{definition}
	Given a consideration set model $(\Ccal,\lambdab)$, the probability to choose product $j$ from an assortment $S$ is computed as follows:
	\begin{equation}
		\label{eq:choice_prob_via_conditional_distribution}
		\mathbb{P}^{(\Ccal,\lambdab)}_j(S)=\sum_{ C' \subseteq S: j \in C'} \frac{\lambda_{C'|S}}{\abs{C'}}, 
	\end{equation} 
	if $j \in S$  and 0, otherwise.
\end{definition}  

In other words, Equation~\eqref{eq:choice_prob_via_conditional_distribution} implies that when an assortment $S$ is offered, a customer forms a consideration set $C'$ with probability $\lambda_{C' \mid S}$ and, within this set, assigns equal preference to all products (as outlined in preference set (a) in Definition~\ref{def:preselected_set_preferences}) while considering every product preferable to the default (no-purchase) option (as specified in preference set (b) in Definition~\ref{def:preselected_set_preferences}). To this end, we assume that a \emph{customer does not rely on a specific second-stage choice mechanism to determine the ``best'' product from $C'$ and therefore would select any product from $C'$ with equal likelihood} which leads to a factor $\lambda_{C' \mid S} / |C'|$ in the summation.
Recall that we intentionally have this assumption to make the second-stage choice mechanism as simple as possible, allowing us to focus exclusively on analyzing the role and power of the consideration set formation layer in choice modeling.
Alternatively, this assumption can be justified by the bounded rationality of customers who, constrained by cognitive and physical limitations, are unable to differentiate and rank all options within the consideration set \citep{simon1955behavioral}.

As mentioned above, we can also model the customer's decision-making process by first sampling $C$ from $\lambdab$ and then choosing a product from $C \cap S$. In this case, we have the following expression to compute the probability of choosing item $j$ from offer set $S$:
\begin{equation}
	\label{eq:choice_prob_via_original_distribution}
	\mathbb{P}^{(\Ccal,\lambdab)}_j(S) =\sum_{ C \subseteq \Ccal: j \in C}    \frac{\lambda_C}{\abs{C \cap S}},
\end{equation}  
if $j \in S$, and $0$, otherwise. Similar to Equation~\eqref{eq:choice_prob_via_conditional_distribution}, the factor $|C \cap S|$ comes from the fact that a customer equally prefers to buy a product from $C \cap S$ after she/he forms the consideration set $C \subseteq N$ and thus chooses one from $C \cap S$ uniformly at random. According to Definition~\ref{def:random_set_distribution}, the choice probabilities $\mathbb{P}^{(\Ccal,\lambdab)}_j(S)$ in Equations~\eqref{eq:choice_prob_via_conditional_distribution} and~\eqref{eq:choice_prob_via_original_distribution} are equivalent (by rearranging the sums). We next present a simple example to illustrate the calculation of choice probabilities $\mathbb{P}^{(\Ccal,\lambdab)}_j(S)$.
\begin{example}
	Consider a universe of $n=5$ products, i.e., $N=\{ 1,2,3,4,5 \}$, and assume that customers make choices in accordance with a consideration set model $(\Ccal,\lambdab)$, where $\Ccal = \{ C_1,C_2,C_3 \}$ such that $C_1 = \{  1,3,5 \}$, $C_2 = \{ 2,3,4,5 \}$, and $C_3 = \{ 3,4,5 \}$, alongside $\left( \lambda_{C_1}, \lambda_{C_2}, \lambda_{C_3} \right) = \left(  0.1,0.6,0.3  \right)$. One way to interpret this modeling setup is that an individual customer, before making a final choice, samples a consideration set from $\{ C_1, C_2, C_3 \}$ with probabilities $0.1$, $0.6$, and $0.3$, respectively. Another interpretation implies that the customer population consists of three customer types $C_1$, $C_2$, and $C_3$, where the probability mass of customer types $C_1$,$C_2$, and $C_3$ is $10\%$, $60\%$, and $30\%$, respectively.
	
	Next, assuming that assortment $S = \{ 1,2 \}$ is offered, we can compute the conditional set distribution $\lambdab_{\cdot | S}$ using Equation~\eqref{eq:conditional_set_distribution} and obtain $\lambda_{\{ 1 \} \mid S} = \lambda_{C_1} = 0.1$, $\lambda_{\{ 2 \} \mid S} = \lambda_{C_2} = 0.6$, $\lambda_{ \{ 1,2 \} \mid S } = 0$, and $\lambda_{ \emptyset \mid S} = 1 - 0.6 - 0.1 = 0.3$. Consequently, we can compute choice probabilities under the assortment  $S$ as follows: $\mathbb{P}^{(\Ccal,\lambda)}_1( S ) = \lambda_{ \{ 1 \}  \mid S } / | \{ 1 \} | = 0.1$, $\mathbb{P}^{(\Ccal,\lambda)}_2(S) = \lambda_{ \{ 2 \}  \mid S} / | \{ 2 \} | = 0.6$, and $\mathbb{P}^{(\Ccal,\lambda)}_0( S) = 1 - 0.1 - 0.6 = 0.3$. Alternatively, we can directly factor the intersections $C \cap S$ into our computations as shown in Equation~\eqref{eq:choice_prob_via_original_distribution} which would result in the same choice probabilities: $\mathbb{P}^{(\Ccal,\lambda)}_1( S ) =  \lambda_{C_1} /{| \{ C_1 \cap S \}| } = 0.1$, $\mathbb{P}^{(\Ccal,\lambda)}_2( S )  =  { \lambda_{C_2} }/{ | \{ C_2 \cap S \} | } = 0.6$, and $\mathbb{P}^{(\Ccal,\lambda)}_0( S ) = 0.3$.
	
	Similarly, when $S' = \{ 1,2,4,5 \}$ is an offered assortment, one can show that $\lambda_{\{ 1,5 \} \mid S'} = 0.1$, $\lambda_{\{ 2,4,5 \} \mid S'}  = 0.6$, and $\lambda_{\{ 4,5 \} \mid S} =  0.3$, leading to choice probabilities $\mathbb{P}^{(\Ccal,\lambda)}_1(  S' ) =\lambda_{\{ 1,5 \} \mid S'}/2 = 0.05$, $\mathbb{P}^{(\Ccal,\lambda)}_2(  S' ) =\lambda_{\{ 2,4,5 \} \mid S'}/3 = 0.2$, $\mathbb{P}^{(\Ccal,\lambda)}_4(  S' ) = \lambda_{\{ 2,4,5 \} \mid S'} / 3 + \lambda_{\{ 4,5 \} \mid S'}/2 = 0.35 $,  and $\mathbb{P}^{(\Ccal,\lambda)}_5(  S' ) = \lambda_{\{ 1,5 \} \mid S'}/2 + \lambda_{\{ 2,4,5 \} \mid S'}/3 + \lambda_{\{ 4,5 \} \mid S'}/2 = 0.4$. By comparing the choice probabilities under the two different assortments $S$ and $S'$, we can clearly see that the consideration set model is not restricted by the independence of irrelevant alternatives (IIA) axiom \citep{luce2012individual,hausman1984specification} and thus not subsumed by the MNL model. Specifically, $\mathbb{P}^{(\Ccal,\lambda)}_2( S ) / \mathbb{P}^{(\Ccal,\lambda)}_1( S ) =  6 \neq 4 = \mathbb{P}^{(\Ccal,\lambda)}_2( S' ) / \mathbb{P}^{(\Ccal,\lambda)}_1( S' )$.
\end{example}

Finally, we highlight that although we simplify the choice mechanism by assuming customers select items from the considered set of products uniformly at random, this does not imply that customers have equal probabilities of purchasing any two offered products that appear in the same consideration set. Moreover, our model can readily accommodate scenarios where a customer decides not to purchase any product from the offered assortment.
Following the previous example, consider a model specification $(\Ccal,\lambdab)$ in which a representative customer is characterized by $\Ccal = \{ C_4, C_5, \emptyset \}$, with $C_4 = \{ 1 \}$, $C_5 = \{ 1,2 \}$, and $(\lambda_{C_4},\lambda_{C_5},\lambda_{\emptyset}) = (0.3,0.4,0.3)$. When an assortment $S = \{ 1,2 \}$ is offered, a customer would choose product $1$ with a probability of $0.5$, product $2$ with a probability of $0.2$, and opt not to purchase any offered product with a probability of $0.3$.  It is important to note that, in this scenario, the customer does not show indifference between products $1$ and $2$ appearing in the same consideration set $C_5 = \{1, 2\}$. In fact, the customer is more likely to choose product $1$ due to the presence of a smaller, nested consideration set $C_4$ within $C_5$ and also has the option not to purchase any product from the assortment. 

\subsection{Connection to Other Choice Models}
\label{subsec:connection_to_other_models}
Upon closer examination, it becomes clear that the consideration set model, while not built on an entirely new foundation, is both sophisticated and highly flexible. This nonparametric model offers up to $2^n-1$ degrees of freedom, precisely matching the number of parameters required to define the distribution $\lambdab$ over the $2^n$ subsets in $N$. Moreover, our model is a special case of several well-established choice models. 

The consideration set model, in fact, can be viewed as a special case of the mixed MNL model. To recap, the mixed MNL model is characterized by $k$ segments, with each segment $\ell \in \{ 1,2,\ldots,k \}$ accounting for a probability mass $\lambda_\ell$ of the market. Customers in segment $\ell$ make purchasing decisions based on an MNL model with parameters $(w_{\ell 1}, w_{ \ell 2}, \ldots, w_{ \ell n})$. For a given assortment $S$, the choice probability under the mixed MNL model is calculated as $\Pbb_j(S) = \sum_{\ell =1}^k \lambda_\ell \cdot \frac{w_{\YCRminor{\ell j}}}{1 + \sum_{i \in S} w_{\YCRminor{ \ell i}}}$. This formulation demonstrates that the consideration set model can be effectively represented within the structure of the mixed MNL model. Specifically, suppose $\Ccal = \{ C_1, C_2, \ldots, C_k \}$ and $\lambdab = (\lambda_{C_1}, \ldots, \lambda_{C_k})$. We can then construct a mixed MNL model as follows. Let $\Mcal$ be a sufficiently large constant. For each $C_\ell$, $\ell = 1,2,\ldots,k$, we define a customer segment in the mixed MNL model with population weight $\lambda_\ell \equiv \lambda_{C_\ell}$, and set $w_{ \ell i} = \Mcal$ for $i \in C_\ell $ and $w_{\ell i} = 0$, otherwise. As $\Mcal$ approaches infinity, it becomes evident that the corresponding choice probabilities in this mixed MNL model converge to those in Equation~\eqref{eq:choice_prob_via_original_distribution}. Furthermore, the consideration set model can also be viewed as a special case of the ranking-based model. This relationship is formally stated in the following lemma.
\begin{lemma}
	\label{lemma:ssm_subclass_of_RUM}
	The class of consideration set models is nested in the class of ranking-based models, and the reverse statement does not hold.
\end{lemma}
We formally prove Lemma~\ref{lemma:ssm_subclass_of_RUM} in Section~\ref{subsec:proof_SS_under_RUM}. Herein, we illustrate the idea of the proof with a straightforward example. Suppose the consideration set model consists of a single consideration set, $C_1 = \{ 1, 2 \}$. This model can be represented as a ranking-based model comprising two rankings, $\sigma_1 = \{ 1 \succ 2 \succ 0 \}$ and $\sigma_2 = \{ 2 \succ 1 \succ 0 \}$, each assigned a weight of $0.5$. In other words, to construct an equivalent ranking-based model, we interpret each consideration set $C$ as an equal-weighted average of $|C|!$ rankings, where each ranking is a permutation of the elements in $C$ followed by the no-purchase option $0$. By averaging over all permutations of $C$, we ensure that customers assign equal preference to each product in $C$. It is worth noting that Lemma~\ref{lemma:ssm_subclass_of_RUM} also establishes that the consideration set model is a member of the random utility maximization (RUM) class \citep{thurstone1927law,mcfadden1973conditional}. In the RUM framework, each alternative is associated with a random utility, and customers select the alternative with the highest utility once the randomness is revealed, effectively maximizing their utility. It is well-known that the RUM class is equivalent to the ranking-based model class, as the revealed utilities of products can be sorted to form a ranking over the products \citep{block1959random,farias2013nonparametric}. Finally, we note that our general consideration set model, where $\lambdab$ is a distribution over consideration sets, subsumes the more restrictive consideration set distribution parameterized by independent attention parameters \citep{manzini2014stochastic}. This directly follows from the parameterization of the distribution $\lambdab$ in the CSM model:	$\lambda_{C} = \prod_{i \in C} \gamma_i \cdot \prod_{i \in N \backslash C} (1 - \gamma_i)$, where $\boldsymbol{\gamma}$ is the vector of attention parameters specifying the independent consideration set formation in the paper by  \cite{manzini2014stochastic}. We formally state this result in the following lemma.
\begin{lemma}
	\label{lemma:include_manzini_2014}
    The distribution over consideration sets proposed by \cite{manzini2014stochastic} is a special case of distribution $\lambdab$ of the CSM model. 
\end{lemma}

\section{Identifiability and Axiomatic Characterization}
\label{sec:model_characterization}
In this section, we establish the identifiability of the consideration set model and present its axiomatic characterization. Additionally, we explore the implications of these findings for general two-stage choice models, highlighting how both the ``consider" and ``choose" steps influence the representational power of these models. For ease of notation, we let $N_j \equiv \{ S \subseteq N: j \in S \}$ denote the collection of assortments that include product~$j$. \YC{Also, throughout this section, we let \emph{choice data} refer to the collection of the ground-truth choice probabilities $\{ \Pbb_j(S) : j \in S^+, S \subseteq N  \}$.} For now, we assume that the exact value of choice probabilities is provided to us, that is, we ignore potential finite sample issues. This is a reasonable assumption if the number of transactions under each assortment is large. We relax this assumption in Section~\ref{subsec:model_estimation} when estimating the consideration set model from real-world grocery retail data.

\subsection{Identification of the Consideration Set Model}
\label{subsec:identifiability}

Recall that the consideration set model is fully characterized by the distribution $\lambdab$ over subsets. Therefore, the identification of the consideration set model reduces to the identification of the distribution~$\lambdab$. For simplicity, we refer to $\lambdab$ as the consideration set model throughout this section. In what follows, we present a collection of results that provide different ways to obtain the distribution $\lambdab$ in closed form. The first result requires the specification of the choice probabilities for selecting an item $j$ across the assortments in $N_j$ to compute $\lambda_C$ for each $C \in N_j$.

\begin{theorem}
	\label{theorem:inference}
	\YC{Suppose that the collection of probabilities to choose product $j$}, i.e., $\{{\mathbb{P}}_j(S) \colon S \subseteq N_j\}$, is available and consistent with an underlying consideration set model. Then, we can reconstruct the model by uniquely computing $\lambda_C$, for every $C \in N_j$, as follows:
	\begin{equation} 
		\lambda_C =\sum_{X \in N_j} \mathbb{I} \bigg[ \abs{C \cup X} \ge n-1 \bigg] \cdot n^{\abs{C \cup X}-n+1}  (-1)^{\abs{C}+\abs{X}-n+1}  \mathbb{P}_j(X). \label{Equation_ConsiderationEqualInferFormula}
	\end{equation}
\end{theorem}
\emph{Proof sketch:} For each $C \in N_j$, we first define three functions as follows: $\chi_C(X) ={1}/{\abs{C \cap X}}$, $\psi_C(X)=\mathbb{I}_{\abs{C \cup X} = n-1} \cdot (-1)^{\abs{C}+\abs{X}-n+1}$, and $\varphi_C(X) =\mathbb{I}_{\abs{C \cup X} = n} \cdot n \cdot (-1)^{\abs{C}+\abs{X}-n+1}$.
The first function $\chi_C(X)$ is used to account for the size of the intersection $C \cap X$ between an assortment $X$ and a consideration set $C$. The remaining two functions $\psi_C(X)$ and $\varphi_C(X)$ jointly form the orthonormal basis to $\chi_C(X)$.
By varying the assortment $X$ and observing the corresponding changes in the choice probability $\Pbb_j(X)$, we can infer the probability weight $\lambda_C$ of a consideration set $C$ that includes product $j$. 
With these three functions in hand, we rewrite the choice probability, defined in Equation~\eqref{eq:choice_prob_via_original_distribution}, by means of the function $\chi_C(X)$ as follows:
\begin{align}
	\label{eq: prob_1_thm1}
	\mathbb{P}_j(X) =\sum_{C \in N_j} \lambda_C  \chi_C(X).
\end{align}
We also notice that $\psi_C(X)$ and $\varphi_C(X)$ share the same factor $(-1)^{|C| + |X| - n + 1}$ and, therefore, their summation can be simplified as follows: 
\begin{align}
    \label{eq: prob_2_thm1}
	\psi_C(X) + \varphi_C(X) = \mathbb{I} \bigg[ \abs{C \cup X} \ge n-1 \bigg] \cdot n^{\abs{C \cup X}-n+1}  (-1)^{\abs{C}+\abs{X}-n+1}.
\end{align}
Next, we claim that for all $C,C' \in N_j$,
\begin{equation} \label{eq: basis}
	\sum_{X \in N_j} \chi_{C'}(X) ( \psi_{C} (X)  +\varphi_{C}(X))=  \mathbb{I} \left[ C' = C  \right].
\end{equation}
Invoking the claim, Equation~\eqref{Equation_ConsiderationEqualInferFormula} in the theorem follows immediately, since
\begin{align}
&  \sum_{X \in N_j} \mathbb{P}_j(X) \cdot \mathbb{I} \bigg[ \abs{C \cup X} \ge n-1 \bigg] \cdot n^{\abs{C \cup X}-n+1}  (-1)^{\abs{C}+\abs{X}-n+1}
\nonumber
\\ = &\sum_{X \in N_j} \sum_{C' \in N_j}  \lambda_{C'} \cdot \chi_{C'}(X) \cdot  (\psi_{C}(X)+\varphi_{C}(X)) \qquad \Big [\text{by Equations~\eqref{eq: prob_1_thm1} and \eqref{eq: prob_2_thm1}} \Big] \nonumber \\
= & \sum_{C' \in N_j}\lambda_{C'} \cdot \mathbb{I} \left[ C' = C  \right] \qquad \Big [\text{by the claim stated above as Equation~\eqref{eq: basis}} \Big] \label{claim: 1} \nonumber \\
= & \lambda_{C}.   \nonumber
\end{align}
Therefore, in order to complete the proof of Equation~\eqref{Equation_ConsiderationEqualInferFormula}, it is sufficient to prove the claim in Equation~\eqref{eq: basis}. Since the proof of the claim is quite involved, we relegate it to Section~\ref{subsec:proof_of_theorem_inference} in the e-companion.

In what follows next, we complete the proof of Theorem~\ref{theorem:inference} by showing that the distribution $\lambdab$ is unique. First, note that Equation~\eqref{eq: prob_1_thm1} relates probability distribution $\lambdab$ over the consideration sets to the choice probability $\mathbb{P} _j\big( X\big)$ through the system of linear equations which can be represented as $\boldsymbol{y}=A  \cdot \boldsymbol{ \lambda}$, where ${\boldsymbol{y}} = \left(\Pbb_j(X) \right)_{X: j \in X}$ and $\lambdab = \left( \lambda_C \right)_{C: C \in N_j}$ are two vectors of length $2^{n-1}$. Then, Equation~\eqref{Equation_ConsiderationEqualInferFormula} provides another relationship between choice frequencies $ \mathbb{P}_j(X)$ and the model parameters $\lambdab$ in a linear form as $\boldsymbol{ \lambda}=B \cdot \boldsymbol{y}$. Given that $A$ is a $2^{n-1} \times 2^{n-1}$ dimensional matrix, proving the uniqueness of $\lambdab$ distribution reduces to showing that $\det(A) \ne 0$. This is easy to see: as we have $\boldsymbol{ \lambda} =B \cdot \boldsymbol{y}= B \cdot A \cdot \lambdab$ for any $\lambdab$, it implies that $I=B \cdot  A$ and $1 = \det(I)=\det(B) \cdot \det(A)$. Consequently, $\det(A)\ne 0$.
We refer the readers to Section~\ref{subsec:proof_of_theorem_inference} in the e-companion  for the complete proof of the theorem.\hfill $\square$

Thus, Theorem~\ref{theorem:inference} allows us to compute the probability mass of each consideration set $C \subseteq N$ such that $C \ne \emptyset$. We can then find $\lambda_{\emptyset}$ directly by using the equation $\sum_{C \subseteq N} \lambda_C =1$. It follows from Theorem~\ref{theorem:inference} that the identification of $\lambda_C$ relies only on the access to the probabilities of choosing an arbitrary item $j$ in $C$ under various assortments, i.e, $j$ can be any item in $C$. As a result, for each $C$, there are at least $\abs{C}$ ways to compute $\lambda_C$. We can also formulate the corollary below that specifies the necessary conditions for the data generation process to be consistent with the consideration set model.
\begin{corollary}
	\label{Corollary:1}
	Consider two products $j$ and $k$. Suppose that the choice data for $j$ and $k$, $\{{\mathbb{P}}_j(S) \colon S \subseteq N_j\}$ and $\{{\mathbb{P}}_k(S) \colon S \subseteq N_k\}$, are consistent with an underlying consideration set model. Then, for every consideration set $C \subseteq N$, the following equation is satisfied:
\begin{align*}    
	\sum_{X \in N_j} \mathbb{I}_{\abs{C \cup X} \ge n-1 } \cdot n^{\abs{C \cup X}-n+1}  (-1)^{\abs{C}+\abs{X}-n+1}  \mathbb{P}_j(X) 
	=\sum_{X \in N_k} \mathbb{I}_{\abs{C \cup X} \ge n-1} \cdot n^{\abs{C \cup X}-n+1}  (-1)^{\abs{C}+\abs{X}-n+1}  \mathbb{P}_k(X). 
\end{align*}
 \end{corollary}

Note that this corollary follows directly from the Theorem~\ref{theorem:inference}. Alternatively, we can recover the underlying parameters of the consideration set model from the
collection of choice probabilities $\{{\mathbb{P}}_0(S) \colon S \subseteq N\}$, where we only need to have access to the probability to choose the default option under assortments $S \subseteq N$. In particular, we have the following result:
\begin{lemma}
	\label{lemma:outside}
	Suppose that the collection of probabilities to choose the default option, i.e., $\{{\mathbb{P}}_0(S) \colon S \subseteq N\}$, is consistent with an underlying consideration set model.  Then, we have that
	\begin{equation}
		\lambda_C =\sum_{X \subseteq C}(-1)^{\abs{C}-\abs{X}} \mathbb{P}_0(N \setminus X). \label{eq: outside option}
	\end{equation}
\end{lemma}
In fact, Lemma~\ref{lemma:outside} follows from a particular form of the inclusion-exclusion principle stated in~\cite{graham1995handbook}.  For any finite set $Z$, if $f \colon 2^Z \to \mathbb{R}$ and $g \colon 2^Z \to \mathbb{R}$ are two real-valued set functions defined on the subsets of $Z$ such that $g(X) = \sum_{Y \subseteq X} f(Y)$, then the inclusion-exclusion principle states that $f(Y) = \sum_{X \subseteq Y} (-1)^{\abs{Y} - \abs{X}} g(X)$.  
Our result then follows from setting $f(Y)$ to $\lambda_C$ and defining $g(X)=\mathbb{P}_0(N \setminus X) = \sum_{C \subseteq X}  \lambda_C$,
where the second equality holds since a customer of type $C$ would choose the outside option $0$ from $N \backslash X$ if and only if $C \subseteq X$. \YCRminor{Importantly, while recovering the distribution $\lambdab$ over consideration sets in the most general case would require observing the default choice probabilities for $2^n$ different assortments, practical evidence suggests that consideration sets typically have limited cardinality \citep{hauser1990evaluation, hauser2014consideration}, which significantly reduces the number of required assortments. Specifically, if the size of consideration sets is bounded by a finite number $m$, then, by Lemma~\ref{lemma:outside}, identifying the distribution over consideration sets $\{ C \subseteq N \mid |C| \leq m \}$ only requires computing the default choice probability $\Pbb_0(N \backslash X)$ for sets $X$ where $|X| \leq m$. This involves only $\sum_{i=0}^m {{n}\choose{n-i}} = O(n^m)$ different assortments.}

Also, note that in real-world retail settings, it can be challenging to observe the probability that customers choose the default option. Consequently, Theorem~\ref{theorem:inference} has higher practical importance in identifying parameters of the consideration set model than Lemma~\ref{lemma:outside}, although Equation~\eqref{eq: outside option} is less involved than Equation~\eqref{Equation_ConsiderationEqualInferFormula}. Lemma~\ref{lemma:outside} also connects to the classic decision theory findings of \cite{block1959random} and \cite{falmagne1978representation}, as it recognizes that the sum in Equation~\eqref{eq: outside option} can be reformulated as a Block-Marshak polynomial. We will revisit these classical results in Section~\ref{subsec:modeling_by_axioms}. Finally, after combining together Theorem~\ref{theorem:inference} and Lemma~\ref{lemma:outside}, we can formulate another set of necessary conditions imposed on the choice data for the data generation process to be consistent with the consideration set model.
\begin{corollary}
	\label{Corollary:2}
	Suppose that the choice data $\{{\mathbb{P}}_j(S) \colon S \subseteq N_j\}$ and $\{{\mathbb{P}}_0(S) \colon S \subseteq N\}$ are consistent with an underlying consideration set model. Then, for every set $C \subseteq N$ such that $j \in C$, we must have 
\begin{align*}    
\sum_{X \subseteq C}(-1)^{\abs{C}-\abs{X}} \mathbb{P}_0(N \setminus X)=\sum_{X \in N_j} \mathbb{I} \bigg[ \abs{C \cup X} \ge n-1 \bigg] \cdot n^{\abs{C \cup X}-n+1}  (-1)^{\abs{C}+\abs{X}-n+1}  \mathbb{P}_j(X). 
\end{align*}
 \end{corollary}

 \subsection{Implications of Theorem~\ref{theorem:inference}}
 \label{subsec:implication_of_identifiability}
 We first highlight that most nonparametric choice models are not identifiable from the choice data $\{ \Pbb_j (S) : j \in S^+, S \subseteq N \}$ alone. The examples include the ranking-based model \citep{farias2013nonparametric} and the decision forest model \citep{chen2022decision}, among others. In particular, \cite{sher2011partial} show that the ranking-based model cannot be identified when $n\ge 4$. Intuitively, the nonidentifiability of the aforementioned nonparametric models can be explained by the fact that the parameter space grows much faster with $n$ than the number of available equations that match predicted and actual choice probabilities $\{ \Pbb_j (S) \}_{j \in S^+, S \subseteq N}$ to identify model parameters. Specifically, these equations can be written in the form $\Pbb^{\texttt{MD}}_j(S) = \Pbb_j(S)$, for $j \in S^+$ and $S \subseteq N$, where $\texttt{MD}$ is a choice model and $\Pbb^{\texttt{MD}}_j(S)$ is the predicted probability to choose item $j$ from assortment~$S$ under the choice model \texttt{MD}. 
 
 For instance, a ranking-based model requires the estimation of $O\left(n!\right)$ parameters, which is the number of all possible rankings over $N^+$. Meanwhile, the number of available equations to identify model parameters is upper bounded by $O \left(n \cdot 2^n\right)$, where the factor $2^n$ is the total number of assortments $S \subseteq N$ and the factor $O(n)$ is the number of choices $j \in S^+$ under an assortment $S$. As $n! \gg n 2^n$, the ranking-based model is not identifiable. In contrast, the number of parameters in the consideration set model is $2^n-1$, which is the number of parameters to specify a distribution $\lambdab = (\lambda_C)_{C \in 2^N}$ over consideration sets. As $2^n < n \cdot 2^n$, it is not very surprising that the consideration set model does not suffer from the overparameterization faced by the ranking-based model. In fact, to the best of our knowledge, the consideration set model is one of the most flexible choice rules (i.e., it has the highest degrees of freedom) which are still identifiable from the choice data alone. Note that identifiability can benefit the downstream applications of the choice model. We empirically demonstrate the value of identifiability in Section~\ref{sec:numerics_identifiability} in the e-companion.

Given the complete identifiability of the first-stage consideration set formation, Theorem~\ref{theorem:inference} suggests that if a two-stage choice model is non-identifiable, 
 the first-stage consideration set formation itself is not responsible for that.  \dmr{Additionally, given that the consideration set model has $2^n-1$ parameters and there are at most $O(n \cdot 2^n)$ equations available to identify a choice model, Theorem~\ref{theorem:inference} indicates that any two-stage choice model characterized both by the general distribution over consideration sets as well as by the second-stage choice mechanism is at high risk of being non-identifiable.} Specifically, any choice mechanism that models a purchase decision based on the first-stage consideration set could significantly increase the degree of freedom by at least $\Omega(n)$ factor, resulting in a choice model with $\Omega(n \cdot 2^n)$ parameters. 
For example, \cite{jagabathula2022demand} demonstrates that the choice model, characterized by a joint distribution function over consideration sets and complete rankings, requires estimating $n! \cdot 2^n$ parameters and cannot be identified solely from sales transaction data.

Finally, we note that Theorem~\ref{theorem:inference} is not only about identifiability but also provides a \emph{closed-form} expression to compute all the parameters of the consideration set model. This further distinguishes the consideration set model from other parametric and non-parametric choice models. The MNL model is one of the very few models that have a similar property. Another example is a consider-then-choose (CTC) model studied by \cite{jagabathula2022demand} where the authors show that the model is \emph{partially identifiable} as the marginal distribution over consideration sets can be identified from the choice data.  \cite{jagabathula2022demand} also investigate a special case of the CTC model, named the GCC model, where all customers follow the same single ranking $\sigma$ to make decisions after sampling a consideration set. They provide closed-form expressions to estimate the parameters of the GCC model. However, this model is quite restrictive and suffers from one-directional cannibalization which implies that all customers have to follow the same preference order \citep{jagabathula2022demand}.

 \subsection{Axiomatic Characterization}
 \label{subsec:modeling_by_axioms}

In the previous section, we outlined various necessary conditions that choice data must satisfy to be consistent with the consideration set model (see Corollaries \ref{Corollary:1} and \ref{Corollary:2}). While these necessary conditions are valuable for ensuring consistency with the consideration set model, Corollaries \ref{Corollary:1} and \ref{Corollary:2} are primarily algebraic in nature and lack significant intuitive interpretation. This encourages us to develop axioms based on customers' revealed preferences that could characterize the consideration set model and provide more intuition. To this end, we propose two axioms, called \emph{default regularity} and \emph{symmetric demand cannibalization}.
 
The axiom of default regularity relates to the classic decision theory developed by \cite{block1959random} and \cite{falmagne1978representation}. Specifically, \cite{block1959random} define the Block-Marshak polynomial of the choice probabilities as follows: 
 \begin{align*}
 	H( i , S  ) =  \sum_{X: S \subseteq X} (-1)^{|X| - |S|} \cdot \Pbb_i(S), \qquad \forall S \subseteq N, \,\, i \in S^+. 
 \end{align*}
 \cite{block1959random} and \cite{falmagne1978representation} jointly show that the choice data belongs to the RUM class, i.e., the choice data is consistent with a ranking-based model, if and only if the inequality $H(i,S) \geq 0$ holds for \emph{all} alternatives $i \in S^+$ and assortments $S \subseteq N$. Our first axiom, the \emph{default regularity}, is specifically equivalent to the nonnegativeness of the Block-Marshak polynomial $H(0,S)$ of only the default (i.e., no-purchase) option $0$. \dmr{Thus, the constraints imposed on the choice probabilities by \emph{default regularity} axiom is no more restrictive than the constraints imposed by the RUM class.}
In what follows we formally state the default regularity condition. 
   \begin{definition}(Default Regularity)
 	\label{def:default_regularity}
 	A collection of choice probabilities $\{ \Pbb_j(S): j \in S^+, S \subseteq N \}$ satisfies the default regularity property if $H( 0 , S  ) \geq 0$ for all assortments $S \subseteq N$.
 \end{definition}

In other words, the axiom of default regularity imposes the RUM type of restriction \emph{only} to the default choice probabilities in the choice data. Our second axiom, symmetric cannibalization, imposes restrictions on the choice data related to purchasing a specific product $j \in N$.
 \begin{definition}(Symmetric cannibalization)
 	\label{def:symmetric_cannibalization}
 	A collection of choice probabilities $\{ \Pbb_j(S): j \in S^+, S \subseteq N \}$ satisfies the symmetric cannibalization property if for all assortments $S \subseteq N$ and $j, k \in S$ such that $j \neq k$, we have $\mathbb{P}_{j}(S \setminus \{ k\}) - \mathbb{P}_{j}(S)=\mathbb{P}_{k}(S \setminus \{ j\}) - \mathbb{P}_{k}(S)$.
 \end{definition}
 
The axiom of symmetric demand cannibalization relates to the concept of \emph{demand cannibalization}, where the sales or market share of one product decreases due to the presence of a competing product. This axiom states that for any pair of products, $j$ and $k$ in $S$, the influence of product $k$ on demand for product $j$ is equal to the influence of product $j$ on demand for product $k$ across all product assortments $S \subseteq N$. This indicates a \emph{symmetric} pattern in how products cannibalize each other's demand. 
  In what follows below, we present our main theorem which characterizes the consideration set model through the axioms defined above. 
 \begin{theorem} \label{theorem: characterization}
 	The collection of choice probabilities $\{\mathbb{P}_j(S) \colon j \in S^+, S \subseteq N\}$ is consistent with a consideration set model with unique distribution $\lambdab$ if and only if it satisfies the axioms of default regularity and symmetric cannibalization. 
 \end{theorem}
 The proof is relegated to the e-companion (see Section~\ref{subsec:proof_axiom_theorem}). As it can be seen therein, establishing necessity is rather straightforward, but establishing sufficiency is more involved as it requires two auxiliary lemmas. We prove sufficiency by constructing the distribution $\lambdab$ and then the uniqueness of $\lambdab$ follows directly from the Theorem~\ref{theorem:inference}. From the proof, we can also notice that the non-negativity of the Block-Marshak polynomial $H(0,S)$ of the no-purchase option also ensures that the parameters of the consideration set model computed from the choice probabilities, as described by Lemma~\ref{lemma:outside}, are well-defined, i.e., $\lambda_C = \sum_{X \subseteq C} (-1)^{|C| - |X|} \Pbb_0(N \backslash X) = \sum_{\bar{C} \subseteq \bar{X}} (-1)^{|\bar{X}| - |\bar{C}|} \Pbb_0(\bar{X}) = H(0,\bar{C})\ge 0 $, where $\bar{X} = N \backslash X$ denotes the set complement for a set $X \subseteq N$. This is because the functional form in the axiom of default regularity resembles the formula used to calculate the probability mass of the consideration set distribution presented in Lemma~\ref{lemma:outside}.

  Interestingly, Theorem~\ref{theorem: characterization} suggests that in a general two-stage choice model, it is the choice mechanism, rather than the consideration set formation, that is responsible for capturing the asymmetric demand cannibalization among products. The theorem reveals a fundamental limitation of the consideration set formation in the first stage. While the consideration set formation can explain the heterogeneity of customers' preferences (as illustrated in Example 1), it falls short of completely capturing demand substitution in the way that the ranking-based model or other general RUM models can. In other words, in order to capture complex inter-product substitution using a two-stage choice model more accurately, one should focus on developing the second-stage choice mechanism.
  Although the symmetric demand cannibalization can be considered as a limitation, in Section~\ref{sec:numerics} we demonstrate that the consideration set model remains competitive in prediction performance compared to the ranking-based model when tested on real-world data. Thus, the effect of demand cannibalization in real-world settings may not be that asymmetric. 
 
 We also note that Theorem~\ref{theorem: characterization} can be used to verify if the choice generation process is indeed consistent with a consideration set model. It can also be observed that the default regularity axiom when combined with the symmetric cannibalization axiom ensures a well-known \emph{regularity} 
 property (or \emph{weak rationality}), i.e, $\mathbb{P}_j(S_1) \geq \mathbb{P}_j(S_2)$ whenever $S_1 \subseteq S_2$. which plays an important role in the economics literature \citep{rieskamp2006extending}. We state it formally as follows.
 \begin{lemma} \label{lemma: regularity}
 	If a choice model $\Pbb$ satisfies the axioms of symmetric cannibalization and default regularity, then for all $S_1, S_2 \subseteq N$ such that $S_1 \subseteq S_2$ we have $\mathbb{P}_j(S_1) \geq \mathbb{P}_j(S_2)$ for any $j \in S_1$.
 \end{lemma}
 Thus, Theorem~\ref{theorem: characterization} and Lemma~\ref{lemma: regularity} jointly imply that the only restriction that the consideration set model imposes on top of the general RUM class of choice models, such as the ranking-based model, is the symmetric demand cannibalization property. %
 Finally, we formulate a lemma that states that if the two axioms are satisfied, then the cannibalization effect of item $k$ on item $j$ diminishes when we enlarge the assortment. 
 \begin{lemma} \label{lemma: monotonicity}
 	If a choice model $\Pbb$ satisfies the axioms of symmetric cannibalization and default regularity, then $\Delta_k \mathbb{P}_j(S_1) \geq \Delta_k \mathbb{P}_j(S_2)$ if $S_1 \subseteq S_2$, where $\Delta_k \mathbb{P}_j(S)=\mathbb{P}_j(S \setminus \{k\}) - \mathbb{P}_j(S)$.
 \end{lemma}
 In other words, it follows from the lemma that if item $k$ cannibalizes item $j$ in assortment $S_1$, then item $k$ also cannibalizes item $j$ in assortment $S_2 \subseteq S_1$. Equivalently, if item $k$ does not cannibalize item $j$ in assortment $S_1$ then item $k$ also does not cannibalize item $j$ in assortment $S_2 \supseteq S_1$. We omit the proofs of Lemmas~\ref{lemma: regularity} and \ref{lemma: monotonicity}, as they follow straightforwardly.

We finish this section by adding two additional remarks. First, we note that the axiomatic characterization of choice models in the economics literature is usually established for \emph{parametric} models. Luce's independence of irrelevant alternatives (IIA) axiom \citep{luce2012individual,hausman1984specification} is one of the most popular examples and is used to demonstrate the limitation of the MNL model both in the economics and operation fields. A more recent example is provided by \cite{echenique2019general} where the authors extend the Luce model by proposing a set of axioms that relax the IIA property. Among the nonparametric models, the ranking-based model is one of a few models that are characterized by axioms. As we discussed above, the choice data under the ranking-based model can be characterized by the Block-Marshal polynomial $H(i,S)$ such that $H(i,S) \geq 0$ for all $i \in S^+$ and $S \subseteq N$; see \cite{barbera1986falmagne} and \cite{mcfadden1990stochastic}. Consequently, our paper contributes to the rare examples of axiomatic nonparametric choice models.

We also note that Theorem~\ref{theorem: characterization} indicates that neither the MNL nor the consideration set model subsumes each other. It is straightforward to check that the MNL model does not satisfy the symmetric cannibalization property unless its attraction parameters are the same for all items, meaning that all items have equal market shares. It is also obvious to see that the MNL model does not subsume the consideration set model, as the latter model is with a much higher degree of freedom. Symmetric cannibalization property also differentiates the consideration set model from the broad class of choice models with a single preference order \citep{manzini2014stochastic,jagabathula2022demand}. In practice, as we will see in Section~\ref{sec:numerics}, the assumption of symmetric demand cannibalization does not seriously impair the predictive performance of the consideration set model, as it performs closely to the mixed MNL model and the ranking-based model.

\section{Assortment Optimization}
\label{sec:assortment_optimization}
In this section, we investigate how accounting for consideration sets in choice modeling might influence the design and complexity of operations strategies in downstream applications. Specifically, we focus on the assortment optimization problem, which aims to identify the optimal set of products to offer to customers to maximize revenue. Throughout this section, let $r_i$ denote the revenue associated with product $i \in N$. Without loss of generality, we assume the products are ordered such that  $r_1 \geq r_2 \geq \ldots \geq r_n > 0$. In this context, the default option generally refers to either a customer leaving without making a purchase (i.e., the no-purchase option) or choosing a product outside the designated set (i.e., the outside option), both of which result in zero revenue. For ease of notation, we let $(\Ccal,\lambdab)$ denote a consideration set model where $\Ccal = \{ C_1,C_2,\ldots,C_k \}$ and $\lambda_{C_j} \equiv \lambda_j > 0$ for $j \in K \equiv \{1,2,\ldots,k\}$. Additionally, for a customer type associated with a consideration set $C$, we denote the expected revenue under assortment $S$ as $\text{Rev}_C(S)$:
\begin{align}
\label{def:revenue_of_a_customer_type}
\text{Rev}_C(S) = \begin{cases}
{ \sum_{i \in C \cap S} r_i}/{| C \cap S  |}, & \text{if $|C \cap S| \geq 1$},\\
0, & \text{if $|C \cap S| = 0$}.
\end{cases}
\end{align}
Hence, $\sum_{j \in K} \lambda_j \cdot \text{Rev}_{C_j}(S)$ is the expected revenue from all the customer segments in the population. We can thus state the assortment optimization problem under the consideration set model as follows:
\begin{align}
\label{problem:AO}
\max_{S \subseteq N} \left\lbrace  \sum_{j \in K} \lambda_j \cdot \text{Rev}_{C_j}(S) \right\rbrace.
\end{align}

In addition, without loss of generality, we assume that $\cup_{j \in K} C_j = N$. Specifically, if a product
$i \notin \cup_{j \in K} C_j$, then $i \notin C_j$ for any $j \in K$, meaning no customer in the market considers purchasing it. As a result, the product has zero demand across all assortments and does not affect the choice probabilities of other products, making it irrelevant to the assortment decision. Therefore, it can be excluded from the product universe $N$.

Finally, note that assortment optimization is an important application of choice modeling, widely used in practical tasks such as menu design and product recommendation. For a comprehensive overview, we refer readers to the monograph by \cite{kok2008assortment}. In this section, we first establish foundational results regarding the optimal solution to Problem~\eqref{problem:AO}, followed by an analysis of its computational complexity.

\subsection{Analysis of Optimal Assortments}
\label{subsec:optimal_assortment}

In this subsection, we provide an exact characterization of the optimal assortment structure. We begin with some definitions. First, we let $b \in \{ 0,1 \}$ be a binary variable. Next, we define $\eta_{b}(C)$ as an operator over a set $C \subseteq N$, such that $\eta_{1}(C) = C$ and $\eta_{0}(C) = \bar{C} = N \backslash C$. In other words, when $b = 1$, the operator $\eta_{b} \left( \cdot \right)$ functions as the identity mapping, whereas for $b = 0$, it returns the complement of the set. Then, given a binary vector $\bb \in \{ 0,1 \}^k$ of length $k$, we define a \emph{block} $I_{\bb}$ in the following way: 
\begin{align}
	I_\bb = \bigcap_{j=1}^k \eta_{b_j} \left(  C_j  \right). \label{eq: definition}
\end{align}
For example, if $\bb = (0,1,0)$, then $I_\bb = \bar{C}_1 \cap C_2 \cap \bar{C}_3$. We also let  $\Bcal = \{ 0,1 \}^k$ and define $\mathcal{I}$ as the collection of all blocks such that $\mathcal{I} = \{ I_{\bb} \mid \bb \in \Bcal \}$. \YCRminor{Consequently, as demonstrated in the proof of the upcoming theorem, the non-empty sets in $\mathcal{I} = \{ I_{\bb} \mid \bb \in \Bcal \}$ form a partition of $N$.}

Second, we define a set of products $S_1$ as \emph{revenue-ordered} within its superset $S_2$ if there exists a threshold $r \in \mathbb{R}$ such that $S_1 = \{ i \in S_2 \mid r_i > r \}$. In other words, a revenue-ordered set $S_1$ consists only of the highest-revenue products in $S_2$. Notably, under this definition, the empty set is also considered revenue-ordered,  as it can be obtained by setting the threshold $r$ arbitrarily high. In fact, the notion of a revenue-ordered structure is well-established in the assortment optimization literature. In particular, the seminar work by \cite{talluri2004revenue} shows that the optimal assortment $S^*$ under the MNL model is revenue-ordered within the product universe $N$. With these definitions in place, we can now present the main theorem that characterizes the optimal assortment for Problem~\eqref{problem:AO}.

\begin{theorem}
	\label{thm:revenue_order_in_block}
If $S^*$ is an optimal assortment for Problem~\eqref{problem:AO}, then for every block $I_\bb \in \mathcal{I}$, the subset $S^*_\bb \equiv  S^* \cap I_\bb $ must be revenue-ordered within $I_{\bb}$.	
\end{theorem}

\YCRminor{ \noindent 
While the formal proof of this theorem is relegated to  Section~\ref{subsec:proof_of_thm_revenue_order} in the e-companion, herein we discuss several intuitive insights derived from it. First, note that 
the products within a block $I_{\bb}$ are always considered together by customers -- if one product in the block is considered by a customer, all other products in that block are considered by the same customer as well. In addition, due to the uniform choice mechanism in the second stage of the consideration set model, each product within the block generates the same demand if offered. Therefore, once deciding to include $n_{\bb}$ products from block $I_{\bb}$ into the assortment, it is optimal to select the $n_{\bb}$ most expensive products. This strategy ensures that the demand is concentrated on the highest-revenue products in each block, maximizing overall revenue.}

Overall, these insights demonstrate that optimal assortments under the consideration set model (CSM) exhibit a clear and well-defined structure -- a characteristic often absent when solving assortment problems under other choice models, such as the ranking-based model or the mixed MNL model. Moreover, leveraging the intuition outlined earlier, Theorem~\ref{thm:revenue_order_in_block} enables a polynomial-time algorithm to solve the assortment problem~\eqref{problem:AO}, provided the number of consideration sets, $k$, is bounded by a constant. We formally state this result below.

\begin{proposition}
	\label{prop:optimal_algo_with_fixed_K}
	There exists a polynomial-time optimal algorithm for the assortment problem~\eqref{problem:AO} if the number of consideration sets in the CSM model, $k$, is bounded by a constant.
\end{proposition}
{\it Proof:} We prove this proposition by invoking Theorem~\ref{thm:revenue_order_in_block}. First, we express each block $I_{\bb}$ as $\{ i^{\bb}_1,i^{\bb}_2,\ldots,i^{\bb}_{|I_{\bb}|}  \}$, where $i^{\bb}_j$ is the $j$th most expensive product in block $I_{\bb}$. Then, by invoking Theorem~\ref{thm:revenue_order_in_block}, the optimal assortment $S^*$ must belong to the following collection of assortments:
\begin{align*}
	\mathcal{S}_{\text{OPT}} = \left\lbrace  S = \bigcup_{\bb \in \Bcal} \{  i^{\bb}_1,i^{\bb}_2,\ldots,i^{\bb}_{n_\bb}  \}   \,\, \bigg| \,\, n_\bb \in \{ 0,1,\ldots, |I_\bb| \}, \,\, \forall \bb \in \Bcal  \right\rbrace.
\end{align*}
Therefore, we can find the optimal solution by enumerating all assortments in $\mathcal{S}_{\text{OPT}}$ and calculating each assortment's expected revenue. Given that $| \Bcal| = 2^k$, the number of assortments in $\mathcal{S}_{\text{OPT}}$ is bounded as follows:
\begin{align*}
	\big| \mathcal{S}_{\text{OPT}}  \big| \leq \prod_{\bb \in \Bcal} \left(  | I_\bb | + 1  \right) \leq  \prod_{\bb \in \Bcal} \left(  n+ 1  \right)   \leq (n+1)^{2^k}.
\end{align*}
Furthermore, computing the expected revenue for each assortment requires a runtime of $O(nk)$. Consequently, the total runtime of the algorithm is $O (nk) \cdot (n+1)^{2^k} = O(k \cdot n^{2^k + 1})$, which remains polynomial in $n$ if $k$ is upper bounded by a constant. \hfill $\square$

In what follows, we discuss several implications of Proposition~\ref{prop:optimal_algo_with_fixed_K}. First, this result can be contrasted with the complexity of assortment optimization under the mixed MNL model. To begin with, recall that the CSM model is a special case of the mixed MNL model, where each customer type follows an MNL model that assigns ``infinite'' weights to considered products (see Section~\ref{subsec:connection_to_other_models} for details). In the mixed MNL setting, it is well-established that the assortment optimization problem is generally intractable. Specifically, no polynomial-time optimal algorithm exists, even for the case of just two customer segments, i.e., $k=2$ \citep{rusmevichientong2014assortment}. To address this, \cite{desir2022capacitated} proposed a fully polynomial-time approximation scheme (FPTAS) that provides a $(1-\epsilon)$-optimal solution for the mixed MNL model when the number of customer segments $k$ is bounded by a constant. In contrast, Proposition~\ref{prop:optimal_algo_with_fixed_K} demonstrates that for the CSM model -- a special case of the mixed MNL model -- the assortment optimization problem becomes more tractable. Specifically, under the same assumption of a constant number of customer types as in \cite{desir2022capacitated}, the optimal solution under the CSM model can be found using a polynomial-time optimal algorithm. This highlights that the CSM model, compared to the general mixed MNL model, results in a more computationally tractable assortment optimization problem.

In the interest of practical implementation, we also demonstrate how the assortment problem~\eqref{prob:AO_IP} can be formulated and solved as a mixed-integer linear program (MILP) in a simpler and more efficient manner. Note that the objective function in the assortment problem~\eqref{problem:AO} takes the form of a linear-fractional sum, enabling the application of standard linearization techniques \citep{charnes1962programming,csen2018conic} to reformulate it as follows:
\begin{subequations}
	\label{prob:AO_IP}
	\begin{alignat}{3}
		& \underset{\xb \in \{0,1\}^n, \hb \geq \zerob,\ub \geq \zerob}{\text{maximize}} & \quad & \sum_{j \in K} \sum_{i \in C_j} \lambda_j \cdot r_i \cdot h_{ij} \\
		& \text{subject to} & &  h_{ij} \leq x_i, \quad & \forall i \in C_j, \,\, j \in K, \\
		& & & h_{ij} \leq u_j, \quad & \forall i \in C_j , \,\, j \in K , \\
		& & & u_j + x_i \leq h_{ij} + 1, \quad & \forall i \in C_j, \,\, j \in K,\\
            & & & \sum_{i \in C_j} h_{ij} \leq 1, \quad & \forall j \in K,
	\end{alignat}
\end{subequations}
where $\xb$ is a binary decision vector such that $x_i = 1$ if and only if product $i$ is included in the assortment. 
Importantly, the four sets of constraints in the optimization problem jointly ensure that $h_{ij}=x_i / \sum_{i \in C_j} x_i$ when $\sum_{i \in C_j} x_i \ne 0$. Also, note that the aforementioned constraints naturally ensure that $h_{ij}=0$ if $\sum_{i \in C_j} x_i = 0$. In the interest of space, we evaluate the scalability and effectiveness of the MILP~\eqref{prob:AO_IP} in Section~\ref{subsec:IP_scalability} of the e-companion.

\subsection{Hardness and Inapproximability}

We further demonstrate that relaxing the assumption of a bounded number of customer segments makes the assortment problem~\eqref{problem:AO} computationally hard, even if the size of each consideration set is restricted. This result is formally stated as follows.
\begin{proposition}
	\label{prop:AO_NP_hardness_with_small_size}
	The assortment problem~\eqref{problem:AO} is NP-hard even if the size of each consideration set is upper bounded by a constant. 
\end{proposition}
We prove this hardness result by constructing a reduction from the vertex cover problem, a well-known NP-hard problem \citep{garey1979computers}. The proof, provided in Section~\ref{subsec:proof_AO_NP_hardness_small_size} of the e-companion, shows that the assortment problem remains NP-hard even under the restricted condition that each consideration set contains at most two products (i.e., $|C| \leq 2$ for all $C \in \Ccal$). Interestingly, Propositions~\ref{prop:optimal_algo_with_fixed_K} and \ref{prop:AO_NP_hardness_with_small_size} jointly imply that it is the variety of customer types, represented by distinct consideration sets, rather than the size of the consideration sets themselves, that fundamentally drives the computational complexity of the assortment problem under the CSM model.

Furthermore, in the general case where neither the size of the consideration sets nor the number of customer types is bounded by a constant, the assortment problem~\eqref{problem:AO} can be shown to be \emph{NP-hard even to approximate}. We formally state this result as follows\footnote{We sincerely thank Danny Segev for helping us develop this theorem.}. 
\begin{theorem}
	\label{thm:impossibility_of_approximation}
	The assortment optimization problem~\eqref{problem:AO} is NP-hard to approximate within factor $O(  n^{\frac{1}{2}-\epsilon})$ for any fixed $\epsilon>0$.
\end{theorem}

We prove this theorem by constructing a reduction that transforms any instance of the maximum independent set problem on an $n$-vertex graph, known to be NP-hard to approximate within an $O(n^{1-\epsilon})$ factor \citep{haastad1999clique}, into an instance of the assortment problem~\eqref{problem:AO} of $n$ products and $n$ consideration sets. We refer the readers to Section~\ref{subsec:proof_AO_inapproximate} of the e-companion for details. Note that \cite{aouad2018approximability} use a reduction from the maximum independent set problem to show that the assortment optimization problem under the ranking-based model is NP-hard to approximate within factor $O(n^{1-\epsilon})$. While our construction of the problem instances resembles that of \cite{aouad2018approximability}, our retrieval procedure to construct an independent set from an assortment solution is quite different and involved, resulting in the $O( n^{\frac{1}{2}-\epsilon} )$ factor (see the proof of Claim EC.2 in Section~\ref{subsec:proof_AO_inapproximate} of the e-companion). 

\YCRminor{In addition, it is worth noting that Theorem~\ref{thm:impossibility_of_approximation} establishes a lower bound of $O\left( \sqrt{n} \right)$ for the inapproximability factor of the assortment problem~\eqref{problem:AO}. An upper bound of $O(n)$ -- matching the inapproximability for the ranking-based model -- can be easily obtained by approximating Problem~\eqref{problem:AO} with an assortment of all products (i.e., when $S = N$). While the exact inapproximability factor of Problem~\eqref{problem:AO} remains an open question, it is striking that the assortment optimization problem under the CSM model already exhibits an inapproximability factor of at least $O\left( \sqrt{n} \right)$. 
}

From an operational perspective, Theorem~\ref{thm:impossibility_of_approximation} highlights the challenges of incorporating consideration sets into choice models. Even with the simplest possible second-stage mechanism, the CSM model leads to an assortment optimization problem that is hard to approximate. This underscores the crucial role of consideration sets in the tractability of assortment optimization problems: \emph{any choice model that explicitly accounts for consideration set formation is likely to be computationally intractable} unless additional structural assumptions are imposed to simplify the consideration set formation process.

\section{Empirical Study: Prediction Performance}
\label{sec:numerics}
In this section, we compare the predictive performance of the consideration set model against state-of-the-art benchmark models using the IRI Academic dataset \citep{bronnenberg2008database}.

\subsection{Data Preprocessing, Performance Metrics}
\label{subsec:IRI_preprocess}
The IRI Academic Dataset consists of consumer packaged goods (CPG) purchase transaction data over a chain
of grocery stores in two large Behavior Scan markets in the USA.
In this dataset, each item is represented by its universal product code (UPC) and we aggregate all the items with the same vendor code (comprising digits 3 through 7 in a 13-digit-long UPC code) into a unique ``product''. Next, to alleviate data sparsity, we first include in our analyses only the products with a relatively high market share (i.e., products with at least 1\% market share) and then aggregate the remaining products into the outside option. In order to streamline our case study, we focus on the top fifteen product categories out of thirty-one that have the highest number of unique products (see Table~\ref{tb:IRI_and_SS_size}) and consider the first four weeks in the year 2007 for our analyses. 

We represent our sales transactions with the set of the tuples $\{ (S_t,i_t) \}_{t \in \Tcal}$, where $i_t$ is the purchased product, $S_t$ is the offered assortment and $\Tcal$ denotes the collection of all transactions. For every purchase instance in the
dataset which is characterized by a tuple $(S_t,i_t)$, we have the week and the store ID of the purchase which allows us to approximately construct the offer set $S_t$ by taking the union of all the products that were purchased within the same category as $i_t$, during the same week, and at the same store.

Next, we split our sales transaction data into the training set, which consists of the first two weeks of our data and is used for the calibration of the choice models, and the test/hold-out set, which consists of the last two weeks of our data and is used to compute the prediction performance scores defined below. 
In what follows, we use the mean absolute percentage error (MAPE) to measure the predictive performance of the choice models:
\begin{align}
	\text{MAPE} = \frac{1}{\sum_{S \in \Scal} \tau_o(S)} \cdot \sum_{S \in \Scal } \tau_o(S) \sum_{i \in S^+} \bigg| \frac{ \hat{p}_{i,S} - \bar{p}_{i,S} }{ \bar{p}_{i,S} } \bigg|,
\end{align}
where $\bar{p}_{i,S}$ is the empirical choice frequency computed directly from the sales transaction data, i.e., $\bar{p}_{i,S} = \tau_o({S,i}) / (\sum_{i \in N^+} \tau_o({S,i}) )$ and $\tau_o({S,i})$ is the number of times alternative $i \in S^+$ was chosen under assortment $S$ in the test dataset, $\hat{p}_{i,S}$ is the predictive probability of choosing item $i \in S^+$ from the offer set $S$ by a specific choice model, $\Scal$ is the set of unique assortments in the test dataset, and $\tau_o(S) = \sum_{i \in S^+} \tau_o(S,i) $ is the number of observed transactions under assortment $S$. In the interest of space, we report the predictive outcome based on out-of-sample KL-divergence in Section~\ref{subsec:numerics_KL_outcome} of the e-companion. For both metrics, a lower score indicates better performance.

\begin{table}[]
	\centering
	\begin{tabular}{lcrrrc} \toprule
		Product Category                & \# products & \# assortments & \# transactions & \# types &  set size \\ \midrule
		\textit{Beer}                    & 19        & 721       & 759,968     & 39            & 1.36         \\
		\textit{Coffee}                  & 17        & 603       & 749,867     & 38            & 1.86         \\
		\textit{Deodorant}               & 13        & 181       & 539,761     & 108           & 2.40         \\
		\textit{Frozen Dinners}          & 18        & 330       & 1,963,025    & 40            & 1.29         \\
		\textit{Frozen Pizza}            & 12        & 138       & 584,406     & 54            & 2.21         \\
		\textit{Household Cleaners}      & 21        & 883       & 562,615     & 41            & 1.30         \\
		\textit{Hotdogs}                 & 15        & 533       & 202,842     & 51            & 1.82         \\
		\textit{Margarine/Butter}        & 11        & 27        & 282,649     & 36            & 1.93         \\
		\textit{Milk}                    & 18        & 347       & 476,899     & 80            & 2.84         \\
		\textit{Mustard/Ketchup}         & 16        & 644       & 266,291     & 38            & 1.87         \\
		\textit{Salty Snacks}            & 14        & 152       & 1,476,847    & 86            & 1.84         \\
		\textit{Shampoo}                 & 15        & 423       & 574,711     & 45            & 1.81         \\
		\textit{Soup}                    & 17        & 315       & 1,816,879    & 44            & 1.37         \\
		\textit{Spaghetti/Italian Sauce} & 12        & 97        & 552,033     & 45            & 1.84         \\
		\textit{Tooth Brush}             & 15        & 699       & 392,079     & 37            & 2.07         \\ \bottomrule
	\end{tabular}
	\caption{%
    Descriptive statistics of the IRI dataset after preprocessing. The first four columns show the category name, the number of products, the number of unique observed assortments, and the number of transactions. \YCRminor{The last two columns report the number of customer types ($|\Ccal|$) and the average consideration set size in the estimated CSM model.}} \label{tb:IRI_and_SS_size}
\end{table}

\subsection{Consideration Set Model Estimation Results}

We begin by discussing the insights gained from calibrating the CSM model. For brevity, a detailed description of the CSM calibration method is provided in the e-companion. Specifically, Section~\ref{sec:estimation_method} introduces an estimation approach based on the maximum likelihood estimation (MLE) framework. The core idea is to reformulate the MLE problem as a large-scale concave maximization problem, where the objective is the log-likelihood function and the linear constraints map the distribution of the consideration sets, $\lambdab$, to the choice probabilities. To solve this problem optimally, we employ the column generation technique. This approach has also been used in previous studies, including \cite{van2014market} for estimating ranking-based models and \cite{chen2022decision} for estimating decision forest models from sales data.
Additional details can be found in Section~\ref{sec:estimation_method} of the e-companion.

After estimating the consideration set model, we can emphasize several key observations. First, the fifth column of Table~\ref{tb:IRI_and_SS_size} reports the number of unique customer types (i.e., $|\Ccal|$) in the estimated model. This column reveals that the number of consideration sets is moderate, ranging from 36 to 108 across all product categories, which suggests sparsity in the number of customer types. Second, the last column of Table~\ref{tb:IRI_and_SS_size} presents the weighted average size of the consideration sets in the estimated model $(\Ccal, \lambdab)$, with each weight corresponding to the probability $\lambda_C$ of a consideration set $C \in \Ccal$. From this column, we observe that the typical customer considers a relatively small number of products, with consideration set sizes ranging from 1.3 to 2.8 across all categories, despite some categories featuring as many as 18–21 products. This finding aligns with prior empirical research in behavioral economics and marketing, which consistently shows that consumers tend to consider only a limited number of alternatives before making their final choice \citep{hauser1990evaluation,hauser2014consideration}.

\subsection{Brand Choice Prediction Results}  
\label{subsec:IRI_performance}

In this subsection, we compare the predictive performance of the CSM model against six benchmark models. The first two benchmarks are the \emph{independent demand model} and the \emph{MNL model}. The independent model, though widely used in practice, does not capture substitution effects. The MNL model, widely used in both academic research and practical applications,  is calibrated using a maximum likelihood estimation (MLE) approach in a straightforward way. The third benchmark, the \emph{mixed MNL model}, is estimated using the expectation-maximization (EM) algorithm \citep{train2009discrete} with $K=10$ latent classes. The fourth benchmark is the \emph{ranking-based model}, which is prominent in the operations management literature \citep{farias2013nonparametric,van2014market}. Like the mixed MNL model, the ranking-based model is estimated via the MLE framework \citep{van2014market,van2017expectation}. Notably, both the mixed MNL and the ranking-based models subsume the CSM model studied in this paper as they are equivalent to the RUM class, and thus they are highly competitive benchmarks. The fifth model is the \emph{Markov chain model} studied by \cite{blanchet2016markov}, which captures demand substitution by Markov chains. We estimate this model by the EM algorithm \citep{csimcsek2018expectation}. While the Markov chain model also belongs to the RUM class, \cite{berbeglia2022comparative} empirically demonstrate that this model has superior predictive performance relative to the mixed MNL and ranking-based models across several datasets. The last benchmark model is the \emph{decision forest model} proposed by \cite{chen2022decision}, which can also be estimated by an MLE approach. The decision forest model is outside of the RUM class as it can subsume any discrete choice model. To alleviate the computational effort required for cross-validation, we estimate the decision forest model using trees with a depth of three.

Table~\ref{tb:IRI_prediction_MAPE} summarizes the out-of-sample predictive performance of the consideration set model and the benchmark models, evaluated using the MAPE score on test data. The models are denoted as follows: ID (independent demand), MNL (MNL model), MMNL (mixed MNL), RBM (ranking-based model), MC (Markov chain), DF (decision forest), and CSM (consideration set model). As expected, the independent demand and MNL models exhibit significantly worse predictive performance compared to the CSM. Although the MNL model is not subsumed by the CSM (see Section~\ref{subsec:modeling_by_axioms}), the latter consistently outperforms it in prediction accuracy.

The consideration set model also achieves comparable predictive performance to the mixed MNL and ranking-based models, both of which represent the general RUM class. Furthermore, the CSM remains competitive with the Markov chain model, which has been shown to have an edge over other RUM-based models \citep{berbeglia2022comparative}. Of all the benchmarks, the decision forest model achieves the best predictive accuracy on real-world transaction data, as measured by the MAPE score, and also performs well in terms of KL-divergence (see Section~\ref{subsec:numerics_KL_outcome}). This result is expected, given that the decision forest model lies outside the RUM class and offers the greatest flexibility in capturing complex customer preferences. While the nonparametric nature and high flexibility of the decision forest model make it powerful for fine-grained predictions, they come at the cost of significantly increased computational complexity. Specifically, its large degree of freedom makes downstream operational tasks, such as assortment optimization, much more challenging compared to the CSM model \citep{akchen2021assortment}.

The key takeaway from this study is the effectiveness of the CSM model in accurately predicting customer choices, even with the simplest uniformly random second-stage choice mechanism. This underscores the pivotal role of first-stage consideration set formation within the two-stage choice framework and highlights its substantial influence on choice modeling.

In Table~\ref{tb:IRI_prediction_MAPE}, we also present two variations of the consideration set model. The first, CSM2, includes only consideration sets with at most two products, i.e., $|C| \leq 2$ for all $C$ in $\Ccal$. Interestingly, CSM2 only slightly underperforms the general CSM in predictive accuracy. This result is consistent with earlier findings on small consideration set sizes (see Table~\ref{tb:IRI_and_SS_size}) and aligns with empirical studies in the literature \citep{hauser1990evaluation,hauser2014consideration}. The second variant, \YCRminor{the CSM model blended with rankings} (CSMR), is a mixture of the CSM model and the ranking-based model, enhancing the latter's predictive performance by accounting for ties between products (see Lemma~\ref{lemma:ssm_subclass_of_RUM}). While CSMR remains within the RUM class, it outperforms the standard ranking-based model, which cannot explicitly handle product ties, and performs comparably to the Markov chain model. These findings are consistent with the study by \cite{desir2021mallows}, which demonstrates that integrating a smoothed mixture of rankings can substantially enhance predictive performance.

\YCR{
\begin{table}[]
	\centering
	\begin{tabular}{lccccccccc}
		\toprule
		Category           & ID   & MNL  & MMNL & RBM  & MC    & DF & CSM  & CSM2 & CSMR \\  \cmidrule(r){1-1} \cmidrule{2-7} \cmidrule(l){8-10}
		\emph{Beer}               & 2.26 & 2.03 & 1.88 & 1.96 & 1.87 &    1.63     & 1.90 & 1.87 & 1.85\\
		\emph{Coffee}             & 3.61 & 3.23 & 2.80 & 2.94 & 2.76 &   2.61      & 2.96 & 2.92 & 2.83\\
		\emph{Deodorant}          & 1.02 & 0.84 & 0.80 & 0.80 & 0.81 &     0.76    & 0.79 & 0.80 & 0.79\\
		\emph{Frozen Dinners}     & 1.90 & 1.65 & 1.49 & 1.47 & 1.46 &      1.24   & 1.49 & 1.46 & 1.44\\
		\emph{Frozen Pizza}       & 2.32 & 1.89 & 1.54 & 1.48 & 1.54 &     1.17    & 1.54 & 1.60 & 1.49\\
		\emph{Household Cleaners} & 1.74 & 1.52 & 1.45 & 1.49 & 1.41 &  1.37    & 1.46 & 1.42 & 1.41\\
		\emph{Hotdogs}            & 3.08 & 2.81 & 2.54 & 2.56 & 2.53 &  2.52   &    2.57 & 2.61 & 2.51 \\
		\emph{Margarine/Butter}   & 1.63 & 1.44 & 1.21 & 1.21 & 1.19 &  0.81   & 1.22 & 1.22 & 1.21\\
		\emph{Milk}               & 3.97 & 3.09 & 2.50 & 2.60 & 2.54 & 2.44    & 2.54 & 2.65 & 2.54 \\
		\emph{Mustard/Ketchup}    & 2.58 & 1.93 & 1.74 & 1.84 & 1.73 &  1.86    & 1.79 & 1.76 & 1.72\\
		\emph{Salty Snacks}       & 2.17 & 1.51 & 1.22 & 1.18 & 1.25 &   1.10   & 1.25 & 1.31 & 1.28  \\
		\emph{Shampoo}            & 1.39 & 1.05 & 0.96 & 1.07 & 0.98 &  0.87    & 0.97 & 0.95 & 0.96 \\
		\emph{Soup}               & 2.14 & 1.85 & 1.74 & 1.78 & 1.70 &  1.34   &    1.71 & 1.71 &  1.69 \\
		\emph{Spaghetti/Italian Sauce}    & 2.29 & 1.72 & 1.46 & 1.43 & 1.47 & 0.98    &     1.47 & 1.47 & 1.42 \\
		\emph{Tooth Brush}        & 1.79 & 1.40 & 1.24 & 1.33 & 1.24 &  1.28   &  1.29 & 1.28 &1.24 \\ \cmidrule(r){1-1} \cmidrule{2-7} \cmidrule(l){8-10}
		Average            & 2.26 & 1.86 & 1.64 & 1.67 & 1.63 &   1.46  & 1.66 & 1.68 & 1.63 \\ \bottomrule
	\end{tabular}
	\caption{Out-of-sample prediction performance results measured by MAPE (in unit of $10^{-1}$).} \label{tb:IRI_prediction_MAPE}
\end{table}
}

\subsection{Additional Analyses}
\label{subsec:numerics_introducing_appendix}
In the interest of space, we relegate additional analyses and experiments to the e-companion. In Section~\ref{subsec:numerics_KL_outcome}, we compare the CSM model with benchmark models using an additional performance metric, KL-divergence. Our findings confirm that the insights from Table~\ref{tb:IRI_prediction_MAPE} remain consistent when evaluated with alternative metrics, highlighting the robustness of our results. In Section~\ref{subsec:numerics_convergence}, we examine the computational efficiency of the CSM model compared to the ranking-based model in the estimation process. Specifically, we analyze how the in-sample log-likelihood of both models evolves over a finite runtime. The results show that the estimation algorithm for the CSM model converges significantly faster to a near-optimal solution, requiring much less time than the ranking-based model.

In Section~\ref{subsec:numerics_symmetry_cannibalization}, we explore the role of the symmetric cannibalization property introduced in Section~\ref{subsec:modeling_by_axioms} in the predictive performance of the CSM model relative to the mixed MNL model. As noted earlier, symmetric cannibalization is a defining feature of the CSM model that enhances its tractability compared to other models in the RUM class. However, this property may also limit its ability to fully capture customer purchasing behavior. Our analysis identifies a correlation between the mixed MNL model’s predictive performance over the CSM model and the degree of demand cannibalization asymmetry, suggesting that deviations from symmetric cannibalization contribute to the CSM model’s occasional underperformance.

Finally, in Section~\ref{sec:numerics_identifiability}, we demonstrate the operational value of model identifiability in choice modeling. Using assortment planning as a revenue management application, we show that non-identifiable choice models can lead to significant variability in the optimal assortments they produce, resulting in reduced average revenue performance. To illustrate this, we compare the CSM model, which is identifiable, with the ranking-based model, which is non-identifiable, using the IRI dataset. The results demonstrate the benefits of choice model identifiability in achieving stable and reliable operational outcomes.

\section{Concluding Remarks}
\label{sec:conclusion}
In this paper, we explore a class of consideration-based choice models that are fully defined by the distribution over consideration sets (i.e., the consideration set model) and examine the fundamental role and power of consideration sets in discrete choice modeling.
We first prove that the consideration set model is identifiable from choice data in closed form and results in symmetric demand cannibalization. Then, we demonstrate the operational significance of consideration sets in choice modeling through the emphasis on assortment planning. To this end, we show that the optimal assortment is blockwise revenue-ordered under the consideration set model, leading to a polynomial-time optimal algorithm for the assortment problem if the number of consideration sets in the model is bounded by a constant. However, in general, the assortment optimization problem under the consideration set model is computationally hard even to approximate, although the model has the simplest possible second-stage choice mechanism. Finally, we empirically highlight the competitive predictive performance of the consideration set model despite its symmetric demand cannibalization property. To conclude, this paper examined the role and power of accounting for consideration sets in choice-based demand modeling, with the hope of motivating further research on consideration-set-based choice models and their applications in operations management.

\section*{Acknowledgments}
We sincerely thank the department editor, the associate editor, and the three anonymous referees for their thoughtful comments that helped significantly improve this work. We are also grateful to Jacob Feldman and Danny Segev for their valuable feedback on the results in the assortment optimization section, which has greatly strengthened our analysis in that part of the study.

\bibliographystyle{ormsv080}

\renewcommand*{\bibfont}{\footnotesize}
\renewcommand*{\bibfont}{\normalfont\footnotesize\linespread{1}\selectfont}
\setlength{\bibsep}{1.0pt}
\bibliography{mybib}

\ECSwitch

\ECHead{Electronic Companion}

\section{Supplementary Proofs for Section~\ref{sec:model}}

\subsection{Proof of the Lemma~\ref{lemma:ssm_subclass_of_RUM}}
\label{subsec:proof_SS_under_RUM}

First, we argue that the consideration set model is nested in the class of the ranking-based models. Let $(\Ccal,\lambdab)$ be a consideration set model. We will construct the ranking-based model in which, each $C \in \Ccal$ is associated with $|C|!$ rankings that have the same probability weight. Overall, the support of the constructed ranking-based model consists of $\sum_{C \in \Ccal} |C|!$ rankings. 
In what follows, we describe the construction of the corresponding ranking-based model in detail. 

For each consideration set $C$, let $\mathcal{P}_C$ denote the collection of all permutations of elements in $C$. Then, for each element of this permutation $\rho = (i_1,i_2,\ldots,i_{|C|}) \in \mathcal{P}_C$, we construct a ranking $\sigma(\rho)$ which can be represented as follows: 
\begin{align*}
	\sigma(\rho) = \{ i_1 \succ i_2 \succ \ldots i_{|C|} \succ 0  \}.
\end{align*}
Next, we assign a probability weight of $\lambda_C / |C|!$ to each ranking $\sigma(\rho)$ for all $\rho \in \mathcal{P}_C$. Note that the rank positions of products less preferred than the outside option $0$ are not important, as these products will not be purchased anyway, since the outside option is assumed to be always available.

Finally, we show that the consideration set model $(\Ccal,\lambdab)$ and the constructed ranking-based model result in the same choice probabilities. To this end, let us compute the probability of purchasing item $j$ from assortment $S$ according to the ranking-based model: 
\begin{align*}
	\mathbb{P}_j(S)&=\sum_{C \subseteq \Ccal}  \sum_{\sigma \in \mathcal{P}_C} \frac{\lambda_C}{|C|!} \cdot \mathbb{I}\left[  \text{$j$ is the most preferred product in $C \cap S$ according to $\sigma$}   \right] \\
    & =\sum_{C \subseteq \Ccal} \frac{\lambda_C}{k!} \cdot \mathbb{I}[j \in C \cap S] \cdot (k-p-1)! \cdot \binom{k}{k-p} \cdot p! \ \ \ \text{[\scriptsize where $k=\abs{C}$ and $k-p=\abs{C \cap S}$]} \\
	&=\sum_{C \subseteq \Ccal} \frac{\lambda_C}{k!} \cdot \mathbb{I}[j \in C \cap S] \cdot (k-p-1)! \cdot \frac{k!}{p! (k-p)!} \cdot p! \\
	&=\sum_{C \subseteq \Ccal} \frac{\lambda_C}{k-p} \cdot \mathbb{I}[j \in C \cap S] \\ & = \sum_{C \subseteq \Ccal} \frac{\lambda_C}{\abs{C \cap S}} \cdot \mathbb{I}[j \in C \cap S],
\end{align*}
where the last expression is used to compute the choice probability under the consideration set model (see Equation~\eqref{eq:choice_prob_via_original_distribution}).

To complete the proof, we now provide an example that shows that there exists a ranking-based model that cannot be represented by any consideration set model. Let $N$ denote the universe of two items plus the default option $0$, i.e., $N=\{1,2\}$. Consider two rankings $\sigma_1=\{1 \succ 2 \succ 0\}$ and $\sigma_2=\{ 2 \succ 1 \succ 0\}$, along with a probability distribution $\mub$ such that $\mu_{\sigma_1} + \mu_{\sigma_2} = 1$ and $\mu_{\sigma_1} \neq \mu_{\sigma_2}$. Under this ranking-based model specified by $\mub$, we have that
$$\mu_{\sigma_1}=\mathbb{P}_2(\{2\})-\mathbb{P}_2(\{1,2\}) \not =\mathbb{P}_1(\{1,2\})-\mathbb{P}_1(\{1\})=\mu_{\sigma_2},$$
which violates the symmetric cannibalization property (Definition~\ref{def:symmetric_cannibalization}) that all consideration set models have to satisfy (Theorem~\ref{theorem: characterization}). \hfill $\square$

\section{Supplementary Proofs for Section \ref{sec:model_characterization}}

\subsection{Proof of Theorem~\ref{theorem:inference}}
\label{subsec:proof_of_theorem_inference}
\noindent We follow the proof sketch in Section~\ref{subsec:identifiability}. For every $C \in N_j$ we define boolean functions $\chi_C:N_j \rightarrow \mathbb{R} $, $\psi_{C}:N_j \rightarrow \mathbb{R}$, and $\varphi_{C}:N_j \rightarrow \mathbb{R}$ by 
\begin{align*}
	\chi_C(X)&=\frac{1}{\abs{C \cap X}}, \\
	\psi_C(X)&=\mathbb{I}\bigg[\abs{C \cup X} = n-1 \bigg]\cdot (-1)^{\abs{C}+\abs{X}-n+1}, \\
	\varphi_C(X)&=\mathbb{I}\bigg[\abs{C \cup X} = n \bigg]\cdot n \cdot (-1)^{\abs{C}+\abs{X}-n+1}, 
\end{align*}
where $\mathbb{I}[A]$ is an indicator function that is equal to 1, if condition $A$ is satisfied, and 0 otherwise. Note that for all $X \in N_j$ we have that 
\begin{align}
	\mathbb{P}_j(X)&=\sum_{C \in N_j} \lambda_C \cdot \chi_C(X), \label{eq: prob} \\
	\lambda_C&=\sum_{X \in N_j} \mathbb{P}_j(X) \cdot (\psi_C(X)+\varphi_C(X)), \label{eq: prob2}
\end{align}
Then, for all $C_1,C_2 \in N_j$  we claim that
\[\sum_{X \in N_j} \chi_{C_1}(X) \cdot ( \psi_{C_2} (X)  +\varphi_{C_2}(X))=
\begin{cases}
1, \text{ if } C_1=C_2, \\ 
0, \text{ otherwise}.  
\end{cases}
\]
Consequently, it follows from the claim that 
\begin{align*}
	&\sum_{X \in N_j} \mathbb{I}\bigg[\abs{C \cup X} \ge n-1 \bigg]\cdot n^{\abs{C \cup X}-n+1} \cdot (-1)^{\abs{C}+\abs{X}-n+1} \cdot \mathbb{P}_j(X) \\
	= & \sum_{X \in N_j} \mathbb{P}_j(X) \cdot (\psi_C(X)+\varphi_C(X))=\sum_{X \in N_j} \sum_{C_1 \in N_j} \lambda_{C_1} \cdot \chi_{C_1}(X) \cdot (\psi_C(X)+\varphi_C(X)) \\
	= & \sum_{C_1 \in N_j} \lambda_{C_1} \cdot  \sum_{X \in N_j} \chi_{C_1}(X) \cdot (\psi_C(X)+\varphi_C(X)) =\lambda_C, \ [\text{\scriptsize by the above claim}].
\end{align*}
Then, to complete the proof of the theorem, it is sufficient to prove the claim and show the uniqueness of the solution. In the following proof, we slightly abuse the notation for the calligraphic $\Ccal$, using $\Ccal^n_m = \frac{ n!  }{m!(n-m)!}$ to denote the binomial coefficient. This choice is made to enhance readability, as the standard notation $\binom{n}{m}$ can become cumbersome when multiple parentheses appear in the same equation. We use either `$-$' or `$\backslash$' to denote the set subtraction. We prove the claim by considering four different cases. \\
\emph{I) First, consider the case $C_1=C_2=C$:} \\
In what follows below, we first claim that $\sum_{X \in N_j} \chi_{C}(X) \cdot  \psi_{C} (X)= \frac{(\abs{C}-n)}{\abs{C}}$ and our second claim is that $\sum_{X \in N_j} \chi_{C}(X) \cdot \varphi_C(X)=\frac{n}{\abs{C}}$, and thus, it follows from those two claims that $\sum_{X \in N_j} \chi_{C_1}(X) \cdot ( \psi_{C_2} (X)  +\varphi_{C_2}(X))=1$. We first prove the first claim in the following way: 
\begin{align*}
	&\sum_{X \in N_j} \chi_{C}(X) \cdot  \psi_{C} (X) \\ & =\sum_{X \in N_j} \mathbb{I}\bigg[\abs{C \cup X} = n-1 \bigg]\cdot (-1)^{\abs{C}+\abs{X}-n+1} \cdot \frac{1}{\abs{C \cap X}} \\
    &=\sum_{X_1 \subseteq N - C }  \sum_{X_2 \subseteq C : j \in X_2 } \mathbb{I}\bigg[\abs{C \cup X_1} = n-1 \bigg]\cdot  (-1)^{\abs{C}+\abs{X_1} + \abs{X_2}-n+1} \cdot \frac{1}{\abs{X_2}} \\
    &\scriptsize \big[ \text{We write $X$ as a disjoint union of $X_1$ and $X_2$, where $X_2 = X \cap C$ and $X_1 = X - C$. Note that $j \in X_2$ since $j \in C$ and $j \in X$.} \big] \\
    &=\sum_{X_1 \subseteq N - C } \mathbb{I}\bigg[\abs{C \cup X_1} = n-1 \bigg]\cdot \sum_{X_2 \subseteq C : j \in X_2 } (-1)^{\abs{C}+\abs{X_1} + \abs{X_2}-n+1} \cdot \frac{1}{\abs{X_2}} \\
	&=\sum_{X_1 \subseteq N - C} \mathbb{I}\bigg[\abs{C \cup X_1} = n-1 \bigg]\cdot \sum_{X_2 \subseteq C : j \in X_2 } (-1)^{\abs{X_2}} \cdot \frac{1}{\abs{X_2}} \\ &= (n-\abs{C}) \cdot \sum_{k=1}^{\abs{C}} \frac{(-1)^k \cdot \mathcal{C}_{k-1}^{\abs{C}-1} }{k} \qquad \big[{\scriptsize \text{$j$ must be in $X_2$. Also, $X_1$ can be any set that consists of all $N \backslash C$ except one element}} \big] \\
	&= (n-\abs{C}) \cdot \sum_{k=1}^{\abs{C}} (-1)^k \cdot \frac{ (\abs{C}-1)! }{k \cdot (k-1)!(\abs{C}-k)!} =\frac{(n-\abs{C})}{\abs{C}} \cdot \sum_{k=1}^{\abs{C}} (-1)^k \cdot \frac{ \abs{C}! }{k!(\abs{C}-k)!} \\
	&= \frac{(n-\abs{C})}{\abs{C}} \sum_{k=1}^{\abs{C}} (-1)^k \cdot \mathcal{C}_{k}^{\abs{C}} = \frac{(n-\abs{C})}{\abs{C}} \bigg(\sum_{k=0}^{\abs{C}} (-1)^k \cdot \mathcal{C}_{k}^{\abs{C}} -1\bigg)=  \frac{(\abs{C}-n)}{\abs{C}}.
\end{align*}
Then, we prove the second claim in the following way: 
\begin{align*}
	&\sum_{X \in N_j} \chi_{C}(X) \cdot \varphi_C(X ) \\ & = \sum_{X \in N_j} \mathbb{I}\bigg[\abs{C \cup X} = n \bigg]\cdot n \cdot (-1)^{\abs{C}+\abs{X}-n+1} \cdot \frac{1}{\abs{C \cap X}} \\
    &=\sum_{X_1 \subseteq N - C} \sum_{X_2 \subseteq C : j \in X_2 } \mathbb{I}\bigg[\abs{C \cup X_1} = n \bigg]\cdot n \cdot (-1)^{\abs{C}+\abs{X_1} + \abs{X_2} -n+1} \cdot \frac{1}{\abs{X_2}} \\
    &\scriptsize \big[ \text{We write $X$ as the disjoint union of $X_1$ and $X_2$, where $X_2 = X \cap C$ and $X_1 = X - C$. Note that $j \in X_2$ since $j \in C$ and $j \in X$.}  \big] \\
    &=\sum_{X_1 \subseteq N - C} \mathbb{I}\bigg[\abs{C \cup X_1} = n \bigg]\cdot n \cdot \sum_{X_2 \subseteq C : j \in X_2 }(-1)^{\abs{C}+\abs{X_1} + \abs{X_2} -n+1} \cdot \frac{1}{\abs{X_2}} \\
	&= (-n) \cdot \sum_{X_2 \subseteq C : j \in X_2 }(-1)^{\abs{X_2}} \cdot \frac{1}{\abs{X_2}} \qquad \big[ {\scriptsize \text{Since $X_1$ can only be $N-C$. Thus, $\abs{X_1} = n - |C|$.} } \big] \\ &= (-n) \cdot \sum_{k=1}^{\abs{C}} \frac{(-1)^k \cdot \mathcal{C}_{k-1}^{\abs{C}-1} }{k} \qquad \big[ {\scriptsize \text{Since $j$ must be in $X_2$.} } \big]\\
	&= (-n) \cdot \sum_{k=1}^{\abs{C}} (-1)^k \cdot \frac{ (\abs{C}-1)! }{k \cdot (k-1)!(\abs{C}-k)!} = -\frac{n}{\abs{C}} \sum_{k=1}^{\abs{C}} (-1)^k \cdot \mathcal{C}_{k}^{\abs{C}} =  \frac{n}{\abs{C}}.
\end{align*}
\emph{II) Second, consider the case $C_1 \not =C_2$, $C_1 \cap C_2 = \{j\}$:} \\
In what follows below we first claim that $\sum_{X \in N_j} \chi_{C_1}(X) \cdot  \psi_{C_2} (X)=0$ and our second claim is that $\sum_{X \in N_j} \chi_{C_1}(X) \cdot \varphi_{C_2}(X)=0$, and thus, it follows from those claims that $\sum_{X \in N_j} \chi_{C_1}(X) \cdot ( \psi_{C_2} (X)  +\varphi_{C_2}(X))=0$. We prove the first claim in the following way:
\begin{align*}
	&\sum_{X \in N_j} \chi_{C_1}(X) \cdot  \psi_{C_2} (X) \\ & =\sum_{X \in N_j} \mathbb{I}\bigg[\abs{C_1 \cup X} = n-1 \bigg]\cdot (-1)^{\abs{C_1}+\abs{X}-n+1} \cdot \frac{1}{\abs{C_2 \cap X}} \\
    &=\sum_{X_1 \subseteq N - C_1} \sum_{X_2 \subseteq C_1 : j \in X_2} \mathbb{I}\bigg[\abs{C_1 \cup X_1} = n-1 \bigg]\cdot    (-1)^{\abs{C_1}+\abs{X_1} + \abs{X_2} -n+1} \cdot \frac{1}{\abs{C_2 \cap X_1} + \abs{C_2 \cap X_2}} \\
    &\scriptsize \big[\text{We write $X$ as disjoint union of $X_1$ and $X_2$, where $X_2 = X \cap C_1$ and $X_1 = X - C_1$. Note that $j \in X_2$ since $j \in C_1$ and $j \in X$.} \big] \\
    &=\sum_{X_1 \subseteq N - C_1}\mathbb{I}\bigg[\abs{C_1 \cup X_1} = n-1 \bigg]\cdot  \sum_{X_2 \subseteq C_1 : j \in X_2}  (-1)^{\abs{C_1}+\abs{X_1} + \abs{X_2} -n+1} \cdot \frac{1}{\abs{C_2 \cap X_1} + \abs{C_2 \cap X_2}} \\
    &=\sum_{X_1 \subseteq N - C_1} \mathbb{I}\bigg[\abs{C_1 \cup X_1} = n-1 \bigg]\cdot \frac{1}{\abs{X_1 \cap C_2}+1} \cdot \sum_{X_2 \subseteq C_1 : j \in X_2}  (-1)^{\abs{C_1}+\abs{X_1} + \abs{X_2} -n+1} \quad \big[ \text{\scriptsize since   $C_1 \cap C_2 = \{j\}, X_2 \subseteq C_1$ } \big] \\
	&=\sum_{X_1 \subseteq N- C_1} \mathbb{I}\bigg[\abs{C_1 \cup X_1} = n-1 \bigg]\cdot \frac{1}{\abs{X_1 \cap C_2}+1} \cdot \sum_{X_2 \subseteq C_1 : j \in X_2}  (-1)^{\abs{X_2}} \qquad \big[ {\scriptsize \text{Since $|C_1| + |X_1| = n-1$}}  \big] \\
	&= \sum_{X_1 \subseteq N - C_1} \mathbb{I}\bigg[\abs{C_1 \cup X_1} = n-1 \bigg] \cdot \frac{1}{\abs{X_1 \cap C_2}+1} \cdot (-1) \cdot  \sum_{k=0}^{\abs{C_1}-1} (-1)^k \cdot \mathcal{C}_{k}^{\abs{C_1}-1}=0.
\end{align*}
Then, we prove the second claim in the following way:
\begin{align*}
	&\sum_{X \in N_j} \chi_{C_1}(X) \cdot \varphi_{C_2}(X) \\ & = \sum_{X \in N_j} \mathbb{I}\bigg[\abs{C_1 \cup X}  = n \bigg]\cdot n \cdot (-1)^{\abs{C_1}+\abs{X}-n+1} \cdot \frac{1}{\abs{C_2 \cap X}} \\
    & = \sum_{X_1 \subseteq N - C_1}  \sum_{X_2 \subseteq C_1 : j \in X_2} \mathbb{I}\bigg[\abs{C_1 \cup X_1}  = n \bigg]\cdot n \cdot  (-1)^{\abs{C_1}+\abs{X_1} + \abs{X_2}-n+1} \cdot \frac{1}{\abs{C_2 \cap X_1} + \abs{C_2 \cap X_2}} \\
    &\scriptsize \big[\text{We write $X$ as the disjoint union of $X_1 = X - C_1$ and  $X_2 = X \cap C_1$. Note that $X_2$ includes $j$.} \big] \\
    & = \sum_{X_1 \subseteq N - C_1}  \mathbb{I}\bigg[\abs{C_1 \cup X_1}  = n \bigg]\cdot n \cdot \sum_{X_2 \subseteq C_1 : j \in X_2} (-1)^{\abs{C_1}+\abs{X_1} + \abs{X_2}-n+1} \cdot \frac{1}{\abs{C_2 \cap X_1} + \abs{C_2 \cap X_2}} \\
	& = \frac{-n}{\abs{ C_2 \cap X_1}+1} \cdot  \sum_{X_2 \subseteq C_1 : j \in X_2} (-1)^{\abs{X_2}} \   \big[ \text{\scriptsize since $C_1 \cap C_2 = \{j\}, X_2 \subseteq C_1$} \big] \\
	&= \frac{-n}{\abs{ C_2 \cap X_1}+1} \cdot (-1) \cdot \sum_{k=0}^{\abs{C_1}-1} (-1)^k \cdot \mathcal{C}_{k}^{\abs{C_1}-1}= 0.
\end{align*}
\emph{III) Third, consider the case $C_1 \not =C_2$, $\abs{C_1 \cap C_2} > 1$ and $C_1 \not \subseteq  C_2$:} \\
In what follows below we first claim that $\sum_{X \in N_j} \chi_{C_1}(X) \cdot  \psi_{C_2} (X)=0$ and our second claim is that $\sum_{X \in N_j} \chi_{C_1}(X) \cdot \varphi_{C_2}(X)=0$. It follows from those two claims that $\sum_{X \in N_j} \chi_{C_1}(X) \cdot ( \psi_{C_2} (X)  +\varphi_{C_2}(X))=0$. We prove the first claim in the following way: 
\begin{align*}
	&\sum_{X \in N_j} \chi_{C_1}(X) \cdot  \psi_{C_2} (X) \\ & =\sum_{X \in N_j} \mathbb{I}\bigg[\abs{C_1 \cup X} = n-1 \bigg]\cdot (-1)^{\abs{C_1}+\abs{X}-n+1} \cdot \frac{1}{\abs{C_2 \cap X}} \\
    &=\sum_{X_1 \subseteq N-C_1} \sum_{X_2 \subseteq C_1: j \in X_2}\mathbb{I}\bigg[\abs{C_1 \cup X_1} = n-1 \bigg]\cdot   (-1)^{\abs{C_1}+\abs{X_1}+\abs{X_2}-n+1} \cdot \frac{1}{\abs{C_2 \cap X_1}+\abs{C_2 \cap X_2}} \\
     &\scriptsize \big[\text{We write $X$ as the disjoint union of $X_1$ and $X_2$, where $X_2 = X \cap C_1$ and $X_1 = X - C_1$. Note that $j \in X_2$.} \big] \\
    &=\sum_{X_1 \subseteq N-C_1} \mathbb{I}\bigg[\abs{C_1 \cup X_1} = n-1 \bigg]\cdot \sum_{X_2 \subseteq C_1: j \in X_2}  (-1)^{\abs{C_1}+\abs{X_1}+\abs{X_2}-n+1} \cdot \frac{1}{\abs{C_2 \cap X_1}+\abs{C_2 \cap X_2}} \\
	&=\sum_{X_1 \subseteq N-C_1} \mathbb{I}\bigg[\abs{C_1 \cup X_1} = n-1 \bigg]\cdot \sum_{X_2 \subseteq C_1: j \in X_2}  (-1)^{\abs{X_2}} \cdot \frac{1}{\abs{C_2 \cap X_1}+\abs{C_2 \cap X_2}} \\
	&\scriptsize \big[\text{where $\abs{C_2 \cap X_2}$ varies from 1 to $\abs{C_2 \cap C_1}$}
	\big] \\
	&=\sum_{X_1 \subseteq N-C_1} \mathbb{I}\bigg[\abs{C_1 \cup X_1} = n-1 \bigg]\cdot \sum_{Y \subseteq C_2 \cap C_1: j \in Y}  \sum_{Z \subseteq C_1 - C_2}  (-1)^{\abs{Y}+\abs{Z}} \cdot \frac{1}{\abs{C_2 \cap X_1}+\abs{C_2 \cap Y}} \\
	&\scriptsize \big[\text{where $X_2=Y \cup Z$ such that $Y$ is any subset of $C_2 \cap C_1$ that includes $j$ and $Z$ is any subset of $C_1-C_2$}
	\big] \\
	&=\sum_{X_1 \subseteq N-C_1} \mathbb{I}\bigg[\abs{C_1 \cup X_1} = n-1 \bigg]\cdot \sum_{Y \subseteq C_2 \cap C_1: j \in Y} \frac{1}{\abs{C_2 \cap X_1}+\abs{C_2 \cap Y}} (-1)^{\abs{Y}} \cdot \sum_{Z \subseteq C_1 - C_2}  (-1)^{\abs{Z}}  \\
	&=\sum_{X_1 \subseteq N-C_1} \mathbb{I}\bigg[\abs{C_1 \cup X_1} = n-1 \bigg]\cdot \sum_{Y \subseteq C_2 \cap C_1: j \in Y} \frac{1}{\abs{C_2 \cap X_1}+\abs{C_2 \cap Y}} (-1)^{\abs{Y}}  \cdot \sum_{k=0}^{\abs{C_1 - C_2}} (-1)^k \cdot \mathcal{C}_{k}^{\abs{C_1-C_2}} \\
	&=\sum_{X_1 \subseteq N-C_1} \mathbb{I}\bigg[\abs{C_1 \cup X_1} = n-1 \bigg]\cdot \sum_{Y \subseteq C_2 \cap C_1: j \in Y} \frac{1}{\abs{C_2 \cap X_1}+\abs{C_2 \cap Y}} (-1)^{\abs{Y}}  \cdot 0=0.
\end{align*}
Then, we prove the second claim in the following way: 
\begin{align*}
	&\sum_{X \in N_j} \chi_{C_1}(X) \cdot \varphi_{C_2}(X) \\ = & \sum_{X \in N_j} \mathbb{I}\bigg[\abs{C_1 \cup X}  = n \bigg]\cdot n \cdot (-1)^{\abs{C_1}+\abs{X}-n+1} \cdot \frac{1}{\abs{C_2 \cap X}} \\
    = & \sum_{X_1 \subseteq N - C_1}  \sum_{X_2 \subseteq C_1 : j \in X_2}\mathbb{I}\bigg[\abs{C_1 \cup X_1}  = n \bigg]\cdot n \cdot  (-1)^{\abs{C_1}+\abs{X_1} + \abs{X_2}-n+1} \cdot \frac{1}{\abs{C_2 \cap X_1} + \abs{C_2 \cap X_2}} \\
    &\scriptsize \big[\text{We write $X$ as disjoint union of $X_1 = X - C_1$ and $X_2 = X \cap C_1$. Note that $j \in X_2$.}\big] \\
    = & \sum_{X_1 \subseteq N - C_1}  \mathbb{I}\bigg[\abs{C_1 \cup X_1}  = n \bigg]\cdot n \cdot \sum_{X_2 \subseteq C_1 : j \in X_2} (-1)^{\abs{C_1}+\abs{X_1} + \abs{X_2}-n+1} \cdot \frac{1}{\abs{C_2 \cap X_1} + \abs{C_2 \cap X_2}} \\
	= & (-n) \cdot \sum_{X_2 \subseteq C_1 : j \in X_2} (-1)^{\abs{X_2}} \cdot \frac{1}{\abs{C_2 \cap X_1}+\abs{C_2 \cap X_2}} \\
	&\scriptsize \big[\text{where $\abs{C_2 \cap X_2}$ varies from 1 to $\abs{C_2 \cap C_1}$}
	\big] \\
	= &  (-n) \cdot \sum_{Y \subseteq C_2 \cap C_1: j \in Y}  \sum_{Z \subseteq C_1 - C_2}  (-1)^{\abs{Y}+\abs{Z}} \cdot \frac{1}{\abs{C_2 \cap X_1}+\abs{C_2 \cap Y}} \\
	&\scriptsize \big[\text{where $X_2=Y \cup Z$ such that $Y$ is any subset of $C_2 \cap C_1$ that includes $j$ and $Z$ is any subset of $C_1-C_2\}$}
	\big] \\
	= & (-n) \cdot \sum_{Y \subseteq C_2 \cap C_1: j \in Y} \frac{1}{\abs{C_2 \cap X_1}+\abs{C_2 \cap Y}} (-1)^{\abs{Y}} \cdot \sum_{Z \subseteq C_1 - C_2}  (-1)^{\abs{Z}}  \\
	= & (-n) \cdot \sum_{Y \subseteq C_2 \cap C_1: j \in Y} \frac{1}{\abs{C_2 \cap X_1}+\abs{C_2 \cap Y}} (-1)^{\abs{Y}}  \cdot \sum_{k=0}^{\abs{C_1-C_2}} (-1)^k \cdot \mathcal{C}_{k}^{\abs{C_1 - C_2}} \\
	= & (-n) \cdot \sum_{Y \subseteq C_2 \cap C_1: j \in Y} \frac{1}{\abs{C_2 \cap X_1}+\abs{C_2 \cap Y}} (-1)^{\abs{Y}}  \cdot 0=0. 
\end{align*}
\emph{IV) Fourth, consider the case $C_1 \not =C_2$, $\abs{C_1 \cap C_2} > 1$ and $C_1 \subseteq C_2 $:} \\
In what follows below we first simplify summations $\sum_{X \in N_j} \chi_{C_1}(X) \cdot  \psi_{C_2} (X)$ and $\sum_{X \in N_j} \chi_{C_1}(X) \cdot \varphi_{C_2}(X)$, and then, show that $\sum_{X \in N_j} \chi_{C_1}(X) \cdot ( \psi_{C_2} (X)  +\varphi_{C_2}(X))=0$. First, we have that
\begin{align*}
	&\sum_{X \in N_j} \chi_{C_1}(X) \cdot  \psi_{C_2} (X) \\ & =\sum_{X \in N_j} \mathbb{I}\bigg[\abs{C_1 \cup X} = n-1 \bigg]\cdot (-1)^{\abs{C_1}+\abs{X}-n+1} \cdot \frac{1}{\abs{C_2 \cap X}} \\	
    &=\sum_{X_1 \subseteq N - C_1} \sum_{X_2 \subseteq C_1 : j \in X_2} \mathbb{I}\bigg[\abs{C_1 \cup X_1} = n-1 \bigg]\cdot (-1)^{\abs{C_1}+\abs{X_1} + \abs{X_2}-n+1} \cdot \frac{1}{\abs{C_2 \cap X_1} + \abs{C_2 \cap X_2}} \\
    &\scriptsize \big[\text{We write $X$ as the disjoint union of $X_1 = X - C_1$ and $X_2 = X \cap C_1$. Note that $j \in X_2$} \big] \\
    &= \sum_{X_1 \subseteq N - C_1} \sum_{X_2 \subseteq C_1 : j \in X_2} \mathbb{I}\bigg[\abs{C_1 \cup X_1} = n-1 \bigg] \cdot  (-1)^{\abs{X_2}} \cdot \frac{1}{\abs{C_2 \cap X_1}+\abs{C_2 \cap X_2}} \\
	& = \sum_{X_1 \subseteq N - C_1} \sum_{X_2 \subseteq C_1 : j \in X_2} \mathbb{I}\bigg[\abs{C_1 \cup X_1} = n-1 \bigg] \cdot (-1)^{\abs{X_2}} \cdot \frac{1}{\abs{C_2 \cap X_1}+\abs{ X_2}} \qquad \big[\text{\scriptsize Since $ C_2 \cap X_2 = C_2 \cap X \cap C_1 = X \cap C_1 = X_2$}\big] \\
	&= \sum_{X_1 \subseteq N - C_1} \sum_{m=1}^{\abs{C_1}} \mathbb{I}\bigg[\abs{C_1 \cup X_1} = n-1 \bigg] \cdot \frac{(-1)^m \cdot \mathcal{C}_{m-1}^{\abs{C_1}-1}}{m+\abs{X_1 \cap C_2}} \qquad \big[\text{\scriptsize since $\abs{X_2}$ varies from 1 to $\abs{C_1}$ }\big] \\
	&= (\abs{C_2}-\abs{C_1}) \cdot \sum_{m=1}^{\abs{C_1}} \frac{(-1)^m \cdot \mathcal{C}_{m-1}^{\abs{C_1}-1}}{m+\abs{C_2}-\abs{C_1}-1} + (n-\abs{C_2}) \cdot \sum_{m=1}^{\abs{C_1}} \frac{(-1)^m \cdot \mathcal{C}_{m-1}^{\abs{C_1}-1}}{m+\abs{C_2}-\abs{C_1}}. \\
	&\scriptsize \big[\text{In the sum over $X_1$, for $\abs{C_2} - \abs{C_1}$ times, $C_1 \cup X_1$ misses an element in $C_2$, then $\abs{X_1 \cap C_2}=\abs{C_2}-\abs{C_1}-1$;} \\ & \scriptsize \text{for $n-|C_2|$ times, $C_1 \cup X_1$ misses an element in $N-C_2$, then $\abs{X_1 \cap C_2} = |C_2|- |C_1|$.}  \big]
\end{align*}
Also, we have that
\begin{align*}
	&\sum_{X \in N_j} \chi_{C_1}(X) \cdot \varphi_{C_2}(X) \\ & = \sum_{X \in N_j} \mathbb{I}\bigg[\abs{C_1 \cup X}  = n \bigg]\cdot n \cdot (-1)^{\abs{C_1}+\abs{X}-n+1} \cdot \frac{1}{\abs{C_2 \cap X}} \\
    & = \sum_{X_1 \subseteq N- C_1} \sum_{X_2 \subseteq C_1: j \in X_2} \mathbb{I}\bigg[\abs{C_1 \cup X_1}  = n \bigg]\cdot n \cdot (-1)^{\abs{C_1}+\abs{X_1} + \abs{X_2}-n+1} \cdot \frac{1}{\abs{C_2 \cap X_1} + \abs{C_2 \cap X_2}} \\
    & \scriptsize \big[ \text{We write $X$ as the disjoint union of $X_1 = X - C_1$ and $X_2 = X \cap C_1$. Note that $j \in X_2$ } \big] \\
    & = \sum_{X_1 \subseteq N- C_1}   \mathbb{I}\bigg[\abs{C_1 \cup X_1}  = n \bigg]\cdot n \cdot \sum_{X_2 \subseteq C_1: j \in X_2} (-1)^{\abs{C_1}+\abs{X_1} + \abs{X_2}-n+1} \cdot \frac{1}{\abs{C_2 \cap X_1} + \abs{C_2 \cap X_2}} \\
	& = (-n) \cdot \mathbb{I}\bigg[ X_1 = N - C_1 \bigg] \cdot \sum_{X_2 \subseteq C_1: j \in X_2}  (-1)^{\abs{X_2}} \cdot \frac{1}{\abs{C_2 \cap X_1}+\abs{C_2 \cap X_2}} \\
	& = (-n) \cdot \mathbb{I}\bigg[ X_1 = N - C_1 \bigg] \cdot \sum_{X_2 \subseteq C_1: j \in X_2}  (-1)^{\abs{X_2}} \cdot \frac{1}{\abs{C_2 \cap X_1}+\abs{X_2}} \\
	& = (-n) \cdot \sum_{X_2} (-1)^{\abs{X_2}} \cdot \frac{1}{\abs{C_2}-\abs{C_1} +\abs{X_2}} \\
	&= (-n) \cdot \sum_{m=1}^{\abs{C_1}} \frac{(-1)^m \cdot \mathcal{C}_{m-1}^{\abs{C_1}-1}}{m+\abs{C_2}-\abs{C_1}} 
	\qquad \scriptsize \big[\text{where $\abs{X_2}$ varies from 1 to $\abs{C_1}$}
	\big].
\end{align*}
After simplifying $\sum_{X \in N_j} \chi_{C_1}(X) \cdot  \psi_{C_2} (X)$ and $\sum_{X \in N_j} \chi_{C_1}(X) \cdot \varphi_{C_2}(X)$ as above, we have
\begin{align*}
	&\sum_{X \in N_j} \chi_{C_1}(X) \cdot ( \psi_{C_2} (X)+\varphi_{C_2}(X) ) \\
	&= (\abs{C_2}-\abs{C_1}) \cdot \sum_{m=1}^{\abs{C_1}} \frac{(-1)^m \cdot \mathcal{C}_{m-1}^{\abs{C_1}-1}}{m+\abs{C_2}-\abs{C_1}-1} -\abs{C_2} \cdot \sum_{m=1}^{\abs{C_1}} \frac{(-1)^m \cdot \mathcal{C}_{m-1}^{\abs{C_1}-1}}{m+\abs{C_2}-\abs{C_1}} \\
	&=p \cdot \sum_{m=1}^{i} \frac{(-1)^m \cdot \mathcal{C}_{m-1}^{i-1} }{m+p-1} -(p+i) \cdot \sum_{m=1}^{i} \frac{(-1)^m \cdot \mathcal{C}_{m-1}^{i-1} }{m+p}, \qquad \big[\text{\scriptsize where we denote $p \equiv \abs{C_2}-\abs{C_1}$, \ $i \equiv \abs{C_1}$ }\big] \\
	&= -\sum_{m=1}^{i} \frac{(-1)^m \cdot (m-1) \cdot \mathcal{C}_{m-1}^{i-1}}{m+p-1}+\sum_{m=1}^{i}(-1)^m \cdot \mathcal{C}_{m-1}^{i-1}-(p+i) \cdot \sum_{m=1}^{i} \frac{(-1)^m \cdot \mathcal{C}_{m-1}^{i-1} }{m+p} \\
    & \big[ \scriptsize \text{letting the $p$ at the beginning expand as $p = (m+p-1) - (m-1)$} \big]\\
	&=-\sum_{m=1}^{i} \frac{(-1)^m \cdot (m-1) \cdot \mathcal{C}_{m-1}^{i-1}}{m+p-1}-(p+i) \cdot \sum_{m=1}^{i} \frac{(-1)^m \cdot \mathcal{C}_{m-1}^{i-1} }{m+p} \\
	&=-\sum_{m=1}^{i} \frac{(-1)^m \cdot (m-1) \cdot \mathcal{C}_{m-1}^{i-1}}{m+p-1}-\sum_{m=1}^{i}(-1)^m \cdot \mathcal{C}_{m-1}^{i-1}+\sum_{m=1}^{i} \frac{(-1)^m \cdot m \cdot \mathcal{C}_{m-1}^{i-1}}{m+p} -i \cdot \sum_{m=1}^{i} \frac{(-1)^m \cdot \mathcal{C}_{m-1}^{i-1} }{m+p} \\
    & \big[ \scriptsize \text{expanding the last term by breaking the factor $(p+i)$} \big]\\
	&=-\sum_{m=1}^{i} \frac{(-1)^m \cdot (m-1) \cdot \mathcal{C}_{m-1}^{i-1}}{m+p-1}+\sum_{m=1}^{i} \frac{(-1)^m \cdot m \cdot \mathcal{C}_{m-1}^{i-1}}{m+p} -i \cdot \sum_{m=1}^{i} \frac{(-1)^m \cdot \mathcal{C}_{m-1}^{i-1} }{m+p} \\
	&=-\sum_{m=1}^{i} \frac{(-1)^m \cdot (m-1) \cdot \mathcal{C}_{m-1}^{i-1}}{m+p-1}+\sum_{m=1}^{i} \frac{(-1)^m \cdot m \cdot \mathcal{C}_{m-1}^{i-1}}{m+p} - \sum_{m=1}^{i} \frac{(-1)^m \cdot m \cdot \mathcal{C}_{m}^{i} }{m+p} \\
	&=\sum_{m=0}^{i-1} \frac{(-1)^m \cdot m \cdot \mathcal{C}_{m}^{i-1}}{m+p}+\sum_{m=1}^{i} \frac{(-1)^m \cdot m \cdot \mathcal{C}_{m-1}^{i-1}}{m+p} - \sum_{m=1}^{i} \frac{(-1)^m \cdot m \cdot \mathcal{C}_{m}^{i} }{m+p} \\
	&=\sum_{m=0}^{i-1} \frac{(-1)^m \cdot m \cdot \mathcal{C}_{m}^{i-1}}{m+p}+\sum_{m=1}^{i} \frac{(-1)^m \cdot m \cdot \mathcal{C}_{m-1}^{i-1}}{m+p} - \sum_{m=1}^{i} \frac{(-1)^m \cdot m \cdot (\mathcal{C}_{m-1}^{i-1} +\mathcal{C}_{m}^{i-1})}{m+p} \\
	&=\sum_{m=0}^{i-1} \frac{(-1)^m \cdot m \cdot \mathcal{C}_{m}^{i-1}}{m+p}+\sum_{m=1}^{i} \frac{(-1)^m \cdot m \cdot \mathcal{C}_{m-1}^{i-1}}{m+p} - \sum_{m=1}^{i} \frac{(-1)^m \cdot m \cdot \mathcal{C}_{m-1}^{i-1}}{m+p} - \sum_{m=1}^{i-1} \frac{(-1)^m \cdot m \cdot  \mathcal{C}_{m}^{i-1}}{m+p} \\
	&=\sum_{m=0}^{i-1} \frac{(-1)^m \cdot m \cdot \mathcal{C}_{m}^{i-1}}{m+p}- \sum_{m=1}^{i-1} \frac{(-1)^m \cdot m \cdot  \mathcal{C}_{m}^{i-1}}{m+p}\\
	&= \sum_{m=0}^{i-1} \frac{(-1)^m \cdot m \cdot \mathcal{C}_{m}^{i-1}}{m+p}- \sum_{m=0}^{i-1} \frac{(-1)^m \cdot m \cdot  \mathcal{C}_{m}^{i-1}}{m+p}=0.
\end{align*}
In order to complete the proof, we show the uniqueness of the probability distribution function $\lambda$ in our setting. First, note that Equation~\eqref{eq: prob} relates probability distribution  $\lambda$
over consideration sets to the choice frequencies $\mathbb{P}_j \big(X\big)$ through the system of linear equations:
\begin{align} 
	\mathbb{P}_j(X)&=\sum_{C \in N_j} \lambda_C \cdot \chi_C(X), \ \forall \ X \subseteq N \Longleftrightarrow \boldsymbol{y}=A \cdot \boldsymbol{ \lambda},  \label{Equation_linear_6}
\end{align} 
where $\boldsymbol{y}=(y_X)_{X	 \subseteq N_j}$ denotes the $\abs{2^{N-1}} \times 1$ vector of choice fractions and $\boldsymbol{\lambda}=(\lambda_{C})_{C \in N_j}$ denotes the $\abs{2^{N-1}} \times 1$ vector that represents the probability distribution function over consideration sets. $A$ is the $\abs{2^{N-1}}\times \abs{2^{N-1}}$ matrix such that 
$A$'s entry corresponding to the row $X$ and column $C$ is equal to $\chi_{C}(X)$. As a result, the relation between the choice probabilities and the underlying model can be represented
in a compact form as $\boldsymbol{y}=A \cdot \boldsymbol{ \lambda}$. Then the proof of the uniqueness of $\lambda$ reduces to showing that $\det(A) \ne 0$. It follows from  Equation~\eqref{eq: prob2} that
\begin{align*}
	\lambda_C&=\sum_{X \in N_j} \mathbb{P}_j(X) \cdot (\psi_C(X)+\varphi_C(X)), \ \forall C \in N_j \Longleftrightarrow \boldsymbol{ \lambda}=B\cdot \boldsymbol{y}, 
\end{align*}
which provides another relationship between choice frequencies $ \mathbb{P}_j(X)$ and the model parameters $\lambda$ in a linear form as $\boldsymbol{ \lambda}=B\cdot \boldsymbol{y}$, where $B$ is the $\abs{2^{N-1}}\times \abs{2^{N-1}}$ matrix such that $B$'s entry corresponding to the row $C$ and column $X$ is equal to $\psi_{C}(X)+\varphi_C(X)$. Therefore, we have the following set of equalities: 
\begin{align*}
	\boldsymbol{ \lambda}&=B\cdot \boldsymbol{y}= B\cdot A \cdot \lambda, \ \ \big[ \text{\scriptsize by Equation (\ref{Equation_linear_6})}\big]  \\  &\Longrightarrow \ I=B \cdot A  \Longrightarrow \ \det(I)=\det(B) \cdot \det(A) \\ &\Longrightarrow 1=\det(B) \cdot \det(A) \ \Longrightarrow \ \det(A)\ne 0.
\end{align*}
\hfill $\square$

\subsection{Proof of Theorem~\ref{theorem: characterization}}
\label{subsec:proof_axiom_theorem}

As in the proof of Theorem~\ref{theorem:inference} in Section \ref{subsec:proof_of_theorem_inference}, we slightly abuse the notation for the calligraphic $\Ccal$, using $\Ccal^n_m = \frac{ n!  }{m!(n-m)!}$ to denote the binomial coefficient. This choice is made to enhance readability, as the standard notation $\binom{n}{m}$ can become cumbersome when multiple parentheses appear in the same equation.

\noindent\underline{The Auxiliary Lemmas}\\
We first present two lemmas with proofs which are used below to prove Theorem~\ref{theorem: characterization}. First, we establish the following combinatorial identity. 
\begin{lemma} \label{lemma: combinatorial}
	For all $n$ and $m$ such that $m \le n$ the combinatorial identity below holds
	$$\sum_{k=0}^{m-1} (-1)^k \cdot \mathcal{C}_k^{m-1} \cdot \bigg(\frac{n-m}{n-m+k}-\frac{n}{n-m+1+k} \bigg)=0.$$
\end{lemma}
\textsc{Proof:}
We prove it by induction on $m$: \\
\emph{\underline{Base case:}} $m=1$.
\begin{align*}
\sum_{k=0}^{m-1} (-1)^k \cdot \mathcal{C}_k^{m-1} \cdot \bigg(\frac{n-m}{n-m+k}-\frac{n}{n-m+1+k} \bigg)=(1-1)=0
\end{align*}
\emph{\underline{Induction hypothesis:}} $m=p$. \\
\begin{align*}
	0&=\sum_{k=0}^{p-1} (-1)^k \cdot \mathcal{C}_k^{p-1} \cdot \bigg(\frac{n-p}{n-p+k}-\frac{n}{n-p+1+k} \bigg) \\
	&= (p-1)! \cdot \left[ \sum_{k=0}^{p-1} (-1)^k \cdot \frac{ (n-p)}{(p-1-k)! \cdot k! \cdot(n-p+k)}-\sum_{k=0}^{p-1} (-1)^k \cdot \frac{ n}{(p-1-k)! \cdot k! \cdot(n-p+1+k)} \right] \\
	&= (p-1)! (n-p) \left[ \sum_{k=0}^{p-1} (-1)^k \cdot \frac{ 1}{(p-1-k)! \cdot k! \cdot(n-p+k)} -\frac{n}{(n-p)} \cdot \sum_{k=0}^{p-1} (-1)^k \cdot \frac{ 1}{(p-1-k)! \cdot k! \cdot(n-p+1+k)} \right].
\end{align*}
Thus, it implies
\begin{align*}
0 = \sum_{k=0}^{p-1} (-1)^k \cdot \frac{ 1}{(p-1-k)! \cdot k! \cdot(n-p+k)} -\frac{n}{(n-p)} \cdot \sum_{k=0}^{p-1} (-1)^k \cdot \frac{ 1}{(p-1-k)! \cdot k! \cdot(n-p+1+k)}.
\end{align*}
\emph{\underline{Induction step:}} $m=p+1$.
\begin{align*}
	&\sum_{k=0}^{p} (-1)^k \cdot \mathcal{C}_k^{p} \cdot \bigg(\frac{n-p-1}{n-p-1+k}-\frac{n}{n-p+k} \bigg) \\
	&= p! \cdot \left[ \sum_{k=0}^{p} (-1)^k \cdot \frac{ (n-p-1)}{(p-k)! \cdot k! \cdot(n-p-1+k)}-\sum_{k=0}^{p} (-1)^k \cdot \frac{ n}{(p-k)! \cdot k! \cdot(n-p+k)} \right] \\
	&= p! \cdot \left[ \sum_{k=0}^{p} (-1)^k \cdot \frac{ (n-p-1)}{(p-k)! \cdot k! \cdot(n-p-1+k)}-\frac{n}{n-p} \cdot \sum_{k=0}^{p} (-1)^k \cdot \frac{ n-p}{(p-k)! \cdot k! \cdot(n-p+k)} \right] \\
	&= p! \cdot \left[ \sum_{k=0}^{p} (-1)^k \cdot \frac{ (n-p-1+k-k)}{(p-k)! \cdot k! \cdot(n-p-1+k)}-\frac{n}{n-p} \cdot \sum_{k=0}^{p} (-1)^k \cdot \frac{ n-p+k-k}{(p-k)! \cdot k! \cdot(n-p+k)} \right] \\
	&=p! \cdot \left[ \sum_{k=0}^{p} (-1)^k \cdot \frac{ (-k)}{(p-k)! \cdot k! \cdot(n-p-1+k)}-\frac{n}{n-p} \cdot \sum_{k=0}^{p} (-1)^k \cdot \frac{-k}{(p-k)! \cdot k! \cdot(n-p+k)} \right]\\
	& \scriptsize \big[\text{since $\sum_{k=0}^{p} (-1)^k \cdot \frac{1}{(p-k)! \cdot k! }= \frac{1}{p!} \cdot \sum_{k=0}^{p} (-1)^k \cdot \mathcal{C}_k^p=0$} \big] \\
	&= p! \cdot \left[ \sum_{k=1}^{p} (-1)^k \cdot \frac{ (-k)}{(p-k)! \cdot k! \cdot(n-p-1+k)}-\frac{n}{n-p} \cdot \sum_{k=1}^{p} (-1)^k \cdot \frac{-k}{(p-k)! \cdot k! \cdot(n-p+k)}\right] \\
	&= p! \cdot \left[ \sum_{k=1}^{p} (-1)^k \cdot \frac{-1}{(p-k)! \cdot (k-1)! \cdot(n-p-1+k)} -\frac{n}{n-p} \cdot \sum_{k=1}^{p} (-1)^k \cdot \frac{-1}{(p-k)! \cdot (k-1)! \cdot(n-p+k)} \right] \\
	&= p! \cdot \left[ \sum_{k=0}^{p-1} (-1)^k \cdot \frac{ 1}{(p-k-1)! \cdot k! \cdot(n-p+k)} -\frac{n}{n-p} \cdot \sum_{k=1}^{p} (-1)^k \cdot \frac{1}{(p-k-1)! \cdot k! \cdot(n-p+k+1)} \right] \\ & =0  \qquad [\text{\scriptsize by the induction hypothesis}].
\end{align*}
\hfill $\square$

Then, we show that if the symmetric cannibalization axiom holds then choice data satisfy the equality below. 
\begin{lemma} \label{lemma: theorem}
	If for all $S \subseteq N$ and $i,j \in S$ we have that 
	$$\mathbb{P}_j(S)-\mathbb{P}_j(S \setminus \{i\})=\mathbb{P}_i(S)-\mathbb{P}_i(S \setminus \{j\})$$
	then for all $S \subseteq N$ and $j \in S$ we have that
	\begin{align*}
		\mathbb{P}_j(S)&=\sum_{Y \subseteq S} \mathbb{P}_0(Y) \cdot \bigg(\mathbb{I}[j \not \in Y] \cdot \sum_{k=0}^{\abs{Y}} (-1)^k \frac{\mathcal{C}_k^{\abs{Y}}}{\abs{S}-\abs{Y}+k} -\mathbb{I}[j \in Y] \cdot \sum_{k=0}^{\abs{Y}-1} (-1)^k \frac{\mathcal{C}_k^{\abs{Y}-1}}{\abs{S}-\abs{Y}+k+1} \bigg).
	\end{align*}
\end{lemma}
\textsc{Proof:}
We prove it by induction on $\abs{S}$: \\
\emph{\underline{Base case:}} $\abs{S}=1$.
$$\mathbb{P}_j(\{j\})=1-\mathbb{P}_0(\{j\})$$
\emph{\underline{Induction hypothesis:}} the equation holds to compute $\mathbb{P}_j(S \setminus \{i\})$ for all $i \in N \setminus \{j\}$.
\begin{align*}
	\mathbb{P}_j(S \setminus \{i\})=\sum_{Y \subseteq S \setminus \{i\}} \mathbb{P}_0(Y) \cdot \bigg(\mathbb{I}[j \not \in Y] \cdot \sum_{k=0}^{\abs{Y}} (-1)^k \frac{\mathcal{C}_k^{\abs{Y}}}{\abs{S}-1-\abs{Y}+k}- \mathbb{I}[j \in Y] \cdot \sum_{k=0}^{\abs{Y}-1} (-1)^k \frac{\mathcal{C}_k^{\abs{Y}-1}}{\abs{S}-\abs{Y}+k} \bigg).\end{align*}
\emph{\underline{Induction step:}} the equation holds to compute $\mathbb{P}_j(S)$.
\begin{align*}
	&\mathbb{P}_j(S)=\mathbb{P}_i(S) + \mathbb{P}_j(S \setminus \{i\})-\mathbb{P}_i(S \setminus \{j\}) \\ 
	\Rightarrow & \sum_{i \in S \setminus \{j\}} \mathbb{P}_j(S) = \sum_{i \in S \setminus \{j\}} \mathbb{P}_i(S) + \sum_{i \in S \setminus \{j\}} \mathbb{P}_j(S \setminus \{i\}) -\sum_{i \in S \setminus \{j\}} \mathbb{P}_i(S \setminus \{j\}) \\
	\Rightarrow & \sum_{i \in S \setminus \{j\}} \mathbb{P}_j(S) =1-\mathbb{P}_0(S)-\mathbb{P}_j(S)+ \sum_{i \in S \setminus \{j\}} \mathbb{P}_j(S \setminus \{i\})-\sum_{i \in S \setminus \{j\}} \mathbb{P}_i(S \setminus \{j\}) \\
	\Rightarrow & \sum_{i \in S \setminus \{j\}} \mathbb{P}_j(S) =1-\mathbb{P}_0(S)-\mathbb{P}_j(S)+ \left( \sum_{i \in S \setminus \{j\}} \mathbb{P}_j(S \setminus \{i\}) \right) -(1-\mathbb{P}_0(S \setminus \{j\})) \\ \Rightarrow
	& \,\, \abs{S} \cdot \mathbb{P}_j(S) = \mathbb{P}_0(S \setminus \{j\})-\mathbb{P}_0(S) + \sum_{i \in S \setminus \{j\}} \mathbb{P}_j(S \setminus \{i\})  \\ \Rightarrow 
	& \sum_{i \in S \setminus \{j\}} \mathbb{P}_j(S \setminus \{i\})= \mathbb{P}_j(S) \cdot \abs{S}+\mathbb{P}_0(S) -\mathbb{P}_0(S \setminus \{j\}).
\end{align*}
Therefore, it is sufficient to prove that 
\begin{align*}
	&\sum_{i \in S \setminus \{j\}} \mathbb{P}_j(S \setminus \{i\})= \sum_{Y \subseteq S} \mathbb{P}_0(Y) \cdot \bigg(\mathbb{I}[j \not \in Y] \cdot \sum_{k=0}^{\abs{Y}} (-1)^k \frac{\mathcal{C}_k^{\abs{Y}}}{\abs{S}-\abs{Y}+k} \\
	&-\mathbb{I}[j \in Y] \cdot \sum_{k=0}^{\abs{Y}-1} (-1)^k \frac{\mathcal{C}_k^{\abs{Y}-1}}{\abs{S}-\abs{Y}+k+1} \bigg) \cdot \abs{S}+\mathbb{P}_0(S) -\mathbb{P}_0(S \setminus \{j\}).
\end{align*}
This way we complete the proof:
\begin{align*}
	&\sum_{i \in S \setminus \{j\}} \mathbb{P}_j(S \setminus \{i\}) \\ = & \sum_{i \in S \setminus \{j\}} \sum_{Y \subseteq S \setminus \{i\}} \mathbb{P}_0(Y) \cdot \bigg(\mathbb{I}[j \not \in Y] \cdot \sum_{k=0}^{\abs{Y}} (-1)^k \frac{\mathcal{C}_k^{\abs{Y}}}{\abs{S}-1-\abs{Y}+k} -\mathbb{I}[j \in Y] \cdot \sum_{k=0}^{\abs{Y}-1} (-1)^k \frac{\mathcal{C}_k^{\abs{Y}-1}}{\abs{S}-\abs{Y}+k} \bigg) \\ & \big[\text{\scriptsize by induction hypothesis} \big] \\
	= & \sum_{Y \subseteq S} \mathbb{P}_0(Y) \cdot \bigg(\mathbb{I}[j \not \in Y] \cdot \sum_{k=0}^{\abs{Y}} (-1)^k \frac{\mathcal{C}_k^{\abs{Y}}}{\abs{S}-1-\abs{Y}+k} \cdot (\abs{S}-\abs{Y}-1) \\
	&\hspace{3cm}-\mathbb{I}[j \in Y] \cdot \sum_{k=0}^{\abs{Y}-1} (-1)^k \frac{\mathcal{C}_k^{\abs{Y}-1}}{\abs{S}-\abs{Y}+k} \cdot (\abs{S}-\abs{Y})  \bigg) +\mathbb{P}_0(S) - \mathbb{P}_0(S \setminus \{j\})\\
	&\scriptsize \hspace{-0.03in}  \big[\text{(1) in the summation $\sum_{i \in S \setminus \{j\}} \sum_{Y \subseteq S \setminus \{i\}}$, $Y$ can be any subset of $S$ but $S$ and $S \setminus \{j\}$; } \\
	& \scriptsize \text{(2) in the summation $\sum_{i \in S \setminus \{j\}} \sum_{Y \subseteq S \setminus \{i\}}$, if $j \in Y$, then $Y$ is assigned to a specific subset of $S$ for $\abs{S}-\abs{Y}$ times; } \\
	&\scriptsize \text{(3) in the summation $\sum_{i \in S \setminus \{j\}} \sum_{Y \subseteq S \setminus \{i\}}$, if $j \not \in Y$, then $Y$ is assigned to a specific subset of $S$ for $\abs{S}-\abs{Y}-1$ times. } \big] \\
	&= \sum_{Y \subseteq S} \mathbb{P}_0(Y) \cdot \bigg(\mathbb{I}[j \not \in Y] \cdot \sum_{k=0}^{\abs{Y}} (-1)^k \frac{\mathcal{C}_k^{\abs{Y}}}{\abs{S}-1-\abs{Y}+k} \cdot (\abs{S}-\abs{Y}-1) \\
	& \hspace{3cm} -\mathbb{I}[j \in Y] \cdot \sum_{k=0}^{\abs{Y}-1} (-1)^k \frac{\mathcal{C}_k^{\abs{Y}-1}}{\abs{S}-\abs{Y}+k+1} \cdot \abs{S} \bigg)+\mathbb{P}_0(S) - \mathbb{P}_0(S \setminus \{j\})  \\
	&\scriptsize \big[\text{by invoking Lemma~\ref{lemma: combinatorial}, where $m=\abs{Y}$ and $n=\abs{S}$} \big] \\
	&= \sum_{Y \subseteq S} \mathbb{P}_0(Y) \cdot \bigg(\mathbb{I}[j \not \in Y] \cdot \sum_{k=0}^{\abs{Y}} (-1)^k \frac{\mathcal{C}_k^{\abs{Y}}}{\abs{S}-\abs{Y}+k} \cdot \abs{S} \\
	& \hspace{3cm} -\mathbb{I}[j \in Y] \cdot \sum_{k=0}^{\abs{Y}-1} (-1)^k \frac{\mathcal{C}_k^{\abs{Y}-1}}{\abs{S}-\abs{Y}+k+1} \cdot \abs{S}  \bigg) +\mathbb{P}_0(S) - \mathbb{P}_0(S \setminus \{j\})\\
	&\scriptsize \big[\text{by invoking Lemma~\ref{lemma: combinatorial}, where $m=\abs{Y}+1$ and $n=\abs{S}$} \big] \\
	&= \sum_{Y \subseteq S} \mathbb{P}_0(Y) \cdot \bigg(\mathbb{I}[j \not \in Y] \cdot \sum_{k=0}^{\abs{Y}} (-1)^k \frac{\mathcal{C}_k^{\abs{Y}}}{\abs{S}-\abs{Y}+k} \\
	&\hspace{3cm} -\mathbb{I}[j \in Y] \cdot \sum_{k=0}^{\abs{Y}-1} (-1)^k \frac{\mathcal{C}_k^{\abs{Y}-1}}{\abs{S}-\abs{Y}+k+1}   \bigg) \cdot \abs{S} +\mathbb{P}_0(S) - \mathbb{P}_0(S \setminus \{j\})\\
\end{align*}
\hfill $\square$

\noindent\underline{Proof of Theorem~\ref{theorem: characterization}}\\
We first prove the ``$\Leftarrow$'' direction, i.e., sufficiency. To simplify the exposition, we also let $\bar{X}:=N \setminus X$ and $X^+:=X \cup \{0\}$. Let $ \langle S \rangle$ denote the power set of $S$, i.e.,  $\langle S \rangle=2^S$, and let $A \uplus B$ denote $\{a \cup b: a \in A, b \in B \}$ for any sets $A,B$.
we claim that a choice model that satisfies the symmetric cannibalization and default regularity is a stochastic set model with a probability distribution function $\lambda$ over preselected sets.
\begin{align*}
	& \mathbb{P}_j(S)\\ = & \sum_{Y \subseteq S} \mathbb{P}_0(Y) \cdot \bigg(\mathbb{I}[j \not \in Y] \cdot \sum_{k=0}^{\abs{Y}} (-1)^k \frac{\mathcal{C}_k^{\abs{Y}}}{\abs{S}-\abs{Y}+k} -\mathbb{I}[j \in Y] \cdot \sum_{k=0}^{\abs{Y}-1} (-1)^k \frac{\mathcal{C}_k^{\abs{Y}-1}}{\abs{S}-\abs{Y}+k+1} \bigg)\\ & [\text{\scriptsize by invoking Lemma~\ref{lemma: theorem}}] \\
	= & \sum_{X_1 \subseteq S} \mathbb{P}_0(S \setminus X_1) \cdot \bigg(\mathbb{I}[j  \in X_1] \cdot \sum_{k=0}^{\abs{S}-\abs{X_1}} (-1)^k \frac{\mathcal{C}_k^{\abs{S} -\abs{X_1}}}{\abs{X_1}+k} -\mathbb{I}[j \not \in X_1] \cdot \sum_{k=0}^{\abs{S}-\abs{X_1}-1} (-1)^k \frac{\mathcal{C}_k^{\abs{S}-\abs{X_1}-1}}{\abs{X_1}+k+1} \bigg) \\& [\text{\scriptsize where $X_1=S \setminus Y$}] \\
	= & \sum_{X_1 \subseteq S} \mathbb{P}_0(S \setminus X_1) \cdot (-1)^{-\abs{X_1}} \cdot \bigg(\mathbb{I}[j  \in X_1] \cdot \sum_{k=0}^{\abs{S}-\abs{X_1}} (-1)^{k+\abs{X_1}} \frac{\mathcal{C}_k^{\abs{S}-\abs{X_1}}}{\abs{X_1}+k} \\
	& \hspace{5cm} +\mathbb{I}[j \not \in X_1] \cdot \sum_{k=0}^{\abs{S}-\abs{X_1}-1} (-1)^{k+\abs{X_1}+1} \frac{\mathcal{C}_k^{\abs{S}-\abs{X_1}-1}}{\abs{X_1}+k+1} \bigg) \\
	= & \sum_{X_1 \subseteq S} \mathbb{P}_0(S \setminus X_1) \cdot (-1)^{-\abs{X_1}} \cdot \bigg(\mathbb{I}[j  \in X_1] \cdot \sum_{k=\abs{X_1}}^{\abs{S}} (-1)^{k} \frac{\mathcal{C}_{k-\abs{X_1}}^{\abs{S}-\abs{X_1}}}{k} \\
	&\hspace{5cm}+\mathbb{I}[j \not \in X_1] \cdot \sum_{k=\abs{X_1}+1}^{\abs{S}} (-1)^{k} \frac{\mathcal{C}_{k-\abs{X_1}-1}^{\abs{S}-\abs{X_1}-1}}{k} \bigg) \\
	= & \sum_{X_1 \subseteq S} \mathbb{P}_0(S \setminus X_1) \cdot (-1)^{-\abs{X_1}} \cdot \bigg(\mathbb{I}[j  \in X_1] \cdot \sum_{\substack{C_1 \supseteq X_1: \\ C_1 \subseteq S}} \frac{ (-1)^{\abs{C_1}}}{\abs{C_1}}  \\
	&\hspace{5cm}+\mathbb{I}[j \not \in X_1] \cdot  \sum_{\substack{C_1 \supseteq \{X_1 \cup \{j\}\}: \\ C_1 \subseteq S}} \frac{ (-1)^{\abs{C_1}}}{\abs{C_1}} \bigg) \\
	= & \sum_{X_1 \subseteq S} \mathbb{P}_0(S \setminus X_1) \cdot (-1)^{-\abs{X_1}} \cdot  \sum_{\substack{C_1 \supseteq X_1: \\C_1 \subseteq S, \\ j \in C_1}} \frac{ (-1)^{\abs{C_1}}}{\abs{C_1}}  \\
	= & \sum_{X_1 \subseteq S} \ \ \sum_{\substack{C_1 \in \langle S \setminus X_1 \rangle \uplus X_1 : \\ j \in C_1}}  \frac{ (-1)^{\abs{C_1}-\abs{X_1}} }{\abs{C_1}}  \cdot  \mathbb{P}_0(S \setminus X_1) \\
	= & \sum_{X_1 \subseteq S} \ \ \sum_{\substack{C_1 \in \langle S \setminus X_1 \rangle \uplus X_1 : \\ j \in C_1}}  \frac{ (-1)^{\abs{C_1}-\abs{X_1}} }{\abs{C_1}}  \cdot  \mathbb{P}_0(S \setminus X_1) \cdot \sum_{X_2 \subseteq N \setminus S} \ \ \sum_{C_2 \in \langle \{N \setminus S\} \setminus X_2 \rangle  } (-1)^{\abs{C_2}} \\
	&\scriptsize  [\text{since the summation over $X_2$ and $C_2$ is equal to 1 if $X_2=N\setminus S$ and 0, otherwise}] \\
	= &\sum_{X_1 \subseteq S} \ \ \sum_{\substack{C_1 \in \langle S \setminus X_1 \rangle \uplus X_1 : \\ j \in C_1}}  \frac{ (-1)^{\abs{C_1}-\abs{X_1}} }{\abs{C_1}}  \cdot  \mathbb{P}_0(S \setminus X_1) \cdot \sum_{X_2 \subseteq N \setminus S} (-1)^{-\abs{X_2}} \cdot  \sum_{C_2 \in \langle \{N \setminus S\} \setminus X_2 \rangle \uplus X_2 } (-1)^{\abs{C_2}} \\
	= &\sum_{X_1 \subseteq S} \ \ \sum_{\substack{C_1 \in \langle S \setminus X_1 \rangle \uplus X_1 : \\ j \in C_1}} \ \ \sum_{X_2 \subseteq N \setminus S}  \ \ \sum_{C_2 \in \langle \{N \setminus S\} \setminus X_2 \rangle \uplus X_2 }  \frac{ (-1)^{\abs{C_1}+\abs{C_2}-\abs{X_1}-\abs{X_2}} }{\abs{C_1}}  \cdot  \mathbb{P}_0(S \setminus X_1)\\
	= & \sum_{X_1 \subseteq S} \ \sum_{X_2 \subseteq N \setminus S}  \ \ \sum_{\substack{C_1 \in \langle S \setminus X_1 \rangle \uplus X_1 : \\ j \in C_1}} \ \  \sum_{C_2 \in \langle \{N \setminus S\} \setminus X_2 \rangle \uplus X_2 }  \frac{ (-1)^{\abs{C}-\abs{X}} }{\abs{C_1}}  \cdot  \mathbb{P}_0(S \setminus X_1)\\
	&\scriptsize [\text{where $X=X_1 \cup X_2$ and $C=C_1 \cup C_2$.}] \\
	= & \sum_{X_1 \subseteq S} \ \sum_{X_2 \subseteq N \setminus S}  \ \ \sum_{\substack{C_1 \in \langle S \setminus X_1 \rangle \uplus X_1 : \\ j \in C_1}} \ \  \sum_{C_2 \in \langle \{N \setminus S\} \setminus X_2 \rangle \uplus X_2 }  \frac{ (-1)^{\abs{C}-\abs{X}} }{\abs{C_1}}  \cdot  \mathbb{P}_0(N \setminus X)\\
	&\scriptsize [\text{since the summation is nonzero only if $X_2=N\setminus S$}] \\
	= & \sum_{\substack{C \subseteq N: \\ j \in C}} \sum_{X \subseteq C} (-1)^{\abs{C}-\abs{X}} \cdot \frac{1}{\abs{C \cap S}} \cdot \mathbb{P}_0(N \setminus X) \\
	= & \sum_{\substack{C \subseteq N: \\ j \in C}}  \frac{1}{\abs{C \cap S}} \cdot \sum_{X \subseteq C} (-1)^{\abs{C}-\abs{X}} \cdot \mathbb{P}_0(N \setminus X) \\
	= & \sum_{ C \subseteq N} \lambda_C \cdot \mathbf{I}[j \in S] \cdot \mathbf{I}[j \in C] \cdot \frac{1}{\abs{C \cap S}}, \text{ where } \lambda_C= \sum_{X \subseteq C} (-1)^{\abs{C}-\abs{X}} \cdot \mathbb{P}_0(N \setminus X), 
\end{align*}
which is exactly the equation to compute the probability of purchasing $j \in S$ under the offer set $S \subseteq N$ under the consideration set model. We note that this model is well-defined since we have that
\begin{align*}
\lambda_C= \sum_{X \subseteq C} (-1)^{\abs{C}-\abs{X}} \cdot \mathbb{P}_0(N \setminus X) = \sum_{ \bar{C} \subseteq \bar{X} } (-1)^{\abs{\bar{X}} - \abs{\bar{C}}  } \cdot \Pbb_0(\bar{X}) = H(0, \bar{C}) \geq 0,
\end{align*}
where the non-negativeness is provided by the default regularity. Moreover, it follows from Theorem~\ref{theorem:inference} that $\lambdab$ is defined uniquely. 

Finally, to prove the necessity of the theorem (the ``$\Rightarrow$'' direction), it suffices to show that $\mathbb{P}_j(S \setminus \{k\}) - \mathbb{P}_j(S)$ is invariant to the exchange of the indexes $j$ and $k$, which is shown below. 
\begin{align*}
	&\mathbb{P}_j(S \setminus \{k\}) - \mathbb{P}_j(S)\\ = & \sum_{\substack{C \subseteq N: \\ j \in C} } \frac{\lambda_C}{\abs{C \cap \{S \setminus \{k\}\}}} - \sum_{\substack{C \subseteq N: \\ j \in C} } \frac{\lambda_C}{\abs{C \cap S }} \\
	= & \sum_{\substack{C \subseteq N: \\ j \in C \\ k \in C} } \frac{\lambda_C}{\abs{C \cap \{S \setminus \{k\}\}}} + \sum_{\substack{C \subseteq N: \\ j \in C \\ k \not \in C} } \frac{\lambda_C}{\abs{C \cap \{S \setminus \{k\}\}}}- \sum_{\substack{C \subseteq N: \\ j \in C} } \frac{\lambda_C)}{\abs{C \cap S }} \\
	= & \sum_{\substack{C \subseteq N: \\ j \in C \\ k \in C} } \frac{\lambda_C}{\abs{C \cap \{S \setminus \{k\}\}}} + \sum_{\substack{C \subseteq N: \\ j \in C \\ k \not \in C} } \frac{\lambda_C}{\abs{C \cap \{S \setminus \{k\}\}}}- \sum_{\substack{C \subseteq N: \\ j \in C \\ k \in C} } \frac{\lambda_C}{\abs{C \cap S }}- \sum_{\substack{C \subseteq N: \\ j \in C \\ k \not \in C} } \frac{\lambda_C}{\abs{C \cap S }} \\
	= & \sum_{\substack{C \subseteq N: \\ j \in C \\ k \in C} } \frac{\lambda_C}{\abs{C \cap \{S \setminus \{k\}\}}} + \sum_{\substack{C \subseteq N: \\ j \in C \\ k \not \in C} } \frac{\lambda_C}{\abs{C \cap S }}- \sum_{\substack{C \subseteq N: \\ j \in C \\ k \in C} } \frac{\lambda_C}{\abs{C \cap S }}- \sum_{\substack{C \subseteq N: \\ j \in C \\ k \not \in C} } \frac{\lambda_C}{\abs{C \cap S }} \\
	= & \sum_{\substack{C \subseteq N: \\ j \in C \\ k \in C} } \frac{\lambda_C}{\abs{C \cap \{S \setminus \{k\}\}}} - \sum_{\substack{C \subseteq N: \\ j \in C \\ k \in C} } \frac{\lambda_C}{\abs{C \cap S }}= \sum_{\substack{C \subseteq N: \\ j \in C \\ k \in C} } \frac{\lambda_C}{\abs{C \cap S }-1} - \sum_{\substack{C \subseteq N: \\ j \in C \\ k \in C} } \frac{\lambda_C}{\abs{C \cap S }}.
\end{align*}

\section{Supplementary Proofs and Results for Section~\ref{sec:assortment_optimization}}

\subsection{Proof of Theorem~\ref{thm:revenue_order_in_block}}
\label{subsec:proof_of_thm_revenue_order}

First, it is worth observing that each product $i$ belongs to exactly one block $I_{\bb} \in \mathcal{I}$. This follows directly from the construction of the blocks $I_{\Bb} \in \mathcal{I}$. Specifically, a product $i$ belongs to a block $I_{\bb}$ if and only if the following two conditions are satisfied: (a)~product~$i$ is included in all consideration sets $C_j$ where $b_j=1$; and (b)~product~$i$ is not included in any consideration sets $C_j$ where $b_j=0$. If either of these two conditions is violated for a specific $I_{\bb} \in \mathcal{I}$, it is clear that product $i$ cannot belong to that block, as determined by Equation~\eqref{eq: definition}. Consequently, each product $i$ uniquely belongs to the block $I_{\bb}$ where $b_j = \mathbb{I}[{i \in C_j}]$ for all $j \in K$. It also implies that \YCRminor{the non-empty sets in $\mathcal{I} = \{  I_{\bb} \mid \bb \in \Bcal  \}$ form a partition of $N$}. Thus, in what follows below, we can rewrite the revenue function $\Rev_{C_j}(S)$ defined in Equation~\eqref{def:revenue_of_a_customer_type}. Specifically, when $|S \cap C_j| \geq 1$, we have the following: 
\begin{align}
	\label{eq:revenue_by_blocks}
	\text{Rev}_{C_j}(S) = \frac{  \sum_{i \in S \cap C_j} r_i  }{  \abs{ S \cap C_j} }   =  \frac{   \sum_{{\bb \in \Bcal  } } \sum_{i \in S \cap C_j \cap I_\bb} r_i  }{ \sum_{{\bb \in \Bcal} } \abs{S \cap C_j \cap I_\bb}   }  
	=  \frac{   \sum_{{\bb \in \Bcal :b_j =1} } \sum_{i \in S \cap I_\bb} r_i  }{ \sum_{ \bb \in \Bcal : b_j = 1 } | S \cap I_{\bb} |   },
\end{align}
where the second equality holds because $\mathcal{I} = \{  I_{\bb} \mid \bb \in \Bcal \}$ partitions $N$, and the third equality follows from the fact that, for any binary vector $\bb \in \Bcal$ and any $j \in K$, we have the following equality: 
\begin{align*}
	S \cap C_j \cap I_\bb = \begin{cases}
		S \cap I_\bb, & \text{if $b_j = 1$},\\
		\emptyset, & \text{if $b_j = 0$}.
	\end{cases}
\end{align*}

In what follows next, we fix an arbitrary assortment $S \subseteq N$. We claim that if $S \cap I_{\bb}$ is not revenue-ordered in any block $I_\bb \in \mathcal{I}$, then $S$ is not an optimal assortment. We prove this claim by construction. To this end, we assume that $S \cap I_{\bb^\dagger}$ is not revenue-ordered in block $I_{\bb^\dagger}$ for a binary vector $\bb^\dagger$. \YCRminor{Next, we construct a new assortment $S'$ such that: (a) $S' \cap I_{{\bb}} = S \cap I_{{\bb}}$ for all ${\bb} \neq \bb^\dagger$; and (b) $S' \cap I_{\bb^\dagger}$ consists of the most expensive $|S \cap I_{\bb^\dagger}|$ products in $I_{\bb^\dagger}$. In other words, under the partition $\mathcal{I}$, the two assortments $S$ and $S'$ coincide except in the block $I_{\bb^\dagger}$.} Note that given $S \cap I_{\bb^\dagger}$ is not revenue-ordered, we know that $S \cap I_{\bb^\dagger}$ is not the empty set. Furthermore, it can be seen that $\sum_{i \in S \cap I_{\bb}} r_i = \sum_{i \in S' \cap I_{\bb}} r_i  $ for any $
\bb \neq \bb^\dagger$ and $\sum_{i \in S \cap I_{\bb^\dagger}} r_i < \sum_{i \in S' \cap I_{\bb^\dagger}} r_i $, while $| S \cap I_{\bb}  | = | S' \cap I_{\bb}  |  $ for all $\bb \in \Bcal$. Those observations imply that $S'$ achieves a higher revenue than $S'$. To formalize this claim, we consider the revenue function $\Rev_{C_j}(S)$ of each customer type $j$ as specified in Equation~\eqref{eq:revenue_by_blocks}. We analyze the following two cases:
\begin{itemize}
    \item If $b^{\dagger}_j = 0$, we have $\Rev_{C_j} (S) = \Rev_{C_j} (S')$. This is because the products in block $I_{\bb^\dagger}$ are not considered by customer type $j$ at all if $b_j^\dagger = 0$.
    \item If $b^{\dagger}_j = 1$, then we have the following expression: 
\begin{align*}
	\Rev_{C_j}(S) = & \frac{ \sum_{i \in S \cap I_{\bb^\dagger}}  r_i +  \sum_{{\bb \in \Bcal :  \bb \neq \bb^\dagger, b_j =1} } \sum_{i \in S \cap I_\bb} r_i  }{ \sum_{ \bb \in \Bcal :  b_j = 1 } | S \cap I_{\bb} |    } \\ < & \frac{ \sum_{i \in S' \cap I_{\bb^\dagger}}  r_i +  \sum_{{\bb \in \Bcal :  \bb \neq \bb^\dagger, b_j =1} } \sum_{i \in S' \cap I_\bb} r_i  }{ \sum_{ \bb \in \Bcal :  b_j = 1 } | S' \cap I_{\bb} |    }   =  \Rev_{C_j}(S'). 
\end{align*}
\end{itemize}
As the non-emptiness of $S \cap I_{\bb^\dagger}$ implies the non-emptiness of $I_{\bb^\dagger}$, we know that there must exist at least one customer type $j^* \in K$ such that $b^\dagger_{j^*} = 1$. Specifically, if $b^\dagger_j = 0$ for all $j \in K$, the non-emptiness of $I_{\bb^\dagger}$ implies that there would exist a product $i \in I_{\bb^\dagger} = \cap_{j \in K} \eta_{b^\dagger_j}(C_j) = \cap_{j \in K} \eta_{0}(C_j) = \cap_{j \in K} \bar{C_j}$, violating the assumption of $\cup_{j \in K} C_j = N$ at the beginning of Section~\ref{sec:assortment_optimization}. Due to the existence of this customer type $j^*$, we have $\Rev(S) = \sum_{j \in K} \lambda_j \cdot \Rev_{C_j}(S) < \sum_{j \in K} \lambda_j \cdot \Rev_{C_j}(S') = \Rev(S')$. Thus, $S$ is not optimal. \hfill $\square$

\subsection{Proof of Proposition~\ref{prop:AO_NP_hardness_with_small_size}}
\label{subsec:proof_AO_NP_hardness_small_size}

We prove the statement by constructing a reduction from the vertex cover problem, which is known to be an NP-hard problem \citep{garey1979computers}, to the assortment optimization problem under the stochastic set model. 
First, we recall how the latter problem is defined.  To start with, let $G$ denote a graph with a collection of nodes $V=\{1,...,n\}$ and edges $E$, i.e., $G = (V,E)$. Next, we say that a subset $U \subseteq V$ is a vertex cover if for every edge $e = (i,j) \in E$ either $i \in U$ or $j \in U$. Then, the vertex cover problem addresses the following question: ``Given a graph $G = (V,E)$ and an integer $k$, is there a vertex cover of size at most $k$?''

In what follows, we describe the reduction $\Phi$ that maps any instance of the vertex cover (VC) problem, i.e., $I_{\text{VC}} = (G,k)$, to the instance of the assortment optimization (AO) problem~\eqref{problem:AO}, i.e., $I_{\text{AO}}$, where $\Phi (I_{\text{VC}})$ is defined as follows:

\begin{itemize}
    \item For each vertex $j \in V$, we introduce a product $j$ with the price of 1 dollar each. 
    \item We introduce an additional product $n+1$ to the instance $I_{\text{AO}}$ with the price of 3 dollars. Therefore, instance $I_{\text{AO}}$ consists of $n+1$ products in total. 
    \item For each edge $(i,j) \in E$, there is a corresponding customer type which is characterized by a preselected set $C^E_{(i,j)}$ such that $C^E_{(i,j)} = \{ i,j\}$. Let us label these customers as ``edge customer types''. We denote the collection of ``edge customer types'' by
    $\mathcal{C}^E$ such that $\mathcal{C}^E=\{ C^E_{(i,j)} \mid (i,j) \in E  \}$.  We assign the same weight of $1/(|E| + |V|/3)$ to every customer in $\mathcal{C}^E$. 
    
     \item For each vertex $j \in V$, there is a corresponding customer type which is characterized by a preselected set $C^V_j $ such that $C^V_j = \{ j, n+1   \}$. Let us label these customers as ``vertex customer types''. We denote the collection of ``vertex customer types'' as $\mathcal{C}^V$ such that  $\mathcal{C}^V=\{ C^V_i \mid i \in V \}$. We assign the weight of $1/(3|E| + |V|)$ to every customer in $\mathcal{C}^V$. As a result, we have $| E | + |V|$ customer types in total, i.e., the  collection $\mathcal{C}$ of customer types is $\mathcal{C} = \mathcal{C}^E \cup \mathcal{C}^V$. 
     Note that the resulting distribution over subsets $\lambda_C$ constructed above satisfies the property that the sum of all the weights of customers in $\mathcal{C}$ is equal to 1.      
\end{itemize}

Let us first denote $L$ as the normalizing constant, which is equal to 1/($|E| + |V|/3$) that we will use to simplify the exposition. Then, following the aforementioned hardness result of the vertex cover problem, to establish the hardness of the 
assortment optimization problem under the consideration set problem it is sufficient to prove that $\Phi$ satisfies two properties: 
\begin{itemize}
    \item \textbf{Claim 1:} For a vertex cover of the size at most $k$ in the instance $I_{\text{VC}}$, there is an optimal assortment in instance $I_{\text{AO}}$ that has expected revenue of at least $(|E| + |V | - k/3)\cdot L$. 
    \item \textbf{Claim 2:} Reciprocally, given the optimal assortment 
 in instance $I_{\text{AO}}$ that has expected revenue of at least $(|E| + |V | - k/3)\cdot L$, there exists an instance $I_{\text{VC}}$ with a vertex cover of the size at most $k$.  
 \end{itemize}
\textbf{Proof of Claim 1:}
Let us assume that $U \subseteq V$ is a vertex cover of the graph $G$ and its cardinality is less than $k$, i.e.,  $|U| \leq k$. Then we state that the assortment $S_U = \{  i \mid i \in U \} \cup \{ n+1 \}$ results into the expected revenue of at least $(|E| + |V | - k/3)\cdot L$.
In what follows below we prove this statement which concludes the proof of Claim~1. 
\begin{align*}
& \bigg[\text{Expected revenue under assortment $S_U$} \bigg]= \sum_{C \in \mathcal{C}} \lambda_C \cdot \text{Rev}_C(S_U) \\
&= \sum_{C \in \mathcal{C}^E} \lambda_C \cdot \text{Rev}_C(S_U)+\sum_{C \in \mathcal{C}^V} \lambda_C \cdot \text{Rev}_C(S_U) \\
&= \sum_{C \in \mathcal{C}^E} \lambda_C \cdot 1 + \sum_{C \in \mathcal{C}^V} \lambda_C \cdot \text{Rev}_C(S_U) \ \ \text{[since for every edge $(i,j) \in E$ either $i$ or $j$ is covered]}\\
& \hspace{1.4in} [\text{and the price is the same for both products $i$ and $j$ and it is equal to 1}] \\
& = \sum_{C \in \mathcal{C}^E} \lambda_C + \sum_{j \in U } \lambda_{C^V_{j}} \cdot {\text{Rev}}_{(C^V_{j})}(S_U) +\sum_{j \in V \setminus U } \lambda_{C^V_{j}} \cdot {\text{Rev}}_{(C^V_{j})}(S_U) \\
&= \sum_{C \in \mathcal{C}^E} \lambda_C + \sum_{j \in U } \lambda_{C^V_{j}} \cdot 1/2 \cdot (1 + 3)  +\sum_{j \in V \setminus U } \lambda_{C^V_{j}} \cdot 3 \\
&= \sum_{C \in \mathcal{C}^E} L + \sum_{j \in U } L/3 \cdot 1/2 \cdot (1 + 3)  +\sum_{j \in V \setminus U } L/3 \cdot 3 \\
&=L \cdot (|E|+2/3\cdot|U|+(|V|-|U|))=L \cdot (|E|+|V|-|U|/3) \ge (|E| + |V | - k/3)\cdot L.
\end{align*}

\textbf{Proof of Claim 2:}
Let $S_+$ denote an optimal assortment in instance $I_{\text{AO}}$ such that its expected revenue is greater than or equal to $|E| + |V| - k/3$. It is easy to see that $n+1$ is part of the optimal assortment $S_+$  because adding product $n+1$ to any assortment would increase the expected revenue. Thus, we assume that $n+1 \in S_+$ and  $S_+=S \cup \{n+1\}$ without loss of generality. 

First of all, it is clear that all customer types in $\mathcal{C}^V$ have a positive contribution to the expected revenue under the optimal assortment $S_+$ because 
$n+1 \in S_+$, i.e., the number of customer types in $\mathcal{C}^V$ that have a positive contribution to the revenue is equal to $n$. Moreover, we state that all the customer types in $\mathcal{C}^E$ have a positive contribution to the expected revenue under the optimal assortment $S_+$ and we prove that statement at the very end. For now, if we assume that the latter statement holds, then for every customer type $C^E_{(i,j)} \in \mathcal{C}^E$ to have a positive contribution to the revenue it should be the case that either $i \in S$ or $j \in S$ and thus we have that $U_S$, such that $U_S = \{  \text{vertex $j$} \mid \text{$j \in S$}  \}$, is a vertex cover. Note that in this case, the number of customer types in the set $\mathcal{C}^E$ that have a positive contribution to the revenue under the optimal assortment is equal to $|E|$. Next, we can obtain the following set of equations and inequalities. 
\begin{align*}
&\bigg[ \text{Expected revenue under assortment $S_+$} \bigg]= \sum_{C \in \mathcal{C}} \lambda_C \cdot \text{Rev}_C(S_+) \\
&= \sum_{C \in \mathcal{C}^E} \lambda_C \cdot \text{Rev}_C(S_+)+\sum_{C \in \mathcal{C}^V} \lambda_C \cdot \text{Rev}_C(S_+) \\
&= |E|\cdot L+\sum_{C \in \mathcal{C}^V} \lambda_C \cdot \text{Rev}_C(S_+) \\
&=|E|\cdot L+\sum_{j \in S} \lambda_{C^V_{j}} \cdot {\text{Rev}}_{(C^V_{j})}(S_+) +\sum_{j \in V \setminus S } \lambda_{C^V_{j}} \cdot {\text{Rev}}_{(C^V_{j})}(S_+) \\
&=|E|\cdot L+ |S|/2 \cdot L/3+ 3 |S|/2 \cdot L/3 +\sum_{j \in V \setminus S } \lambda_{C^V_{j}} \cdot {\text{Rev}}_{(C^V_{j})}(S_+) \\
&=|E|\cdot L+ 2|S| \cdot L/3 + (|V|-|S|) \cdot L \\
&=(|E| + |V| - |S|/3) \cdot L \\
&\ge (|E| + |V| - k/3) \cdot L \ \ [\textbf{by the assumption in the Claim~2}] \\
&\Rightarrow |U_S| = |S| \leq k. 
\end{align*}
Then, to conclude the proof of Claim~2 it is sufficient to prove the aforementioned statement that all the customer types in $\mathcal{C}^E$ have a positive contribution to the expected revenue under the optimal assortment $S_+$. We prove that statement by contradiction. Suppose that there exists a customer type $C^E_{(i',j')} \in \mathcal{C}^E$ which does not contribute to the expected revenue, i.e., $i' \notin S_+$ and $j' \notin S_+$. Then, adding $i'$ to $S_+$ will have a combination of two effects: (1) it will make customer type $C^E_{(i',j')}$ positively contribute to the expected revenue and increase the expected revenue by $L$; and (2) contribution of the customer of type $C^V_{i'}$ to the expected revenue will decrease by $(3-2)\cdot L/3=L/3$. As a result, adding item $i'$ to the assortment $S_+$ has a net positive effect on the expected revenue (i.e., it increases the revenue by $2L/3$) which contradicts the fact that $S_+$ is an optimal assortment. \hfill $\square$

\subsection{Proof of Theorem~\ref{thm:impossibility_of_approximation}}
\label{subsec:proof_AO_inapproximate}

In what follows, we first describe a reduction $\Phi$ from any instance $\mathcal{I}$ of the Maximal Independent Set (Max-IS) to an instance $\Phi(\mathcal{I})$ of the assortment problem~\eqref{problem:AO}, which consists of $n$ products and $n$ consideration sets. We then utilize the inapproximability result of Max-IS \citep{haastad1999clique}, which states that the Max-IS problem is NP-hard to approximate within factor $O(n^{1-\epsilon})$, to show that the assortment problem~\eqref{problem:AO} is also NP-hard to approximate. In this proof, we use $[n]$ to denote $\{ 1,2,\ldots,n \} $.

Let a Max-IS instance $\mathcal{I}$ be defined on a graph $G = (V,E)$, where $V = \{ v_1,\ldots,v_n \}$ is a set of vertices and $E$ is a set of edges. For each vertex $v_i \in V$, we use $N^-(i)$ to denote the indices of $v_i$'s neighbors whose indices are smaller than $i$, i.e.,
\begin{align}
	N^-(i) = \{  j \in [n] \,\, \big| \,\, (v_i,v_j) \in E \text{ and } j < i \}.
\end{align}
Now we describe the mapping $\Phi$. For each vertex $v_i \in V$, we introduce a product $i$ with price $r_i = n^{2i} / \alpha$, where $\alpha = 1 / \left( \sum_{i=1}^n n^{-2i} \right)$. For each vertex $v_i \in V$, we also construct a consideration set $C_i = \{ i \} \cup N^-(i)$, which has weight $\lambda_i = \alpha / n^{2i}$. Next, we state the following two claims: 
\begin{claim}
	For any independent set $U \subseteq V$ in the instance $\mathcal{I}$, there exists a corresponding assortment $S_U$ in the instance $\Phi(\mathcal{I})$ such that $\text{Rev}(S_U) \geq |U|$.
\end{claim}

\begin{claim}
	Given any assortment $S$ in the instance $\Phi(\mathcal{I})$ such that $\text{Rev}(S) = \Omega ( n^{\frac{1}{2} + \frac{\epsilon}{2} } )$, there exists a corresponding independent set $U_S$ of size $\Omega(n^{ \frac{\epsilon}{2}  })$ in the instance $\mathcal{I}$.
\end{claim}

Before proving those two claims, we make the following two statements. First, the construction of the mapping $\Phi$ in our proof follows the one in \cite{aouad2018approximability}, except that we do not need to specify the preference over products in each $C_i$. Claim 1 also appears in a similar form in the proof of inapproximability of the assortment optimization problem under the ranking-based model in \cite{aouad2018approximability}. However, Claim 2 is significantly different from the counterpart in \cite{aouad2018approximability}, as it requires a very different argument to construct the independent set $U_S$, leading to the narrowed $O \left(\sqrt{n}\right)$ gap instead of an $O\left(n \right)$ gap.

Second, the two claims jointly lead to the inapproximability of the assortment problem. Our argument proceeds as follows. Based on the analyses provided by \cite{haastad1999clique}, we know that there does not exist a polynomial time algorithm to find an independent set of size at least $n^{\frac{\epsilon}{2}}$ given that the $n$-size vertex graph instance $\mathcal{I}$ has an independent set of size $n^{1-\frac{\epsilon}{2}}$ \citep{khot2010inapproximability}. Now,  let us focus on such a Max-IS instance $\mathcal{I}$. Suppose by contradiction that there exists a polynomial-time approximation algorithm $\mathcal{A}$ which solves the assortment problem within factor $O(n^{\frac{1}{2}-\epsilon})$. We can then use this algorithm $\mathcal{A}$ to obtain an assortment $S'$ in instance $\Phi(\mathcal{I})$ with expected revenue $\Omega(n^{ \frac{1}{2} + \frac{\epsilon}{2}  })$, since
\begin{align*}
	\text{Rev}(S') \geq \frac{R(S^*)}{ c_1 \cdot n^{\frac{1}{2}-\epsilon}}  \geq \frac{  n^{1-\frac{\epsilon}{2}}   }{ c_1 \cdot n^{\frac{1}{2}-\epsilon}} = (1/c_1) \cdot n^{  \frac{1}{2} + \frac{\epsilon}{2} },
\end{align*}
where $c_1 > 0$ is an absolute constant. Here, the first inequality follows from the definition of $\Acal$, and the second inequality follows from Claim~1. In particular, since we assume that $\mathcal{I}$ has an independent set of size $n^{1/2 - \epsilon}$, it follows from Claim 1 that there exists an assortment resulting in revenue that is greater than $n^{1/2-\epsilon}$. This implies that $R({S^*}) \geq n^{1/2-\epsilon}$. Given that, for $S'$ such that $\text{Rev}(S') = \Omega (  n^{  \frac{1}{2} + \frac{\epsilon}{2}}  )$, it follows from Claim~2 that we can construct an independent set $U_{S'}$ that is of the size $\Omega(n^{  \frac{\epsilon}{2} })$. Therefore, it implies that we are able to use a polynomial-time algorithm to construct a $\Omega(n^{ \frac{\epsilon}{2} })$-size independent set which leads to a contradiction to the inapproximability result of the Max-IS problem.

In what follows below we prove the aforementioned two claims which completes the proof of the theorem.

{\bf Proof of Claim 1:} Let $U$ be the independent set in the claim and let us   define $S_U \equiv \{ i \mid v_i \in U  \}$. Since $N^-(i) \cap U = \emptyset$ if $v_i \in U$, we know that $C_i \cap S_U = \{ i \}$. Therefore,
\begin{align*}
	\text{Rev}\left(S_U\right) = \sum_{i =1}^n \lambda_i \cdot \text{Rev}_{C_i}\left( S_U \right) \geq \sum_{i \in S_U} \lambda_i \cdot r_i = |U|.
\end{align*}

{\bf Proof of Claim 2:} Let $S$ be the assortment in the claim. We define two collections of the consideration sets,
\begin{align*}
	G = \left\lbrace i \in [n] \quad \bigg| \quad \text{Rev}_{C_i}(S) \geq \frac{r_i}{\sqrt{n}}  = \frac{n^{2i}}{\alpha \sqrt{n}}  \right\rbrace \quad \text{ and } \quad B =  \left\lbrace i \in [n] \quad \bigg| \quad \text{Rev}_{C_i}(S) < \frac{r_i}{\sqrt{n}}  = \frac{n^{2i}}{\alpha \sqrt{n}}  \right\rbrace,
\end{align*}
which we denote as the ``Good'' and ``Bad'' collections of consideration sets, respectively. We make the following three statements about the ``Good'' collection of consideration sets $G$.

First, we argue that $G \subseteq S$. Let $i$, by contradiction, be a product such that $i \in G$ but $i \notin S$. Then, we have that
\begin{align}
	\label{eq:inapprox_G_property_1}
	\text{Rev}_{C_i}(S) \leq \frac{n^{2(i-1)}}{\alpha} = \frac{n^{2i}}{\alpha \cdot n^2}  < \frac{n^{2i}}{\alpha \cdot \sqrt{n}},
\end{align}
which contradicts the definition of the ``Good'' collection of consideration sets $G$. In the aforementioned chain of equalities and inequalities~\eqref{eq:inapprox_G_property_1}, the first inequality follows because of the fact that all other products in $C_i \cap S$ have a price of at most $n^{2(i-1)} / \alpha$.

Second, we argue that $|G| = \Omega(n^{ \frac{1}{2} + \frac{\epsilon}{2}  })$. To this end, the assumption imposed on the expected revenue of $S$ in Claim 2 leads to the following set of equalities and inequalities:
\begin{align}
	\label{eq:inapprox_G_property_2}
	c_2 \cdot n^{  \frac{1}{2} + \frac{\epsilon}{2}  } \leq \text{Rev}(S)  = \sum_{i \in G} \lambda_i \cdot \text{Rev}_{C_i}(S) + \sum_{i \in B} \lambda_i \cdot \text{Rev}_{C_i}(S) < \sum_{i \in G} 1 + \sum_{i \in B} \frac{1}{\sqrt{n}} \leq |G| + \sqrt{n},
\end{align}
where $c_2 > 0$ is an absolute constant. In the aforementioned chain of equalities and inequalities, the second inequality follows because $\text{Rev}_{C_i}(S) \leq r_i \leq n^{2i}/ \alpha$ and $\lambda_i = \alpha/n^{2i}$. Overall, inequality~\eqref{eq:inapprox_G_property_2} implies that $|G| = \Omega(n^{ \frac{1}{2} + \frac{\epsilon}{2}  })$. In other words, asymptotically, only consideration sets belonging to the ``Good'' collection of consideration sets $G$ are contributing to the revenue $\text{Rev}(S)$.

Third, we argue that for each $i \in G$, we have that $| C_i \cap S | \leq 2\sqrt{n}$. We prove this statement by contradiction. Assume by contradiction that $|C_i \cap S | \geq 2 \sqrt{n} + 1$, then we have that
\begin{align*}
	\text{Rev}_{C_i}(S) = \frac{ \sum_{j \in C_i \cap S} r_j }{ | C_i \cap S |}  \leq \frac{ \sum_{j \in C_i \cap S} r_j }{ 2 \sqrt{n} + 1} <  \frac{ 2 \cdot n^{2i} / \alpha  }{2 \sqrt{n}} = \frac{n^{2i}}{\alpha\sqrt{n}},
\end{align*}
which contradicts the fact that $i \in G$. In the aforementioned expression, the second inequality follows because of the following chain of equalities and inequalities: 
\begin{align*}
	\sum_{j \in C_i \cap S} r_j < \frac{n^{2i}}{\alpha} + \frac{n^{2i-2}}{\alpha} + \frac{n^{2i-4}}{\alpha} + \ldots = \frac{n^{2i}}{\alpha} \cdot \left(  1 + \frac{1}{n^2} +  \frac{1}{n^4} + \ldots \right) < \frac{n^{2i}}{\alpha}  \cdot 2,
\end{align*}
which is a valid expression as long as $n > 1$.

Next, taking into account all three statements related to the ``Good'' collection of the consideration sets $G$, we construct the independent set $U_S$ from $S$ as follows. We begin with an empty set, i.e., $U_S = \emptyset$. Then, starting from the largest index $i_1$ in $G$, we add $i_1$ to $U_S$ and delete the elements in $C_{i_1}$ from $G$, i.e., $G \leftarrow G \backslash C_{i_1}$. We then proceed to the next largest index $i_2$ in the updated $G$, add $i_2$ to $U_S$, and remove the elements in $C_{i_2}$ from $G$. This process is repeated until $G$ becomes empty. The resulting $U_S$ forms an independent set because the neighbors of each vertex are removed from $G$ once the vertex is added to $U_S$. 
We also observe that $|U_S| = \Omega(n^{\frac{\epsilon}{2}})$, because each time a vertex is added, we remove at most $|C_i \cap G| \leq |C_i \cap S| \leq 2 \sqrt{n}$ elements from $G$.
Thus, we at least can add
\begin{align*}
	\frac{|G|}{2 \sqrt{n}} = \frac{\Omega ( n^{  \frac{1}{2} + \frac{\epsilon}{2} } )}{2 \sqrt{n}} = \Omega \left( n^{ \frac{\epsilon}{2} }  \right)
\end{align*}
items to the set $U_S$ which shows that $|U_S| = \Omega \left( n^{ \frac{\epsilon}{2}  }  \right)$. \hfill $\square$

\subsection{Scalability of the Mixed-Integer Linear Program~\eqref{prob:AO_IP}}
\label{subsec:IP_scalability}

To numerically test the scalability of the mixed-integer linear program~\eqref{prob:AO_IP} when solving the assortment problem~\eqref{problem:AO}, we first generate random problem instances as follows. We vary the number of products $n$ and the number of consideration sets $k = |\Ccal|$ while fixing a constant $s$ which is the size of consideration sets. Specifically, for each $C \in \Ccal$, we sample the consideration set $C$ from the set $N = \{ 1,2,\ldots,n \}$ restricting its size by $s$, uniformly at random. We repeat this sampling for all $k$ consideration sets in $\Ccal$. We further let $\lambda_C = 1 / k$ and set $s = 5$, motivated by the empirical evidence from the literature \citep{hauser1990evaluation,hauser2014consideration} and from Table~\ref{tb:IRI_and_SS_size} where we can see that customers usually consider only a few products. We let $n \in  \{ 250,500, 750,1000 \}$ and $k \in \{ n, 3n, 5n \}$, where the scale factor between $n$ and $k$ is consistent with our empirical estimation outcomes in Table~\ref{tb:IRI_and_SS_size}. We set the time limit to 20 minutes when solving the mixed-integer linear program (MILP) using Gurobi. If Gurobi fails to find the optimal solution within the time limit, we report the bound of the optimality gap returned by the solver. 
Thus, for each pair $(n,k)$, we randomly generate ten instances and calculate the average runtime (in minutes) and the optimality gap. 
The scalability results are reported in Table~\ref{tb:MILP_scalability}.
	
	\begin{table}[h!]
		\centering
		\begin{tabular}{rrrr} \toprule
			$n$    & $k$    & Time (min) & Gap (\%) \\ \midrule
			250  & 250  & 0.00      & 0.00     \\
			250  & 750  & 5.99      & 0.00     \\
			250  & 1250 & 17.71     & 0.68     \\
			500  & 500  & 0.08     & 0.00     \\
			500  & 1500 & 20.03     & 0.57     \\
			500  & 2500 & 20.05     & 1.28     \\
			750  & 750  & 0.30     & 0.01     \\
			750  & 2250 & 20.07     & 0.87     \\
			750  & 3750 & 20.13     & 1.30     \\
			1000 & 1000 & 0.53     & 0.01     \\
			1000 & 3000 & 20.14     & 1.16     \\
			1000 & 5000 & 20.23     & 1.59    \\ \bottomrule
		\end{tabular}
		\caption{Scalability of the MILP approach when solving the assortment problem.}
		\label{tb:MILP_scalability}
	\end{table}

It follows from this table that within a twenty-minute time limit, even for large instances like $(n,k) = (1000,5000)$, the MILP achieves a solution with an optimality gap of no more than 2\%. This underscores our claim that MILP~\eqref{prob:AO_IP} is efficient and can be useful for practical applications. Additionally, the performance of the MILP could be further enhanced by adopting the mixed-integer conic reformulation suggested by \cite{csen2018conic}.

\section{Model Estimation}
\label{sec:estimation_method}

\subsection{Estimation Methodology}
\label{subsec:model_estimation}

We propose a maximum likelihood estimation (MLE) procedure based on the expectation maximization (EM) algorithm and the column generation algorithm method. Overall, our estimation methodology follows \cite{van2014market,van2017expectation}, who estimate the ranking-based model using the EM algorithm and column generation method. A similar framework has also been adapted to estimate the decision forest model \citep{chen2022decision}.

We assume we have the sales transactions denoted by $\{ (S_t,i_t)  \}_{t =1,\ldots,T}$, which consist of purchase records over $T$ periods. A purchase record at time $t$ is characterized by a tuple $(S_t,i_t)$, where $S_t \subseteq N$ denotes the subset of products on offer in period $t$ and $i_t \in S_t \cup \{ 0 \}$ denotes the product purchased in period $t$. In addition, we let $\Scal$ denote the set of unique assortments observed in the data and let $m$ denote the cardinality of this set, i.e.,  $m = | \Scal |$. We term each assortment in $\Scal$ as a \emph{historical assortment}, i.e., an assortment that was offered in the past. For each historical assortment $S \in \Scal$ and $i \in S^+ \equiv S \cup \{ 0 \}$, we let $\tau(S,i)$ denote the number of times the tuple $(S,i)$ appears in the sales transactions $\{ (S_t,i_t)  \}_{t =1,\ldots,T}$. 

\subsubsection{Maximum Likelihood Estimation (MLE).}
\label{subsubsec:MLE}

To estimate the model from data $\{ (S_t,i_t)  \}_{t =1,\ldots,T}$, we first write down the following optimization problem with respect to a fixed collection $\bar{\Ccal}$ of consideration sets:
\begin{subequations}
	\label{problem:MLE_w_fixed_set}
	\begin{alignat}{3}
		P^{\text{MLE}}\left(  \bar{\Ccal}  \right): \quad \underset{\lambdab \ge 0, \vb}{\text{maximize}}  \quad & \mathcal{L}(\vb)  \\   \text{such that}  \quad & \sum_{C \in \bar{\Ccal}} \left(\Ab_S\right)_{i,C} \lambda_C =  {v}_{i,S} \qquad \forall S \in \Scal, \,\, i \in N^+, \label{constr:1}
		\\   &  \sum_{C \in \bar{\Ccal}} \lambda_C = 1, 
	\end{alignat}
\end{subequations}
where $\lambda_C$ is an element of the distribution $\lambdab \in \mathbb{R}^{|\bar{C}|}$ over the sets in $\bar{\Ccal}$, ${v}_{i,S}$ is the aggregate likelihood that customers would pick the alternative $i$ from the offer set $S$, and the objective is the log-likelihood function $\mathcal{L}( \vb) \equiv \sum_{S \in \Scal} \sum_{i \in S^+} \tau(S,i) \cdot \log \left(v_{i,S}\right)$, which is a concave function. The matrix $\Ab_S$ is of size $(n+1) \times |\bar{\Ccal}|$ and it maps from the model $\left( \bar{\Ccal}, \lambdab \right)$ to the its choice probability $v_{i,S}$ of item $i$ under historical assortment $S$. The element $\left( \Ab_S \right)_{i,C}$ is equal to $1/|S \cap C|$ if $i \in S \cap C$ and 0, otherwise.

First, it is worth emphasizing that when $\bar{\Ccal}$ includes all subsets of $N$ (i.e., $\bar{\Ccal}$ is the power set of $N$) then optimization problem~\eqref{problem:MLE_w_fixed_set} solves the MLE problem under the consideration set model. When $\bar{\Ccal}$ does not include all the subsets in $N$, we refer the problem $P^{\text{MLE}}\left(  \bar{\Ccal}  \right)$ specified above as a \emph{restricted} optimization problem. Second, we note that the maximization problem~\eqref{problem:MLE_w_fixed_set} is concave and thus we can use a convex programming solver to find the optimal solution. Alternatively, we will exploit an expectation-maximization (EM) algorithm to obtain the optimal solution to the restricted problem $P^{\text{MLE}}\left(  \bar{\Ccal}  \right)$. We find that this EM algorithm is more efficient than existing general convex programming solvers. For now, we assume $\bar{\Ccal}$ is fixed and we will come back to discuss how to enlarge it during the estimation procedure.

\subsubsection{Solving MLE with the EM Algorithm.}
\label{subsubsec:EM}
The EM algorithm is a method for finding the maximum likelihood estimates of parameters in statistical models where the data are incomplete or there are unobserved latent variables. The algorithm proceeds in two steps: the expectation step (E-step), where the expectation of the complete-data log-likelihood is calculated given the current parameter estimates, and the maximization step (M-step), where the parameters are updated to maximize the expected complete-data log-likelihood. In the context of our MLE problem $P^{\text{MLE}}\left(  \bar{\Ccal}  \right)$ specified above, the unobserved latent variable is the probability mass $\lambda_C$ of each customer type $C \in \bar{\Ccal}$, since customer types are not directly observed from the sales transaction data. We start our EM algorithm with arbitrary initial parameter estimate ${\lambdab}^{\text{est}}$. Then, we repeatedly apply the ``E'' and ``M'' steps, which are described below, until convergence. 

\underline{\emph{E-step:}} If customer types associated with each transaction are known to us then we would be able to represent the complete-data log-likelihood function as $\Lcal^{\text{complete}} = \sum_{C \in \bar{\Ccal}} \tau(C) \cdot \log \lambda_C + \text{constant}$, where $\tau(C)$ is the number of transactions made by customers of the type $C$. However, in the context of our problem, the customer types are not observed in the data and we thus replace $\tau(C)$ from the aforementioned complete-data log-likelihood function by its conditional expectation $\Eb \left[ \tau(C) \mid {\lambdab}^{\text{est}} \right]$ which allows us to obtain $\Eb \left[ \Lcal^{\text{complete}} \mid {\lambdab}^{\text{est}} \right]$. To obtain the conditional expectation, we first apply the Bayes' rule
\begin{align}
	\label{sq:EM_E_step_Bayes}
	\Pbb\left( C \mid S,i,{\lambdab}^{\text{est}} \right) = \frac{ \Pbb \left( i \mid C,S,\lambdab^{\text{est}} \right) \cdot \Pbb \left( C \mid S , {\lambdab}^{\text{est}} \right) }{\sum_{C \in \bar{\Ccal}}  \Pbb \left( i \mid C,S,\lambdab^{\text{est}} \right) \cdot \Pbb \left( C \mid S, {\lambdab}^{\text{est}} \right) } =  \frac{ {\Pbb \left( i \mid C,S \right)} \cdot {\lambda}^{\text{est}}_{C} }{\sum_{C \in \bar{\Ccal}}  {\Pbb \left( i \mid C,S \right)} \cdot {\lambda}^{\text{est}}_C},
\end{align}
where $\Pbb\left( C \mid S,i,{\lambdab}^{\text{est}} \right)$ is the probability that an item $i$ from the offer set $S$ is purchased by a customer of type $C$ when ${\lambdab}^{\text{est}}$ represents the probability mass of different customer types and 
$\Pbb \left( i \mid C,S \right)$ is the probability to choose item $i$ from the offer set $S$ by a customer of type $C$. Consequently, the aforementioned Equation~\eqref{sq:EM_E_step_Bayes} allows us to compute $\hat{\tau}(C)$, which is the expected number of transactions made by customers of the type $C$ by $\hat{\tau}(C) = \Eb \left[ \tau(C) \mid {\lambdab}^{\text{est}} \right]=  \sum_{S \in \Scal} \sum_{i \in S^+} \tau(S,i) \cdot \Pbb\left( C \mid S,i,{\lambdab}^{\text{est}} \right)$. We thus compute the expectation of the complete-data log-likelihood function as $\Eb \left[ \Lcal^{\text{complete}}  \mid {\lambdab}^{\text{est}} \right] = \sum_{C} \hat{\tau}(C) \cdot \log \lambda^{\text{est}}_C$.

\underline{\emph{M-step:}} By maximizing the expected complete data log-likelihood function $\Eb \left[ \Lcal^{\text{complete}}  \mid {\lambdab}^{\text{est}} \right] $, we obtain the optimal solution 
$\lambda^*_C =  \hat{\tau}(C)  \big\slash  \sum_{C}  \hat{\tau}(C)$ 
which is used to update $\lambdab^{\text{est}}$.

Finally, we note that the EM algorithm has been widely used in the operations management (OM) field to estimate various choice models when customer types are not directly observed in the data. Examples include the mixed MNL model \citep{train2009discrete}, the ranking-based model \citep{van2017expectation}, the Markov chain choice model \citep{csimcsek2018expectation}, the consider-then-choose (CTC) model \citep{jagabathula2022demand}, and DAG-based choice models \citep{jagabathula2022personalized}. 

\subsubsection{Consideration Set Discovery Algorithm.}
\label{subsubsec:CG_sub}
In theory, as mentioned above, to solve the MLE problem one can exploit the aforementioned EM algorithm and apply it to the optimization problem~$P^{\text{MLE}}\left(  \bar{\Ccal}  \right)$  when the collection of customer segments is equal to the power set of $N$, i.e., $\bar{\Ccal} = 2^N$. However, this approach becomes highly intractable as the number of products $n$ increases. In this case, the number of decision variables would grow exponentially with $n$.  

Therefore, we propose an alternative way to solve the MLE problem which is based on the \emph{column generation} (CG) procedure \citep{bertsimas1997introduction}, a widely used procedure to solve large-scale optimization problems with linear constraints. In accordance with the CG framework, instead of solving the full-scale optimization problem $P^{\text{MLE}}(2^N)$ directly, we will repeatedly execute the following two steps: (i) find the optimal solution $(\vb,\lambdab)$ to the restricted problem $P^{\text{MLE}}(\bar{\Ccal})$ by the EM algorithm described in Section~\ref{subsubsec:EM}; (ii) solve a subproblem (which will be described later) to find a new column and expand the restricted problem $P^{\text{MLE}}(\bar{\Ccal})$ by concatenating the column. Notice that each column in the complete problem $P^{\text{MLE}}(2^N)$ corresponds to a consideration set. Therefore, when we introduce a new column $C^*$ to the restricted problem $P^{\text{MLE}}(\bar{\Ccal})$, we equivalently augment the \emph{support} of the distribution $\lambdab$ from $\bar{\Ccal}$ to $\bar{\Ccal} \cup \{ C^* \}$. We will repeat these two steps until we reach optimality or meet a stopping criterion such as a runtime limit. In the following, we provide the details on how we augment the support of the distribution $\lambdab$ by solving a subproblem, which will complete the description of our MLE framework for calibrating the consideration set model.

Let $(\vb,\lambdab)$ be the optimal primal solution of the optimization problem $P^{\text{MLE}}\left(  \bar{\Ccal}  \right)$, which is defined for a fixed collection $\bar{\Ccal}$ of consideration sets. Let $(\alphab, \beta)$ denote the dual solution of the optimization problem $P^{\text{MLE}}\left( \bar{\Ccal} \right)$, where $\alphab = \left( \alphab_S \right)_{S \in \Scal}$ corresponds to the first set of constraints, with each $\alphab_S$ being a $(n+1)$-dimensional vector, and $\beta$ corresponds to the unit-sum constraint (see Section~\ref{subsec:estimation_additional_discussion} for further notes). We then solve the following CG subproblem:
\begin{align}
	\label{problem:CG_subproblem}
	\max_{C \subseteq N} \left[  \sum_{S \in \Scal} \sum_{i \in N^+} \alpha_{S,i} \cdot \left(\Ab_{S}\right)_{i,C} + \beta \right] = \max_{C \subseteq N} \left[ \sum_{S \in \Scal} \left( \alpha_{S,0} \cdot \mathbb{I}\left[ C \cap S = \emptyset \right] +  \sum_{i \in S} \frac{ \alpha_{S,i} \cdot \mathbb{I}\left[ i \in C \right] }{| C \cap S |} \right) + \beta \right],
\end{align}
where $(\Ab_{S})_{i,C}$ is defined in Section~\ref{subsubsec:MLE} and $\alpha_{S,i}$ is the element of $\alphab_S$ for $i \in S^+$. In Problem~\eqref{problem:CG_subproblem}, we assume ``$0/0 = 0$'' to simply the notation. Let $C^*$ be the optimal solution to the CG subproblem~\eqref{problem:CG_subproblem}. We add $C^*$ to the set $\bar{\Ccal}$, which is the support of the $\lambdab$, if and only if the optimal objective value of the subproblem is positive. This indicates that adding $C^*$ to the set $\bar{\Ccal}$ improves the objective value of $P^{\text{MLE}}\left(  \bar{\Ccal}  \right)$. Consequently, if the optimal objective value of the subproblem~\eqref{problem:CG_subproblem} is not positive, we terminate our algorithm, as it indicates that the current set $\bar{\Ccal}$, along with the distribution $\lambdab$, already represents the consideration set model with maximum likelihood. 

Finally, it remains to solve the subproblem~\eqref{problem:CG_subproblem}. We exploit a mixed-integer linear optimization (MILP) formulation similar to the assortment problem~\eqref{prob:AO_IP}. Here, in addition to the variables $u_S$ and $q_{S,i}$ introduced for the linearization, we also introduce binary variable $z_S$ to present the indicator $\mathbb{I}[C \cap S]$. Along with binary variable $x_i$ that represent $\mathbb{I}\left[  i \in C \right]$, we have the following MILP that solves the CG subproblem~\eqref{problem:CG_subproblem}
\begin{subequations}
	\label{problem:CG_sub_by_IP}
	\begin{alignat}{3}
		P^{\text{CG-sub}}(\alphab): \quad \underset{\xb,\zb,\ub,\qb}{\text{maximize}}  \quad & \sum_{S \in \mathcal{S}} \left[ \alpha_{S,0} \cdot z_s  +  \sum_{i \in S} \alpha_{S,i} \cdot q_{S,i} \right], \quad
		\\ \text{such that} \quad & 0 \leq q_{S,i} \leq x_i, && \forall S \in \mathcal{S}, i \in S, \label{constraint_a_CG_sub_by_IP}\\
		& 0 \leq q_{S,i} \leq u_S, && \forall S \in \mathcal{S}, i \in S,\\
		& u_S + x_i \leq q_{S,i} + 1, && \forall S \in \mathcal{S}, i \in S,\\
		& z_S + \sum_{i \in S} q_{S,i} =  1, && \forall S \in \mathcal{S}, \label{constraint:CG_sub_IP_4}\\
		& x_i \in \{ 0,1 \}, \,\, z_S \in \{ 0,1 \}, \,\, u_S\in [1/n,1], \quad && \forall i \in N, S \in \mathcal{S}. \label{constraint:CG_sub_IP_5}%
	\end{alignat}
\end{subequations}

Overall, the proposed estimation procedure consists of solving a finite sequence of MILPs. The size of each MILP scales in $O(n+m)$ in the number of variables and in $O(nm)$ in the number of constraints, where $n$ is the number of products and $m$ is the number of historical assortments. Thus, not surprisingly, estimating a consideration set model can be more tractable than estimating a ranking-based model \cite{van2014market}, as the corresponding CG subproblem of the latter model scales in $O(n^2 + m)$ in the number of variables and in $O(n^3 + nm)$ in the number of constraints. We relegate additional discussion to Section~\ref{subsec:estimation_additional_discussion}. 

Additionally, we note that the consideration sets we estimate can be any subset of $N$, meaning any set of products. To enhance interpretability and enable potential applications in new product development and pricing, we also demonstrate how product features (such as price) can be incorporated into our model calibration. In Section~\ref{subsec:feature-based_models}, we introduce a model estimation framework that integrates contextual information through conjunctive models, disjunctive models, and compensatory models -- well-known heuristic rules for constructing consideration sets. For an overview, see Section~\ref{subsec:consideration_set_literature}.

\subsection{Additional Discussion}
\label{subsec:estimation_additional_discussion}

In this section, we discuss the estimation procedure described in Section~\ref{subsec:model_estimation}.

\emph{Obtaining the optimal dual variables $\alphab$ and $\beta$}. We can obtain the optimal dual solution once we know the optimal primal solution. Notice that our MLE problem~\eqref{problem:MLE_w_fixed_set} is of the same form as Problem (7) in~\cite{van2014market}, where the difference is only in the entries in the constraint matrix $\Ab$. In particular, each column of our constraint matrix $\Ab = (\Ab_S)_{S \in \Scal}$ encodes the decision of a consideration set under historical assortments $S \in \Scal$. In contrast, each column of the constraint matrix $\Ab$ in \cite{van2014market} encodes the decision of a ranking. Since our MLE problem takes the same form as \cite{van2014market}, the dual variables $\alphab$ and $\beta$ satisfy the same KKT conditions except that the constraint matrix has different coefficients. In the end, given the optimal primal variable, the optimal dual variable can be obtained following Equations (8) and (9) in \cite{van2014market}.

\emph{Optimality of the EM algorithm.} Note that given a collection of consideration sets $\bar{\Ccal}$, we use the EM algorithm to find the optimal distribution $\lambdab$ over sets in $\bar{\Ccal}$ that maximizes the log-likelihood objective, i.e., to solve $P^{\text{MLE}}(  \bar{\Ccal}  )$. Recall that $P^{\text{MLE}}(  \bar{\Ccal}  )$ is a concave maximization problem with a strictly concave objective. Therefore, any locally optimal solution of $P^{\text{MLE}}(  \bar{\Ccal}  )$ implies the globally optimal solution of $P^{\text{MLE}}(  \bar{\Ccal}  )$. As the EM algorithm guarantees that each M-step improves the objective value of $P^{\text{MLE}}(  \bar{\Ccal}  )$ and thus finds the local maximum \citep{mclachlan2007algorithm}, we know that the EM approach of solving the optimization problem~\eqref{problem:MLE_w_fixed_set} guarantees global optimality.

\emph{Only solving a finite number of times of the subproblem/MILP.} This is because at each iteration, we introduce a new variable $\lambda_C$ that corresponds to a consideration set $C$ to the full program $P^{\text{MLE}}(  \bar{\Ccal}  )$ for $\bar{\Ccal} = 2^N$. As there are only a finite number of consideration sets (theoretically at most $2^n$ even though we have a much smaller set of consideration sets in practice according to numerous literature in marketing and Table~\ref{tb:IRI_and_SS_size}), the estimation procedure thus only solves a finite number of MILPs.

\subsection{Feature-based Model Estimation}
\label{subsec:feature-based_models}

In Section~\ref{subsec:model_estimation}, we demonstrated how the consideration set model $(\Ccal,\lambdab)$ can be calibrated from sales transaction data in its most general form. Specifically, each consideration set $C \in \Ccal$ in the model can be any subset of the product universe $N$, where each product is characterized by a unique feature vector. This approach eliminates the need to explicitly incorporate product features into our estimation framework. It also aligns with existing literature in operations research on choice modeling estimation. Notably, the estimation of nonparametric choice models, including ranking-based models \citep{farias2013nonparametric,van2014market}, the Markov chain model \citep{blanchet2016markov,csimcsek2018expectation}, the consider-then-choose model \citep{jagabathula2022demand}, and the decision forest model \citep{chen2022decision}, follows a similar convention.

In contrast, as discussed in Section~\ref{subsec:consideration_set_literature}, customers may form their consideration sets based on heuristic models that explicitly incorporate product features. In this section, we integrate several well-documented feature-based heuristic rules from the marketing science literature into our estimation framework. Specifically, we assume that each product $i \in N$ is represented by a feature vector $\chib_{i} \in \mathbb{R}^d$ in a $d$-dimensional vector space. We define $[a] \equiv {1,2,\ldots,a}$, where $a$ is a positive integer.

\subsubsection{Conjunctive Rule.}
\label{subsubsec:conjunctive_models}

The conjunctive rule heuristics assumes that a product is included in the consideration set if it satisfies a sequence of screening criteria. Each criterion is defined by a feature $p \in [d]$ and a threshold value $t_p$. A product $i$ satisfies the screening criterion for feature $p$ if and only if $\chi_{i,p} \leq t_p$. Mathematically, the conjunctive model is parameterized by $(E,\tb_E)$, where $E \subseteq [d]$ is a set of features, and $\tb_E = (t_e)_{e \in E}$ is the vector of threshold values associated with the features in $E$. A conjunctive model $(E,\tb_E)$ leads to a consideration set $C$ in the following way:
\begin{align}
	\label{eq:conjunctive_set_intersection}
	C = \bigcap_{e \in E} \left\lbrace  i \in N  \mid \chi_{i,e} \leq t_e \right\rbrace.
\end{align}
Note that without loss of generality, we only need to consider the screening rules with the smaller-or-equal-to sign (i.e., $\leq$) since one can always introduce an additional feature $e'$ such that $\chi_{i,e'} = - \chi_{i,e}$ for $i \in N$. Conjunctive models are among the most popular non-compensatory models in the marketing science literature and have been supported by abundant empirical studies \citep{pras1975comparison,brisoux1981evoked,laroche2003decision, gilbride2004choice, jedidi2005probabilistic}. 

Let $\mathcal{C}_{\text{conjunctive}}$ be the collection of all consideration sets that can be specified by the conjunctive models. %
Since our goal is to calibrate a consideration set model by modeling the consideration set formation by means of the conjunctive rule, we solve the following MLE problem:
\begin{align}
	\label{problem:MLE_conjunctive}
	\underset{}{\text{maximize}}  \left[  P^{\text{MLE}} \left(  \bar{\Ccal} \right)  \mid \bar{\Ccal} \subseteq \mathcal{C}_{\text{conjunctive}}  \right],
\end{align}
where $P^{\text{MLE}}$ is defined as in Problem~\eqref{problem:MLE_w_fixed_set} and $\bar{\Ccal}$ is the collection of customer segments. As discussed in Section~\ref{subsec:model_estimation}, if one can enumerate all sets in $\mathcal{C}_{\text{conjunctive}}$, we can simply let  $\bar{\Ccal}=\mathcal{C}_{\text{conjunctive}}$ and apply the EM algorithm to obtain the probability mass of every consideration set from the set $\bar{\Ccal}$. However, given that, in theory, the number of customer segments can be exponential in the number of items in the product universe, we also propose solving Problem~\eqref{problem:MLE_conjunctive} by the column generation approach. To this end, instead of solving the subproblem~\eqref{problem:CG_subproblem} where a candidate consideration set can be any subset in $N$, we ensure that the candidate consideration set belongs to the collection $\mathcal{C}_{\text{conjunctive}}$, i.e., we solve the following optimization problem:
\begin{align}
	\label{problem:CG_subproblem_conjunctive}
	\max_{C \subseteq N } \left[  \sum_{S \in \Scal} \sum_{i \in N^+} \alpha_{S,i} \cdot \left(\Ab_{S}\right)_{i,C} + \beta \Bigm\vert C \in \mathcal{C}_{\text{conjunctive}} \right].
\end{align}

We introduce a MILP formulation to solve Problem~\eqref{problem:CG_subproblem_conjunctive}. In order to avoid using the Big-$M$ parameter in the formulation, which usually results in a weaker system of constraints, we use relative values instead of the exact values of product features. Let $L_p$ be the number of unique values in $\{ \chi_{i,p}  \}_{i \in N}$, i.e., the number of unique values of the feature $p$ observed in the sales data \citep{mivsic2020optimization,akccakucs2021exact}. Next, we let $v_{p,\ell}$ denote the $\ell$th lowest value in $L_p$ such that $v_{p,1} < v_{p,2} < \cdots < v_{p,L_p-1} < v_{p,L_p}$. Then, we let $v_{p,0} = - \infty$ and $\tau_{p,i}$ be the value of $\ell \in \{ 1,\ldots,L_p \}$ such that $\chi_{i,p} = v_{p,\ell}$, i.e., $\tau_{p,i}$ indicates the position of product $i$ in the ranking of all unique values of feature $p$.

After specifying $(\tau_{p,\ell})_{p \in [d], \ell \in [L_p]}$, in what follows below, we propose an MILP formulation to solve Problem~\eqref{problem:CG_subproblem_conjunctive}. We further assume that the conjunctive model consists of at most $R$ screening rules, i.e., $|E| \leq R$ in Equation~\eqref{eq:conjunctive_set_intersection}. The value of $R$ can be either set to be $d$ for the most general conjunctive model or to be a value smaller than $d$, as it is widely assumed that consumers have cognitive and physical limitations and cannot take into account an unlimited set of features when making purchasing decisions. In the latter case, one can view $R$ as a fixed parameter of the model and determine its value by cross-validation.

As above, $\xb,\zb,\ub$, and $\qb$ are our decision variables that have the same interpretation as in Problem~\eqref{problem:CG_sub_by_IP}. In addition, we introduce a new binary decision variable $\lambda_{r,p,\ell}$ which is equal to $1$ if the $r$th screening rule is $\chi_{i,p} \leq v_{p,\ell}$ and $0$, otherwise. Then, we provide our MILP formulation as follows:

\begin{subequations}
	\label{problem:CG_sub_conjunctive_IP}
	\begin{alignat}{3}
		P^{\text{CG-sub}}_{\text{conjunctive}}(\alphab): \quad \underset{\xb,\zb,\ub,\qb,\lambdab}{\text{maximize}}  \quad & \sum_{S \in \mathcal{S}} \left[ \alpha_{S,0} \cdot z_s  +  \sum_{i \in S} \alpha_{S,i} \cdot q_{S,i} \right] \quad
		\\ \text{such that} \quad & \text{Constraint \eqref{constraint_a_CG_sub_by_IP} - \eqref{constraint:CG_sub_IP_5}}  \label{constraint:conjunctive_same_ones}  \\
		& \sum_{p=1}^d \sum_{\ell = 1}^{L_p} \lambda_{r,p,\ell} = 1 , \quad && \forall r \in [R],\label{constraint:conjunctive_one_cut_point}\\
		& x_i \leq \sum_{p=1}^d \sum_{\ell : \ell \geq \tau_{p,i}} \lambda_{r,p,\ell}, && \forall i \in N,  r \in [R],\label{constraint:conjunctive_left}\\
		& \sum_{r=1}^R \sum_{p=1}^d \sum_{\ell : \ell \geq \tau_{p,i}} \lambda_{r,p,\ell} \leq x_i + d -1, \quad && \forall i \in N, \label{constraint:conjunctive_right}\\
		& \lambda_{r,p,\ell} \in \{0,1\}, \quad && \forall r \in [R], p \in [d], \ell \in L_p,
	\end{alignat}
\end{subequations}
where constraint~\eqref{constraint:conjunctive_same_ones} ensures that variables $\xb,\zb,\ub$, and $\qb$ satisfy the same constraints as in Problem~\eqref{problem:CG_sub_by_IP}. Constraint~\eqref{constraint:conjunctive_one_cut_point} ensures that each screening rule is associated with exactly one feature and one threshold value. Constraints~\eqref{constraint:conjunctive_left} and \eqref{constraint:conjunctive_right} ensure that the consideration sets are formed by a conjunctive rule. In particular, Constraint~\eqref{constraint:conjunctive_left} states that product $i$ is in the consideration set ``only if'' it satisfies all screening rules. Constraint~\eqref{constraint:conjunctive_right} completes the ``if'' direction of the statement. Notice that we allow the screening rules to be repeated. Therefore, one can apply up to $|R|$ screening rules for the conjunctive model by solving Problem~\eqref{problem:CG_sub_conjunctive_IP}. While we inevitably have to introduce new binary decision variables $\lambda_{r,p,\ell}$ to characterize the conjunctive model, we can further decrease the number of binary variables by relaxing the binary restriction imposed on the consideration set decisions $\xb$, since they remain integral after relaxation. 

\subsubsection{Disjunctive Rule.} Similarly to the conjunctive rule, the disjunctive rule heuristic is also defined by a set of screening criteria. Differently, the disjunctive rule implies that a product belongs to a customer's consideration set if at least one of the screening criteria is satisfied. Mathematically, a consideration set $C$ can be represented by a disjunctive model if there exists a collection of features $E$ and threshold values such that
\begin{align}
	\label{eq:disjunctive_set_intersection}
	C = \bigcup_{e \in E} \left\lbrace  i \in N  \mid \chi_{i,e} \leq t_e \right\rbrace.
\end{align}
While the disjunctive model is usually benchmarked as an alternative to the conjunctive model, it receives much less attention in the literature. For more details, we refer the reader to the empirical studies conducted by \cite{pras1975comparison, gilbride2004choice, jedidi2005probabilistic}. To calibrate the consideration set model under this heuristic, we follow the aforementioned procedure described in Section~\ref{subsubsec:conjunctive_models} with the only difference that we solve the column generation subproblem in the following way: 
\begin{align*}
	\max_{C \subseteq N } \left[   \sum_{S \in \Scal} \sum_{i \in N^+} \alpha_{S,i} \cdot \left(\Ab_{S}\right)_{i,C} + \beta  \Bigm\vert C \in \mathcal{C}_{\text{disjunctive}} \right],
\end{align*}
where $\mathcal{C}_{\text{disjunctive}}$ is the collection of all consideration sets that can be represented by a disjunctive rule. One can solve this subproblem by formulating it as the following MILP, which is similar to Problem~\eqref{problem:CG_sub_conjunctive_IP}.

\begin{subequations}
	\label{problem:CG_sub_disjunctive_IP}
	\begin{alignat}{3}
		P^{\text{CG-sub}}_{\text{disjunctive}}(\alphab): \quad \underset{\xb,\zb,\ub,\qb,\lambdab}{\text{maximize}}  \quad & \sum_{S \in \mathcal{S}} \left[ \alpha_{S,0} \cdot z_s  +  \sum_{i \in S} \alpha_{S,i} \cdot q_{S,i} \right] \quad
		\\ \text{such that} \quad & \text{Constraint \eqref{constraint_a_CG_sub_by_IP} - \eqref{constraint:CG_sub_IP_5}}  \label{constraint:disjunctive_same_ones}  \\
		& \sum_{p=1}^d \sum_{\ell = 1}^{L_p} \lambda_{r,p,\ell} = 1 , \quad && \forall r \in [R],\label{constraint:disjunctive_one_cut_point}\\
		& x_i \leq \sum_{r=1}^R \sum_{p=1}^d \sum_{\ell: \ell \geq \tau_{p,i}} \lambda_{r,p,\ell}, && \forall i \in N,\label{constraint:disjunctive_left}\\
		& \sum_{p=1}^d \sum_{\ell : \ell \geq \tau_{p,i}} \lambda_{r,p,\ell} \leq x_i, \quad && \forall i \in N, r \in [R], \label{constraint:disjunctive_right}\\
		& \lambda_{r,p,\ell} \in \{0,1\}, \quad && \forall r \in [R], p \in [d], \ell \in L_p.
	\end{alignat}
\end{subequations}
The difference between Problem~\eqref{problem:CG_sub_disjunctive_IP} and Problem~\eqref{problem:CG_sub_conjunctive_IP} is in how the consideration set decision $\xb$ and the screening rule decision $\lambdab$ factor in the system of the constraints in the optimization problem. In the former problem, constraint~\eqref{constraint:disjunctive_left} states that a product is considered ``only if'' there exists a screening criteria which is satisfied, and Constraint~\eqref{constraint:disjunctive_right} completes the argument by adding the ``if'' direction. Similarly to Problem~\eqref{problem:CG_sub_conjunctive_IP}, one can relax the binary restriction imposed on $\xb$ while solving Problem~\eqref{problem:CG_sub_disjunctive_IP} as $\xb$ remains integral after this relaxation. 

\subsubsection{Compensatory Rule.}
Both conjunctive and disjunctive rules belong to the class of non-compensatory decision processes, i.e., they assume that customers do not consider the trade-off between product features (e.g., not willing to trade higher price for higher quality). On the other hand, it is common in conjoint analysis \citep{srinivasan1973linear,evgeniou2005generalized} and in discrete-choice modeling \citep{ben1985discrete,feldman2022customer} to assume that product utility is a composite of the contributions from each product feature. In particular, one can assume that the utility function is linear in product features. In this section, we follow this common approach and assume that a compensatory model is a part-worth vector $\pib \in \Rbb^{d+1}$. A product is in the consideration set if and only if its part-worth utility is nonnegative, i.e.,
\begin{align*}
	C = \{  i \in N \mid \pi_0 + \sum_{p=1}^d \pi_{i,p} \geq 0  \}.
\end{align*}

To estimate the consideration set model under this compensatory heuristics, we follow the aforementioned procedure described in Section~\ref{subsubsec:conjunctive_models} with the only difference that we solve the column generation subproblem in the following way: 
\begin{align*}
	\max_{C \subseteq N } \left[  \sum_{S \in \Scal} \sum_{i \in N^+} \alpha_{S,i} \cdot \left(\Ab_{S}\right)_{i,C} + \beta \Bigm\vert C \in \mathcal{C}_{\text{linear}} \right],
\end{align*}
where $\Ccal_{\text{linear}}$ is the collection of all consideration sets that can be represented by a linear compensatory rule. This subproblem can be formulated as the following MILP:
\begin{subequations}
	\begin{alignat}{3}
		P^{\text{CG-sub}}_{\text{linear}}(\alphab): \quad \underset{\xb,\zb,\ub,\qb,\pib}{\text{maximize}}  \quad & \sum_{S \in \mathcal{S}} \left[ \alpha_{S,0} \cdot z_s  +  \sum_{i \in S} \alpha_{S,i} \cdot q_{S,i} \right] \quad
		\\ \text{such that} \quad & \text{Constraint \eqref{constraint_a_CG_sub_by_IP} - \eqref{constraint:CG_sub_IP_5}}\\
		& \sum_{p=1}^d \chi_{i,p} \pi_p + \pi_0 \leq x_i, \quad  && \forall i \in N, \label{constraint:positive_utility}\\
		& x_i - 1 + \epsilon \leq \sum_{p=1}^d \chi_{i,p} \pi_p + \pi_0 , \quad && \forall i \in N \label{constraint:negatvie_utility},
	\end{alignat}
\end{subequations}
where $\epsilon > 0$ is a sufficiently small constant. Constraints~\eqref{constraint:positive_utility} and~\eqref{constraint:negatvie_utility} ensure that a product is included in the consideration set if and only if the part-worth utility is nonnegative. Note that the subproblem $P^{\text{CG-sub}}_{\text{linear}}(\alphab)$ is invariant if one multiplies the vector $\pib$ by a positive number. Therefore, we do not need to introduce the big-$M$ constant in constraints~\eqref{constraint:positive_utility} and~\eqref{constraint:negatvie_utility}.

\section{Additional Results on the Predictive Performance and Model Calibration}
\label{sec:additional_results_estimation}

\subsection{Prediction Performance Measured by KL Divergence}
\label{subsec:numerics_KL_outcome}

We formally define the KL-divergence metric as follows:
\begin{align}
	\text{KL} = - \frac{1}{\sum_{S \in \Scal} \tau_o(S)} \cdot \sum_{S \in \Scal}  \tau_o(S)    \sum_{i \in S^+} \bar{p}_{i,S} \log \left(  \frac{ \hat{p}_{i,S} }{ \bar{p}_{i,S} }  \right),
\end{align}
where parameters $\tau_o(S), \bar{p}_{i,S},$ and $\hat{p}_{i,S}$ are defined as in Section~\ref{sec:numerics}. Table~\ref{tb:IRI_prediction_KL} reports the predictive performance of each choice model introduced in Section~\ref{sec:numerics}, measured by the out-of-sample KL divergence. We observe qualitatively similar results in Table~\ref{tb:IRI_prediction_KL} as those presented in Table~\ref{tb:IRI_prediction_MAPE}. In both tables, the decision forest model has the best predictive performance while the MNL and the independent demand models provide the worst performance. Although it is a special case, the consideration set model demonstrates predictive performance comparable to the ranking-based and mixed MNL models, with all three achieving similar KL divergence scores of approximately $3.30 \times 10^{-2}$. Furthermore, when combined with rankings, the consideration set model achieves performance comparable to the Markov chain model, with KL divergence scores of $3.15 \times 10^{-2}$ for the CSMR (consideration set model with rankings) and $3.16 \times 10^{-2}$ for the Markov chain model.

\begin{table}[h!]
	\centering
	\begin{tabular}{lrrrrrrrrr}
		\toprule
		Category & ID    & MNL  & MMNL & RBM  & MC   & DF   & CSM  & CSM2 & CSMR \\  \cmidrule(r){1-1} \cmidrule{2-7} \cmidrule(l){8-10}
		\emph{Beer}               & 5.61  & 4.45 & 3.87 & 4.03 & 3.75 & 2.98 & 3.97 & 3.86 & 3.73 \\
		\emph{Coffee}            & 12.58 & 9.49 & 7.68 & 7.71 & 7.29 & 6.54 & 8.20 & 8.25 & 7.42 \\
		\emph{Deodorant}          & 1.82  & 1.00 & 0.94 & 0.99 & 0.94 & 0.89 & 0.93 & 0.97 & 0.93 \\
		\emph{Frozen Dinners}     & 3.54  & 2.94 & 2.51 & 2.41 & 2.36 & 1.89 & 2.46 & 2.44 & 2.40 \\
		\emph{Frozen Pizza}      & 6.26  & 3.84 & 2.82 & 2.63 & 2.85 & 1.82 & 2.84 & 3.04 & 2.56\\
		\emph{Household Cleaners}  & 3.44  & 2.67 & 2.37 & 2.46 & 2.21 & 2.23 & 2.38 & 2.35  & 2.26\\
		\emph{Hotdogs}            & 8.61  & 6.51 & 5.39 & 5.44 & 5.28 & 5.56 & 5.50 & 5.89 & 5.29\\
		\emph{Margarine/Butter}   & 3.47  & 2.31 & 1.80 & 1.82 & 1.84 & 1.21 & 1.80 & 1.82 & 1.80\\
		\emph{Milk}               & 16.77 & 8.85 & 6.19 & 6.37 & 6.28 & 5.94 & 6.40 & 6.52 & 6.22\\
		\emph{Mustard/Ketchup}    & 7.20  & 3.87 & 3.27 & 3.48 & 3.11 & 3.65 & 3.33 & 3.31 & 3.15\\
		\emph{Salty Snacks}       & 4.22  & 2.28 & 1.56 & 1.45 & 1.67 & 1.37 & 1.64 & 1.79 & 1.55 \\
		\emph{Shampoo}            & 2.78  & 1.52 & 1.27 & 1.53 & 1.25 & 1.09 & 1.32 & 1.35 & 1.28\\
		\emph{Soup}               & 5.03  & 3.70 & 3.19 & 3.13 & 2.99 & 2.10 & 3.04 & 3.15  & 2.95\\
		\emph{Spaghetti/Sauce}    & 6.98  & 4.19 & 3.35 & 3.25 & 3.34 & 1.54 & 3.42 & 3.51 & 3.29\\
		\emph{Tooth Brush}        & 5.08  & 2.69 & 2.29 & 2.76 & 2.25 & 1.73 & 2.54 & 2.62 & 2.42\\  \cmidrule(r){1-1} \cmidrule{2-7} \cmidrule(l){8-10}
		Average            & 6.23  & 4.02 & 3.25 & 3.29 & 3.16 & 2.77 & 3.32 & 3.39 & 3.15\\ \bottomrule
	\end{tabular}
	\caption{Out-of-sample prediction performance results measured by KL-divergence (in unit of $10^{-2}$).} \label{tb:IRI_prediction_KL}
\end{table}

\subsection{Faster Convergence in CSM Model Calibration}
\label{subsec:numerics_convergence}
In this subsection, we aim to compare the consideration set model with the ranking-based model, as the two share structural similarities. Both models are nonparametric: the consideration set model is characterized by a distribution over subsets, while the ranking-based model is defined by a distribution over rankings. As discussed earlier in Section~\ref{sec:numerics}, neither model consistently outperforms the other in terms of predictive performance, as demonstrated in our case study in Sections~\ref{sec:numerics} and~\ref{subsec:numerics_KL_outcome}.

Then, it is worth emphasizing that scalability is a common challenge when calibrating nonparametric choice models. In particular, solving the likelihood maximization problem under the ranking-based model using the column generation method can be computationally demanding, as its subproblem must be solved via an integer program \citep{van2014market}. However, compared to the ranking-based model, the consideration set model is significantly more tractable. As discussed in Section~\ref{subsec:model_estimation}, the column generation subproblem of the consideration set model can be formulated as an integer program with $O(n + m)$ binary variables and $O(n^2 + nm)$ constraints, where $n$ is the number of products and $m = |\Scal|$ is the number of the historical assortments in the dataset. This is computationally much simpler than the subproblem for the ranking-based model, which requires solving an integer program with $O(n^2 + m)$ binary variables and $O(n^3 + nm)$ constraints.

\YCRminor{
Table~\ref{tb:runtime_comparison_of_CSM-RBM} provides numerical evidence of the computational tractability of the consideration set model. The second to fourth columns in the table present the number of products ($n$), assortments ($m$), and transactions ($|\Tcal|$) for each product category following data preprocessing. The fifth and sixth columns present the estimation runtimes for the consideration set model ($T_{\CSM}$) and the ranking-based model ($T_{\RBM}$), respectively. A runtime limit of 30 minutes was imposed, and if the estimation procedure failed to terminate within this limit, the runtime is recorded as $1800.00$ in Table~\ref{tb:runtime_comparison_of_CSM-RBM}.

Across the five product categories where the estimation procedures for both models fully converged to the optimal solution, the consideration set model demonstrated an average estimation speed that was 380\% faster than that of the ranking-based model. In one category (\emph{Milk}), the consideration set model successfully converged before the time limit, while the ranking-based model failed to do so. For the remaining categories, the estimation procedures for both models did not fully converge within the runtime limit due to the large number of assortments $(m)$. However, we expect the consideration set model would still terminate earlier. In these cases, both models exhibit the tailing-off effect of the column generation method \citep{desrosiers2005primer}, where convergence slows significantly as the algorithm approaches a sufficiently small optimality gap, with each iteration yielding only marginal improvements to the objective value.}

\begin{table}[]
	\centering
        \YCRminor{
	\begin{tabular}{lccccc}
		\toprule
		Product Category                 & $n$  & $m$   & $|\Tcal|$ & $T_{\CSM}$ & $T_{\RBM}$ \\ \midrule
		\textit{Beer}                    & 19 & 721 & 759,968         & 1800.0      & 1800.0      \\
		\textit{Coffee}                  & 17 & 603 & 749,867         & 1800.0      & 1800.0      \\
		\textit{Deodorant}               & 13 & 181 & 539,761         & 365.7  & 1355.4 \\
		\textit{Frozen Dinners}          & 18 & 330 & 1,963,025       & 1800.0      & 1800.0      \\
		\textit{Frozen Pizza}            & 12 & 138 & 584,406         & 111.6  & 439.7  \\
		\textit{Household Cleaners}      & 21 & 883 & 562,615         & 1800.0      & 1800.0      \\
		\textit{Hotdogs}                 & 15 & 533 & 202,842         & 1800.0      & 1800.0      \\
		\textit{Margarine/Butter}        & 11 & 27  & 282,649         & 10.4   & 20.5   \\
		\textit{Milk}                    & 18 & 347 & 476,899         & 1231.2 & 1800.0      \\
		\textit{Mustard/Ketchup}         & 16 & 644 & 266,291         & 1800.0     & 1800.0      \\
		\textit{Salty Snacks}            & 14 & 152 & 1,476,847       & 268.2  & 1786.4 \\
		\textit{Shampoo}                 & 15 & 423 & 574,711         & 1800.0     & 1800.0      \\
		\textit{Soup}                    & 17 & 315 & 1,816,879       & 1800.0      & 1800.0      \\
		\textit{Spaghetti/Italian Sauce} & 12 & 97  & 552,033         & 64.9   & 509.4  \\
		\textit{Tooth Brush}             & 15 & 699 & 392,079         & 1800.0      & 1800.0     \\ \bottomrule
	\end{tabular}
        }
	\caption{\YCRminor{Runtime comparison (in seconds) for estimating the consideration set model (CSM) and the ranking-based model (RBM) across fifteen product categories in the IRI dataset. The runtime is capped at 30 minutes (1,800 seconds).}}
	\label{tb:runtime_comparison_of_CSM-RBM}
\end{table}

\begin{figure}
	\centering
	\includegraphics[scale=1.0]{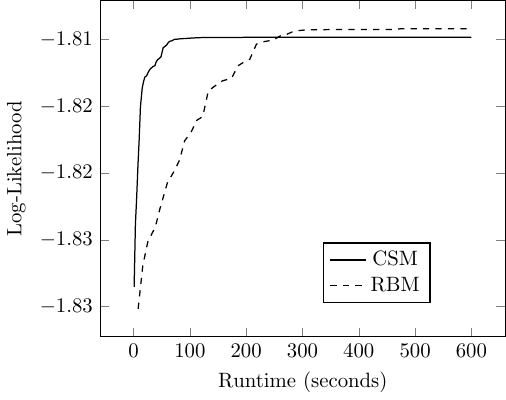}	
	\caption{Change in the in-sample log-likelihood value over time during the MLE algorithm. The y-axis represents the log-likelihood, while the x-axis shows the runtime (in seconds) for the consideration set model (solid line) and the ranking-based model (dashed line) for the \emph{Salty Snacks} category.} \label{fig:runtime_curves_saltsnck}
\end{figure}

We further highlight the computational efficiency advantage of using the consideration set model in the estimation process. In Figure~\ref{fig:runtime_curves_saltsnck}, we illustrate the changes in the log-likelihood value while estimating both the consideration set and ranking-based models for the \emph{Salty Snacks} category. Similar qualitative results are observed when considering other product categories. Specifically, we show how the in-sample log-likelihood for both models improves over time as the estimation algorithms run. From Figure~\ref{fig:runtime_curves_saltsnck}, we observe that the consideration set model’s estimation algorithm converges to a nearly optimal solution in a very short time (around 70 seconds), while the ranking-based model takes approximately four minutes to achieve the same log-likelihood value. However, as shown in the figure, the estimation algorithm of the ranking-based model eventually reaches a higher log-likelihood value than that of the consideration set model, although the improvement is marginal.

Notably, for better illustration, we extend the trajectory of the consideration set model beyond the termination point as a horizontal line. In contrast, the estimation process for the ranking-based model fully terminates at $T_{\RBM} = 1786$. seconds. As the improvement in log-likelihood per data point falls below 0.0005 after $T = 600$ seconds for the ranking-based model, this marginal change is not noticeable at the scale of Figure~\ref{fig:runtime_curves_saltsnck}. Therefore, for clarity, we plot the trajectory only up to $T = 600$ seconds. This further illustrates the tailing-off effect of the column generation method, where a substantial portion -- approximately two-thirds -- of the runtime is spent narrowing the final 0.1\% of the optimality gap.

\subsection{Symmetry of the Demand Cannibalization}
\label{subsec:numerics_symmetry_cannibalization}
As discussed in Section~\ref{subsec:modeling_by_axioms}, the symmetric cannibalization property is a key characteristic of the consideration set model. On one hand, this property differentiates the model from other members of the RUM class (e.g., the mixed MNL model) and enhances its tractability compared to the ranking-based model. On the other hand, it may limit the model's ability to fully capture customers' purchase behavior.

In this section, we empirically examine whether the symmetric cannibalization property adversely affects the predictive performance of the consideration set model when compared to the state-of-the-art mixed MNL model. First, recall that in Section~\ref{subsec:modeling_by_axioms}, we defined a choice model as consistent with the symmetric cannibalization property if and only if the following equation holds:
\begin{align}
	\label{eq:sym_cannibalization_one_side}
	\left[\mathbb{P}_{j}(S \setminus \{a_k\}) - \mathbb{P}_{j}(S) \right] -  \left[ \mathbb{P}_{k}(S \setminus \{a_j\}) - \mathbb{P}_{k}(S) \right] = 0,
\end{align}
for all assortments $S \subseteq N$ such $|S| \geq 2$ and pairs of products $j,k \in S$. To quantify the extent to which the symmetric cannibalization property is violated, we propose a \emph{cannibalization asymmetry index}, defined as:
\begin{align}
	\label{eq:asym_indx}
	\frac{1}{ | \{ S: |S| \geq 2 \} |   } \sum_{S: |S| \geq 2} \frac{1}{ {{|S|}\choose{2}}  } \sum_{ j\neq k} \bigg| \frac{ \left[\mathbb{P}_{j}(S \setminus \{a_k\}) - \mathbb{P}_{j}(S) \right] -  \left[ \mathbb{P}_{k}(S \setminus \{a_j\}) - \mathbb{P}_{k}(S) \right] }{ \mathbb{P}_{j}(S) + \mathbb{P}_{k}(S)  } \bigg|.
\end{align} 
Intuitively, this index measures the degree to which the symmetric cannibalization property, as expressed in Equation~\eqref{eq:sym_cannibalization_one_side}, is violated across all assortments $S \subseteq N$ such $|S| \geq 2$ and all pairs of products $j,k \in S$. The denominator, $\mathbb{P}_{j}(S) + \mathbb{P}_{k}(S)$, is included to ensure the comparability of violations across assortments of different sizes. For larger assortments, $\mathbb{P}_{j}(S)$ and $\mathbb{P}_{k}(S)$ tend to be smaller than the corresponding purchase probabilities in smaller assortments. This adjustment normalizes the index and mitigates the impact of assortment size on the measurement of violations. A higher cannibalization asymmetry index indicates a greater degree of violation of the symmetric cannibalization property in the choice data. Notably, if the choice data are fully consistent with the consideration set model, the cannibalization asymmetry index will be zero.

We further note that the cannibalization asymmetry index defined in Equation~\eqref{eq:asym_indx} requires access to choice probabilities for nearly all assortments in the product universe. However, as shown in Table~\ref{tb:IRI_and_SS_size}, our sales transaction data contain only a limited number of unique assortments for each product category, making it impossible to compute the index directly from the data. To address this limitation, we assume that the choice data generation process follows the mixed MNL model estimated from data (see Section~\ref{sec:numerics}). By calibrating the mixed MNL model, we can estimate the choice probabilities and compute the cannibalization asymmetry index for any collection of assortments. Note that the mixed MNL model is not constrained by the symmetric cannibalization property. Second, even when using the mixed MNL model as the ground truth for choice probabilities, calculating the cannibalization asymmetry index remains computationally challenging for product categories with a large number of items. To overcome this, we employ Monte Carlo simulations to approximate the index. Specifically, for each instance, we randomly sample an assortment $S$ such that $|S| \geq 2$, then randomly select two indices $j$ and $k$ from $S$. We perform this process for a total of 10,000 instances and compute the index as the average over these simulations.

\begin{figure}
	\centering
	\includegraphics[scale=1]{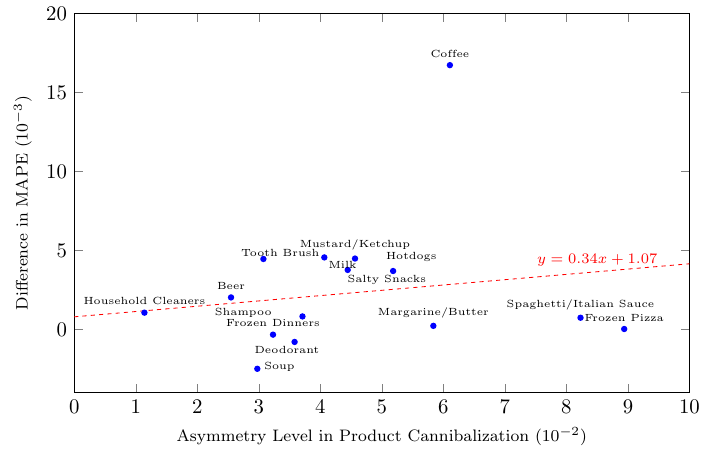}	
	\caption{Scatter plot of the improvement of mixed MNL model over consideration set model in the prediction task against the cannibalization asymmetry index, for the fifteen product categories. The higher the improvement of the mixed MNL model over the consideration set model the higher the difference in their MAPE score.} \label{fig:asym_vs_performance}
\end{figure}

Figure~\ref{fig:asym_vs_performance} presents a scatterplot showing the improvement of the mixed MNL model over the consideration set model, measured by the difference in MAPE scores, against the cannibalization asymmetry index across fifteen product categories. The plot reveals significant variation in the predictive performance improvement of the mixed MNL model over the consideration set model across different product categories. To better understand this variation, we include a linear regression trendline, represented by a red dashed line. The trend suggests a correlation between the improvement in the predictive performance of the mixed MNL model and the cannibalization asymmetry index. This finding implies that violations of the symmetric cannibalization assumption may contribute to the underperformance of the consideration set model in prediction tasks. However, the results also highlight exceptions. For example, the consideration set model performs comparably to the mixed MNL model in categories such as \emph{Spaghetti/Italian Sauce} and \emph{Frozen Pizza}, despite their high asymmetry indices. Moreover, it outperforms the mixed MNL model in categories like \emph{Soup} and \emph{Deodorant}, which have intermediate asymmetry indices. What Figure~\ref{fig:asym_vs_performance} conveys is actually a positive message: when demand cannibalization is not highly asymmetric, the consideration set model remains competitive in its predictive power and is practical for real-world use.

\section{Empirical Analysis of Revenue Performance}
\label{sec:numerics_identifiability}

In this section, we evaluate the performance of the consideration set model in a revenue management task using the IRI Dataset \citep{bronnenberg2008database}. Specifically, we focus on assortment planning as a classical revenue management application, as discussed in Section~\ref{sec:assortment_optimization}. Additionally, we explore the connection between end-to-end performance and the identifiability property established in Section~\ref{sec:model_characterization}. We demonstrate how the identifiability of a choice model limits the number of maximum-likelihood estimates that fit the data equally well, resulting in more stable operational decisions. To this end, we compare the performance of the consideration set model with the ranking-based model, which is non-identifiable.

\subsection{Uncertainty Set and Identifiability}
\label{subsec:uncertainty_set}

Following the notation from Section~\ref{subsubsec:MLE} and given the data $\{ (S_t,i_t) \}_{t \in \Tcal}$, we assume that we have the estimated parameters of the consideration set model using the MLE framework (see Section~\ref{subsec:model_estimation}). We let $\vb^*_{\CSM,S}$ be the choice probability vector over a historical assortment $S \in \Scal = \{ S_1,\ldots,S_m \}$ under this consideration set model. Herein, historical assortments are those appearing in the training dataset. The vector $\vb^*_{\CSM,S}$ has length $n+1$, where its $i$-th component represents the probability of purchasing product $i \in N^+$ from assortment $S^+$. If $i \notin S$, then the $i$-th component is zero. Next, we define $\vb^*_{\CSM}$ as the vertical concatenation of $\vb^*_{\CSM,S_1}$, ..., and $\vb^*_{\CSM,S_m}$, resulting in a vector of length $m(n+1)$. This vector can also be expressed as follows:
$$\vb^*_{\text{CSM}}  = \arg\max_{\vb} \left\lbrace \mathcal{L}(\vb) \mid \exists \, \lambdab_{\text{CSM}} \geq \zerob: \,\, \oneb^T \lambdab_{\text{CSM}} = 1, \,\, \Ab^{\text{CSM}}_S \lambdab_{\text{CSM}} = \vb_S, \,\, \forall S \in \Scal \right\rbrace,$$
where, with a slight abuse of notation, $\Ab^{\text{CSM}}_S$ is a $(n+1) \times 2^n$ matrix, which is the same as $\Ab_S$ in Problem~\eqref{problem:MLE_w_fixed_set} for $\bar{\Ccal} = 2^N$. Note that the matrix $\Ab^{\text{CSM}}_S$ maps the distribution $\lambdab_{\text{CSM}} \in \mathbb{R}^{2^n}_+$, representing the consideration set model, to the choice probability vector $\vb_S$ for each historical assortment $S \in \Scal$. The objective $\mathcal{L}$ in the given expression corresponds to the log-likelihood function described in Section~\ref{sec:estimation_method}.  We further define matrix $\Ab^{\text{CSM}}$ as the vertical concatenation of $\Ab^{\text{CSM}}_{S_1}$, $\Ab^{\text{CSM}}_{S_2}$, ..., $\Ab^{\text{CSM}}_{S_m}$, resulting in $\Ab^{\text{CSM}}$ being an $m(n+1) \times 2^n$ matrix.

Similarly, we assume that we have the estimated parameters of the ranking-based model using the MLE framework \citep{van2014market,van2017expectation}. To this end, we let $\vb^*_{\RBM,S}$ denote the choice probability vector over a historical assortment $S \in \Scal$ under this ranking-based model. Then, the vector $\vb^*_{\RBM,S}$, of length $n+1$, has its $i$-th component indicating the probability of purchasing product $i \in N^+$ from the assortment $S$, or zero if $i \notin S$. Next, we define $\vb^*_{\RBM}$ as the vertical concatenation of $\vb^*_{\RBM,S_1}, \ldots, \vb^*_{\RBM,S_m}$, resulting in a vector of length $m(n+1)$, which can be equivalently expressed as follows: 
$$\vb^*_{\text{RBM}}  = \arg\max_{\vb} \left\lbrace \mathcal{L}(\vb) \mid \exists \, \lambdab_{\text{RBM}} \geq \zerob: \,\, \oneb^T \lambdab_{\text{RBM}} = 1, \,\ \Ab^{\text{RBM}}_S \lambdab_{\text{RBM}} = \vb_S, \,\ \forall S \in \Scal \right\rbrace,$$ 
where $\Ab^{\text{RBM}}_S$ is a $(n+1) \times (n+1)!$ matrix. Note that each column of $\Ab^{\RBM}_S$ corresponds to a ranking-based purchasing decision under assortment $S$. Particularly, the matrix element $\left( \Ab^{\text{RBM}}_S  \right)_{i,\sigma} = \mathbb{I}\left[   i  = \arg\min_{j \in S^+} \sigma(j) \right]$ indicates whether product $i \in N^+$ is the most preferred one under ranking $\sigma$ in assortment $S$. Then,  $\Ab^{\text{RBM}}$ is formed by vertically concatenating $\Ab^{\text{RBM}}_{S_1}, \ldots, \Ab^{\text{RBM}}_{S_m}$. Notably, both vectors $\vb^*_{\text{CSM}}$ and $\vb^*_{\text{RBM}}$ are uniquely determined due to the strict concavity of the log-likelihood function $\Lcal$, ensuring well-defined MLE solutions for both models.

Next, we define the uncertainty sets as follows: 
\begin{align*}
\mathcal{ U  }_{\text{CSM}} & \equiv  \left\lbrace \lambdab_{\text{CSM}} \geq \zerob \, \big| \,   \Ab^{\text{CSM}} \lambdab_{\text{CSM}} = \vb^*_{\text{CSM}}  , \,\, \oneb^T \lambdab_{\text{CSM}} = 1  \right\rbrace,\\
\mathcal{ U  }_{\text{RBM}} & \equiv \left\lbrace \lambdab_{\text{RBM}} \geq \zerob \, \big| \,   \Ab^{\text{RBM}} \lambdab_{\text{RBM}} = \vb^*_{\text{RBM}}  , \,\, \oneb^T \lambdab_{\text{RBM}} = 1  \right\rbrace.
\end{align*}
These sets represent all MLE solutions for their respective models. Importantly, although $\vb^*_{\text{CSM}}$ is unique, $\Ab^{\CSM}$ may have a rank lower than $2^n$, resulting in multiple distributions $\lambdab_{\CSM}$ satisfying $\Ab^{\CSM} \lambdab_{\CSM} = \vb^*_{\text{CSM}}$. Similarly, $\Ucal_{\RBM}$ can contain multiple solutions due to the lower rank of $\Ab^{\RBM}$.

Although the dimensions of $\lambdab_{\CSM}$ and $\lambdab_{\RBM}$ differ, making it difficult to directly compare the sizes of the two uncertainty sets $\Ucal_{\CSM}$ and $\Ucal_{\RBM}$, Theorem~\ref{theorem:inference} indicates that $\Ucal_{\CSM}$ is effectively ``smaller'' due to the identifiability of the consideration set model.
As $m$ increases, $\Ucal_{\CSM}$ converges to a single solution. In contrast, the ranking-based model remains non-identifiable for $n \geq 4$ \citep{sher2011partial}, and $\Ucal_{\RBM}$ generally contains multiple solutions. This is reflected in the fact that the rank of the constraint matrix $\Ab^{\text{RBM}}$ satisfies $\text{rank}(\Ab^{\text{RBM}}) < (n+1)!$ whenever $n \geq 4$, making the system $\Ab^{\text{RBM}} \lambdab_{\RBM} = \vb^*_{\RBM}$ under-determined.

\YCRminor{
\subsection{Experiment Setup}
Notably, the multiplicity in the uncertainty set (either $\Ucal_{\text{CSM}}$ or $\Ucal_{\text{RBM}}$) can introduce significant variability when the corresponding choice model is used to make assortment decisions. Although all models within an uncertainty set achieve the same in-sample likelihood value, the optimal assortments identified by these models can differ substantially. To illustrate this phenomenon, we conduct a numerical experiment using the IRI Dataset. Below, we provide a high-level overview of the experiment, followed by a detailed explanation of each step.

{\bf Overview.} The experiment consists of the following three steps:
\begin{enumerate}
	\item[\underline{Step 1: Model Sampling:}] We sample (with replacement) a set of consideration set models, denoted as $\lambdab^{(1)}_{\CSM}, \lambdab^{(2)}_{\CSM}, \dots, \lambdab^{(\xi)}_{\CSM}$, from the uncertainty set $\Ucal_{\CSM}$. Similarly, we sample ranking-based models, $\lambdab^{(1)}_{\RBM}, \lambdab^{(2)}_{\RBM}, \dots, \lambdab^{(\xi)}_{\RBM}$, from the uncertainty set $\Ucal_{\RBM}$.
    \item [\underline{Step 2: Optimal Assortment Selection:}] For each $i \in [\xi]$, we compute the optimal assortment $S^{(i)}_{\CSM}$ under the consideration set model $\lambdab^{(i)}_{\CSM}$ and the optimal assortment $S^{(i)}_{\RBM}$ under the ranking-based model $\lambdab^{(i)}_{\RBM}$.
	\item [\underline{Step 3: Revenue Evaluation:}] We evaluate the expected revenue associated with the assortments $S^{(1)}_{\CSM}, \dots, S^{(\xi)}_{\CSM}$ and $S^{(1)}_{\RBM}, \dots, S^{(\xi)}_{\RBM}$.
\end{enumerate}

Recall that each consideration set model $\lambdab^{(i)}_{\CSM}$ is guaranteed to maximize the in-sample likelihood within the class of consideration set models. Consequently, if the MLE is used to fit a consideration set model to data, then $S^{(i)}_{\CSM}$ is a potential downstream assortment decision. Similarly, $S^{(i)}_{\RBM}$ represents a potential assortment derived from a ranking-based model estimated using the MLE approach.

We compare the revenue performance of the assortments $\{ S^{(i)}_{\CSM} \}_{i \in [\xi]}$ and $\{ S^{(i)}_{\RBM} \}_{i \in [\xi]}$. We will show that assortments derived from ranking-based models exhibit greater variability in expected revenue, with some assortments achieving substantially lower revenue compared to those derived from consideration set models.

Before presenting the results, in what follows below, we provide a detailed description of the three steps outlined above:

\emph{Details on Step 1}: Note that the uncertainty set $\Ucal_{\text{CSM}}$ is a polyhedron, with each vertex representing a consideration set model. We will sample vertices of $\Ucal_{\text{CSM}}$ using the following procedure, where $\Scal$ denotes the collection of historical assortments in the dataset:

\vspace{1em}

\begin{center}
	\fbox{
		\parbox{0.8\textwidth}{
			\paragraph{Sample a model from the uncertainty set $\Ucal_{\text{CSM}}$}
			\begin{enumerate}
				\item Select an assortment $\tilde{S} \notin \Scal$ uniformly at random.
				\item Sample a simplex vector $\tilde{\vb} \in \Delta^{|\tilde{S}|+1}$ uniformly.
				\item Solve the linear program:
                    \begin{align}
                    \underset{\lambdab}{\text{maximize}} \quad \left\lbrace  \sum_{i \in \tilde{S}^+ } \sum_{C \subseteq \Ccal} \tilde{v}_{i} \left( \Ab^{\CSM}_{\tilde{S}} \right)_{i,C} \lambda_C \,\, \Big| \,\, \lambdab \in \Ucal_{\text{CSM}} \right\rbrace. \label{problem:sampling_LP}
                    \end{align}
				\item Return the optimal solution of the linear program, denoted as $\tilde{\lambdab}_{\CSM}$.
			\end{enumerate}
		}
	}
\end{center}

\vspace{1em}

Notably, the assortment $\tilde{S}$ and its associated choice vector $\tilde{\vb}$ represent a random perturbation of the maximum likelihood solutions. When the new assortment $\tilde{S} \notin \Scal$ and its choice vector $\tilde{\vb}$ are introduced, the returned model $\tilde{\lambdab}_{\CSM} \in \Ucal_{\CSM}$ is the one that best aligns with this new information $(\tilde{S}, \tilde{\vb})$ while maintaining the in-sample likelihood.

Next, by replacing $\Ab^{\CSM}$ and $\Ucal_{\CSM}$ in the optimization problem~\eqref{problem:sampling_LP} with $\Ab^{\RBM}$ and $\Ucal_{\RBM}$, respectively, we can similarly sample a ranking-based model $\tilde{\lambdab}_{\RBM}$ from the uncertainty set $\Ucal_{\RBM}$. 

Finally, we note that the optimization problem~\eqref{problem:sampling_LP} is actually a large-scale linear program. When applied to the uncertainty set $\Ucal_{\CSM}$, the problem involves $O(2^n)$ variables; whereas for $\Ucal_{\RBM}$, the number of variables grows at a rate of $O(n!)$. Given this scale, we use the column generation method to solve the optimization problem~\eqref{problem:sampling_LP} efficiently, as directly formulating and solving the problem using commercial solvers like Gurobi \citep{gurobi} is computationally impractical. In fact, the column generation approach used for this application is analogous to the method detailed in Section~\ref{subsubsec:CG_sub}, and we omit further details for brevity.

\emph{Details on Step 2}: Finding the optimal assortments is rather straightforward. For each consideration set model $\lambdab^{(i)}_{\CSM}$, we solve the mixed-integer linear program \eqref{prob:AO_IP} to obtain its optimal assortment. Similarly, for each ranking-based model $\lambdab^{(i)}_{\RBM}$, we find the optimal assortment via a mixed-integer programming approach \citep{feldman2019assortment,bertsimas2019exact}.

\emph{Details on Step 3}: As mentioned in Section~\ref{subsec:numerics_symmetry_cannibalization}, the IRI dataset comprises real-world transaction data but obviously does not fully reveal a ground truth choice model. Consequently, we cannot directly evaluate the revenue performance of $S^{(i)}_{\CSM}$ and $S^{(i)}_{\RBM}$ for each $i \in [\xi]$. To address this limitation, we assume that the mixed MNL model estimated in Section~\ref{subsec:IRI_performance} is our ground truth choice model. Then, we let $\Pbb^{\text{MM}}(j \mid S)$ denote the predicted probability of choosing product $j$ from assortment $S$ under the estimated mixed MNL model. Consequently, the expected revenue under an assortment $S$ is computed as $\Rev^{\text{MM}} \left( S  \right) = \sum_{j \in S}  r_j \cdot \Pbb^{\text{MM}}( j \mid S) $, where $r_j$ is the revenue of product $j$. We obtain $r_j$ from the IRI dataset by averaging the selling price of product $j$ across all transactions. Using this definition, we obtain the revenues $\Rev^{\text{MM}} ( S^{(i)}_{\CSM}  )$ and $\Rev^{\text{MM}} ( S^{(i)}_{\RBM}  )$ for assortments $S^{(i)}_{\CSM}$ and $S^{(i)}_{\RBM}$, respectively. We further denote the expected revenue of the optimal assortment
$S^*_{\text{MM}}$ under the ground-truth mixed MNL model as $\Rev^* \equiv \Rev^{\text{MM}} ( S^*_{\text{MM}} ) $. As a robustness check, we will later use the estimated decision forest model as the ground truth to evaluate the expected revenue of each assortment.
}

\YCRminor{
\subsection{Results: Revenue Performance}

To begin with, we first follow the three steps outlined above to generate assortments $ S^{(i)}_{\CSM} $ and $ S^{(i)}_{\RBM} $ along with their corresponding revenues $\Rev^{\text{MM}} ( S^{(i)}_{\CSM}  )$ and $\Rev^{\text{MM}} ( S^{(i)}_{\RBM}  )$ for $i \in [\xi]$, where we set $\xi = 100$. Then, to reduce the variance when comparing the performance of these assortments, we use the same random perturbation $(\tilde{S},\tilde{v})$ to generate $ S^{(i)}_{\CSM} $ and $ S^{(i)}_{\RBM} $ for each $i$. This experiment is repeated across five product categories from the IRI Dataset:  \emph{Deodorant}, \emph{Frozen Pizza}, \emph{Margarine Butter}, \emph{Salty Snacks}, and \emph{Spaghetti/ Italian Sauce}. We chose these five categories because they have the smallest numbers of products after preprocessing (see Table~\ref{tb:IRI_and_SS_size}). Note that solving the large-scale linear program \eqref{problem:sampling_LP} under the ranking-based models via the column generation method is still computationally intensive, particularly since it is repeated $\xi = 100$ times. Therefore, we had to restrict our analysis to those five categories with smaller numbers of products.

We present the results of the experiment in Table~\ref{tb:sampling_revenue_mmnl}. Each pair of rows in the table provides the descriptive statistics of the collections of revenues, $\Rcal_{\CSM} \equiv \{  \Rev^{\text{MM}}(S^{(i)}_{\CSM}) \mid i \in [\xi] \}$ and $\Rcal_{\RBM} \equiv \{  \Rev^{\text{MM}}(S^{(i)}_{\RBM}) \mid i \in [\xi] \}$, under a specific product category. The statistics include the minimum (third column), maximum (fourth column), average (fifth column), and standard deviation (last column) for each revenue set. For example, the second row shows that for the \emph{Deodorant} category, the minimum revenue of the $\xi$ assortments generated by the ranking-based model (i.e., the minimum of $\Rcal_{\RBM}$) is 3.058, the maximum is 3.126, the average is 3.085, and the standard deviation is 0.006. 

Several key observations arise across these five product categories. First, the ranking-based model produces assortments with significantly higher variability in revenue compared to the consideration set model. For all categories, the minimum revenue of $\Rcal_{\RBM}$ is lower than that of $\Rcal_{\CSM}$, and the maximum of $\Rcal_{\RBM}$ is higher. This is further evidenced by the standard deviation of $\Rcal_{\RBM}$, which is consistently much larger than that of $\Rcal_{\CSM}$. This increased variability is expected, as the ranking-based model subsumes the consideration set model, making it more flexible. However, this flexibility can have mixed consequences. For instance, in the \emph{Spaghetti/Italian Sauce} category, the ranking-based model achieves a similar minimum revenue as the consideration set model but attains a higher maximum, resulting in better overall performance. In contrast, for categories such as \emph{Frozen Pizza}, \emph{Margarine Butter}, and \emph{Salty Snacks}, this flexibility proves detrimental. In these cases, the minimum revenue of $\Rcal_{\RBM}$ is substantially lower than that of $\Rcal_{\CSM}$, leading to a noticeably lower average revenue. For example, in the \emph{Frozen Pizza} category, the minimum revenue of $\Rcal_{\RBM}$ is less than one-third of the minimum of $\Rcal_{\CSM}$, causing its average revenue to be 25\% lower than that of $\Rcal_{\CSM}$. Overall, in four out of the five categories, the consideration set model outperforms the ranking-based model in average revenue, leading to more profitable assortments. On average, across all five categories, the mean of $\Rcal_{\CSM}$ is 9.1\% higher than that of $\Rcal_{\RBM}$.

Finally, we note that except for the \emph{Deodorant} category, the optimal assortments obtained by the consideration set model or the ranking-based model exhibit a noticeable gap when compared to the maximum revenue $\Rev^*$. It is important to clarify that this gap should not be interpreted as the standard optimality gap in the mathematical optimization literature, as each assortment is solved \emph{optimally} with respect to its corresponding choice model. Rather, this gap highlights the inherent challenges of data-driven assortment optimization in an end-to-end (i.e., data-to-decision) setting. Since the number 
$m$ of historical assortments is typically much smaller than $2^n$, the total number of possible assortments, the available data are often underspecified, which exacerbates the under-determined nature of non-parametric choice models like the ranking-based model. For instance, in the \emph{Salty Snacks} category, a ranking-based model that maximizes the likelihood of the data produces an assortment $S^{(i)}_{\RBM}$ with revenue of 1.036 -- only 40\% of $\Rev^*$.

\begin{table}[]
	\centering
	\begin{tabular}{lccccc} \toprule
		Category & Assortments & Min   & Max   & Avg.   & Std.   \\ \midrule
		Deodorant  {\tiny ($\Rev^* = 3.137$)}   &  $S^{(i)}_{\CSM}$  & 3.117 & 3.117 & 3.117 & 0.000 \\
		& $S^{(i)}_{\RBM}   $   & 3.058 & 3.126 & 3.085 & 0.006 \\ \midrule
		Frozen Pizza {\tiny ($\Rev^* = 3.714$)} & $S^{(i)}_{\CSM}$   & 3.555 & 3.555 & 3.555 & 0.000 \\
		& $S^{(i)}_{\RBM}   $   & 1.187 & 3.676 & 2.842 & 0.526 \\ \midrule
		Margarine Butter {\tiny ($\Rev^* = 2.907$)}  & $S^{(i)}_{\CSM}$   & 2.308 & 2.375 & 2.354 & 0.031 \\
		& $S^{(i)}_{\RBM}   $   & 1.689 & 2.669 & 2.102 & 0.099 \\ \midrule
		Salty Snacks {\tiny ($\Rev^* = 2.690$)} & $S^{(i)}_{\CSM}$   & 2.109 & 2.162 & 2.114 & 0.014 \\
		& $S^{(i)}_{\RBM}   $    & 1.036 & 2.208 & 1.871 & 0.400 \\ \midrule
		Spaghetti/ Italian Sauce {\tiny ($\Rev^* = 2.907$)}&$S^{(i)}_{\CSM}$  & 2.139 & 2.242 & 2.152 & 0.035 \\
		& $S^{(i)}_{\RBM}   $   & 2.131 & 2.691 & 2.272 & 0.170 \\ \bottomrule
	\end{tabular}
	\caption{Descriptive statistics (minimum, maximum, average, standard deviation) of the revenues for assortments $\{  S^{(i)}_{\CSM} \mid i \in [\xi] \}$ and $\{  S^{(i)}_{\RBM} \mid i \in [\xi] \}$, across five product categories in the IRI dataset, evaluated using the estimated mixed MNL model. \label{tb:sampling_revenue_mmnl}}
\end{table}

In summary, the consideration set model, as an identifiable and therefore more constrained class of choice models compared to the non-identifiable ranking-based model, exhibits lower variance in revenue performance. When adopting an estimate-then-optimize approach -- estimating a choice model using maximum likelihood estimation and then solving the assortment optimization problem -- the average performance of assortments generated by the consideration set model is 9\% higher than those from the ranking-based model, as measured using the mixed MNL model ground truth. These findings highlight that in data-to-decision strategies, increased model flexibility may come at the cost of greater variation in decision quality, underscoring the trade-offs inherent in using more flexible and potentially non-identifiable models. In fact, to mitigate the risks of an overly flexible ranking-based model, one approach is to adopt a robust assortment optimization method \citep{sturt2021value}, though this comes with significant computational challenges.

\subsection{Robustness Check}

As a robustness check, we also evaluate the performance of assortments $ S^{(i)}_{\CSM}$ and $ S^{(i)}_{\RBM}$ ($i \in \left[ \xi \right]$) using the decision forest model estimated in Section~\ref{subsec:IRI_performance} as the ground truth choice model. Specifically, we let $\Pbb^{\text{DF}}(j \mid S)$ be the predicted probability of choosing product $j$ from the assortment $S$ under the estimated decision forest model. Consequently, the revenue function is defined as $\Rev^{\text{DF}}(S) = \sum_{j \in S} r_j \cdot \Pbb^{\text{DF}}(j \mid S)$. Table~\ref{tb:sampling_revenue_DF} presents the descriptive statistics for $\Rcal'_{\CSM} \equiv \{ \Rev^{\text{DF}}( S^{(i)}_{\CSM}  ) \mid  i \in [\xi] \}$ and $\Rcal'_{\RBM} \equiv \{ \Rev^{\text{DF}}( S^{(i)}_{\RBM}  ) \mid  i \in [\xi] \}$, which are the collections of revenues for assortments obtained under the consideration set model and the ranking-based model, respectively, for each product category. Similarly to the previous section, we further denote the expected revenue of the optimal assortment
$S^*_{\text{DF}}$ under the ground-truth decision forest model as $\Rev^* \equiv \Rev^{\text{DF}} ( S^*_{\text{DF}} ) $.

\begin{table}[]
	\centering
	\begin{tabular}{lccccc} \toprule
		Category & Assortments & Min   & Max   & Avg.   & Std.   \\ \midrule
		Deodorant {\tiny ($\Rev^* = 3.128$)}    &  $S^{(i)}_{\CSM}$  & 2.939 & 2.939 & 2.939 & 0.000 \\
		& $S^{(i)}_{\RBM}   $   & 2.180 & 3.106 & 2.946 & 0.266 \\ \midrule
		Frozen Pizza {\tiny ($\Rev^* = 3.649$)} & $S^{(i)}_{\CSM}$   & 2.902 & 2.902 & 2.902 & 0.000 \\
		& $S^{(i)}_{\RBM}   $   & 0.294 & 2.930 & 1.621 & 0.569 \\ \midrule
		Margarine Butter {\tiny ($\Rev^* = 2.173$)} & $S^{(i)}_{\CSM}$   & 1.660 & 1.730 & 1.682 & 0.032 \\
		& $S^{(i)}_{\RBM}   $   & 0.194 & 2.139 & 1.751 & 0.468 \\ \midrule
		Salty Snacks {\tiny ($\Rev^* = 2.230$)} & $S^{(i)}_{\CSM}$   & 1.882 & 2.069 & 2.043 & 0.065 \\
		& $S^{(i)}_{\RBM}   $    & 0.115 & 2.094 & 1.455 & 0.773 \\ \midrule
		Spaghetti/ Italian Sauce  {\tiny ($\Rev^* = 2.180$)} &$S^{(i)}_{\CSM}$  & 1.324 & 1.970 & 1.886 & 0.218 \\
		& $S^{(i)}_{\RBM}   $   & 0.098 & 2.023 & 1.244 & 0.663 \\ \bottomrule
	\end{tabular}
	\caption{Descriptive statistics (minimum, maximum, average, standard deviation) of the revenues for assortments $\{  S^{(i)}_{\CSM} \mid i \in [\xi] \}$ and $\{  S^{(i)}_{\RBM} \mid i \in [\xi] \}$, across five product categories in the IRI dataset, evaluated using the estimated decision forest model. \label{tb:sampling_revenue_DF}}
\end{table}

As expected, the results in Table~\ref{tb:sampling_revenue_DF}, using the decision forest model as the ground truth, align closely with those in Table~\ref{tb:sampling_revenue_mmnl}, which use the mixed MNL model as the ground truth. Consistent with our previous findings, the ranking-based model continues to generate assortments with significantly greater revenue variability compared to the consideration set model. In fact, across all categories, the minimum revenue of $\Rcal'_{\RBM}$ is lower than that of $\Rcal'_{\CSM}$, while its maximum is higher. Notably, the minimum of $\Rcal'_{\RBM}$ is substantially lower than the minimum of $\Rcal_{\RBM}$ from Table~\ref{tb:sampling_revenue_mmnl}. 
As a result, when averaged across all five categories, the mean of $\Rcal'_{\CSM}$ is 33.4\% higher than that of $\Rcal'_{\RBM}$, representing a significant increase from the 9.1\% reported in Table~\ref{tb:sampling_revenue_mmnl}.
}

\end{document}